\documentclass[a4paper,11pt]{article}
\pdfoutput=1
\usepackage{./auxi/jheppub}
\usepackage[utf8]{inputenc}
\usepackage{comment}

\usepackage{slashed} 
\usepackage{mathtools} 
\usepackage{tikz}
\usepackage{dsfont} 
\usetikzlibrary{positioning,arrows,calc}
\usetikzlibrary{arrows.meta}
\usetikzlibrary{decorations.pathmorphing}
\usetikzlibrary{decorations.markings}
\usetikzlibrary{shapes.geometric}
\usetikzlibrary{patterns}
\usepackage{xparse}
\usepackage{diagbox} 
\usepackage{tensor} 
\usepackage{subcaption} 

\newcommand{\Op}{\mathcal{O}}

\newcommand{\Wl}{\mathcal{W}_\ell}

\newcommand{\Wm}{\mathcal{W}}

\newcommand{\vvev}[1]{\langle\!\langle\, #1 \, \rangle\!\rangle}
\newcommand{\vev}[1]{\langle\, #1 \, \rangle}

\newif\ifstartcompletesineup
\newif\ifendcompletesineup
\pgfkeys{
    /pgf/decoration/.cd,
    start up/.is if=startcompletesineup,
    start up=true,
    start up/.default=true,
    start down/.style={/pgf/decoration/start up=false},
    end up/.is if=endcompletesineup,
    end up=true,
    end up/.default=true,
    end down/.style={/pgf/decoration/end up=false}
}
\pgfdeclaredecoration{complete sines}{initial}
{
    \state{initial}[
        width=+0pt,
        next state=upsine,
        persistent precomputation={
            \ifstartcompletesineup
                \pgfkeys{/pgf/decoration automaton/next state=upsine}
                \ifendcompletesineup
                    \pgfmathsetmacro\matchinglength{
                        0.5*\pgfdecoratedinputsegmentlength / (ceil(0.5* \pgfdecoratedinputsegmentlength / \pgfdecorationsegmentlength) )
                    }
                \else
                    \pgfmathsetmacro\matchinglength{
                        0.5 * \pgfdecoratedinputsegmentlength / (ceil(0.5 * \pgfdecoratedinputsegmentlength / \pgfdecorationsegmentlength ) - 0.499)
                    }
                \fi
            \else
                \pgfkeys{/pgf/decoration automaton/next state=downsine}
                \ifendcompletesineup
                    \pgfmathsetmacro\matchinglength{
                        0.5* \pgfdecoratedinputsegmentlength / (ceil(0.5 * \pgfdecoratedinputsegmentlength / \pgfdecorationsegmentlength ) - 0.4999)
                    }
                \else
                    \pgfmathsetmacro\matchinglength{
                        0.5 * \pgfdecoratedinputsegmentlength / (ceil(0.5 * \pgfdecoratedinputsegmentlength / \pgfdecorationsegmentlength ) )
                    }
                \fi
            \fi
            \setlength{\pgfdecorationsegmentlength}{\matchinglength pt}
        }] {}
    \state{downsine}[width=\pgfdecorationsegmentlength,next state=upsine]{
        \pgfpathsine{\pgfpoint{0.5\pgfdecorationsegmentlength}{0.5\pgfdecorationsegmentamplitude}}
        \pgfpathcosine{\pgfpoint{0.5\pgfdecorationsegmentlength}{-0.5\pgfdecorationsegmentamplitude}}
    }
    \state{upsine}[width=\pgfdecorationsegmentlength,next state=downsine]{
        \pgfpathsine{\pgfpoint{0.5\pgfdecorationsegmentlength}{-0.5\pgfdecorationsegmentamplitude}}
        \pgfpathcosine{\pgfpoint{0.5\pgfdecorationsegmentlength}{0.5\pgfdecorationsegmentamplitude}}
}
    \state{final}{}
}

\definecolor{bgbox}{RGB}{255,254,230}
\definecolor{setupplane}{RGB}{230,230,230}
\definecolor{gluoncolor}{RGB}{207,54,108}
\definecolor{vertexcolor}{RGB}{53,152,219}
\definecolor{SEcolor}{RGB}{176,156,255}
\definecolor{blobcolor}{RGB}{190,180,230}
\tikzset{
corner/.style={line width=1pt,dashed,draw=black,dash pattern=on 6pt off 4pt},
scalar/.style={line width=1pt,draw=black},
gluon/.style={line width=1pt,decorate, draw=gluoncolor,
    decoration={complete sines,aspect=0,amplitude=1.25mm,segment length=1.5mm,start up,end up}},
ghost/.style={line width=1pt,loosely dotted,draw=black},
wilson/.style={line width=2pt,draw=black},
 }
\NewDocumentCommand\semiloop{O{black}mmmO{}O{above}}
{%
\draw[#1] let \p1 = ($(#3)-(#2)$) in (#3) arc (#4:({#4+180}):({0.5*veclen(\x1,\y1)})node[midway, #6] {#5};)
}

\usepackage{bbm}

\newcommand\x{.6}
\newcommand\scale{.7}
\newcommand\scaleline{1}
\newcommand\scalephi{.4}
\newcommand\scaleU{.7}
\newcommand\scaleCC{1.5}
\newcommand\scaleCB{1.8}

\newcommand{\twopointpr}{\scalebox{\scale}{\begin{tikzpicture}[baseline={([yshift=-2ex]current bounding box.center)}]
\draw[wilson] (-1.5,0) -- (1.5,0);
\draw[scalar] (-.75,.3) arc (180:0:.75);
\draw[scalar] (-.75,0) -- (-.75,.3);
\draw[scalar] (.75,0) -- (.75,.3);
\draw[fill=white] (-.75,0) circle (.1);
\draw[fill=white] (.75,0) circle (.1);
\end{tikzpicture}}}

\newcommand{\twopointnpr}{\scalebox{\scale}{\begin{tikzpicture}[baseline={([yshift=-2ex]current bounding box.center)}]
\draw[wilson] (-4.5,0) -- (1.25,0);
\draw[scalar] (-1,.3) arc (180:0:.75);
\draw[scalar] (-1,0) -- (-1,.3);
\draw[scalar] (.5,0) -- (.5,.3);
\draw[scalar] (-3.75,.3) arc (180:0:.75);
\draw[scalar] (-3.75,0) -- (-3.75,.3);
\draw[scalar] (-2.25,0) -- (-2.25,.3);
\draw[fill=white] (-1,0) circle (.1);
\draw[fill=white] (-2.25,0) circle (.1);
\end{tikzpicture}}}

\newcommand{\wilsonline}{\scalebox{\scaleline}{\begin{tikzpicture}[baseline={([yshift=-2ex]current bounding box.center)}]
\draw[wilson] (-5,0) -- (3,0);
\draw[fill=white] (-4,0) circle (.1);
\draw[fill=white] (-2.5,0) circle (.1);
\draw[fill=white] (0.5,0) circle (.1);
\draw[fill=white] (2,0) circle (.1);
\node[] at (-1,.4) {$\dots$};
\node[] at (-4,.4) {$\tau_1$};
\node[] at (-2.5,.4) {$\tau_2$};
\node[] at (0.5,.4) {$\tau_{n-1}$};
\node[] at (2,.4) {$\tau_{n}$};
\node[] at (-4,-.6) {$\phi^{I_1}$};
\node[] at (-2.5,-.6)  {$\phi^{I_2}$};
\node[] at (0.5,-.6)  {$\phi^{I_{n-1}}$};
\node[] at (2,-.6)  {$\phi^{I_n}$};
\node[] at (3.5,0) {$\Wl$};
\end{tikzpicture}}}

\newcommand{\twopttree}{\scalebox{\scale}{\begin{tikzpicture}[baseline={([yshift=-2.5ex]current bounding box.center)}]
\draw[wilson] (-2,0) -- (2,0);
\draw[scalar] (-1,0) arc (180:0:1);
\draw[fill=white] (-1,0) circle (.1);
\draw[fill=white] (1,0) circle (.1);
\end{tikzpicture}}}
\newcommand{\twoptSE}{\scalebox{\scale}{\begin{tikzpicture}[baseline={([yshift=-3ex]current bounding box.center)}]
\draw[wilson] (-2,0) -- (2,0);
\draw[scalar] (-1,0) arc (180:0:1);
\draw[fill=white] (-1,0) circle (.1);
\draw[fill=white] (1,0) circle (.1);
\draw[fill=SEcolor,SEcolor] (0,1) circle (.15);
\end{tikzpicture}}}
\newcommand{\twoptY}{\scalebox{\scale}{\begin{tikzpicture}[baseline={([yshift=-4ex]current bounding box.center)}]
\draw[wilson] (-2,0) -- (2,0);
\draw[scalar] (-1,0) arc (180:0:1);
\draw[fill=white] (-1,0) circle (.1);
\draw[fill=white] (1,0) circle (.1);
\draw[gluon] (0,1) -- (0,1.5);
\draw[fill=vertexcolor,vertexcolor] (0,1) circle (.075);
\draw[fill=gluoncolor,gluoncolor] (-1.5,0) circle (.08);
\draw[fill=gluoncolor,gluoncolor] (0,0) circle (.08);
\draw[fill=gluoncolor,gluoncolor] (1.5,0) circle (.08);
\end{tikzpicture}}}

\newcommand{\twoptUone}{\scalebox{\scale}{\begin{tikzpicture}[baseline={([yshift=-3.5ex]current bounding box.center)}]
\draw[wilson] (-2,0) -- (2,0);
\draw[scalar] (-1.6,0) arc (150:30:1.85);
\draw[scalar] (-.2,0) arc (210:-20:.23);
\draw[fill=white] (.2,0) circle (.1);
\draw[fill=white] (1.6,0) circle (.1);
\end{tikzpicture}}}
\newcommand{\twoptUtwo}{\scalebox{\scale}{\begin{tikzpicture}[baseline={([yshift=-2ex]current bounding box.center)}]
\draw[wilson] (-2,0) -- (2,0);
\draw[scalar] (-1.6,0) arc (180:0:.7);
\draw[scalar] (.2,0) arc (180:0:.7);
\draw[fill=white] (-.2,0) circle (.1);
\draw[fill=white] (1.6,0) circle (.1);
\end{tikzpicture}}}
\newcommand{\twoptUthree}{\scalebox{\scale}{\begin{tikzpicture}[baseline={([yshift=-2ex]current bounding box.center)}]
\draw[wilson] (-2,0) -- (2,0);
\draw[scalar] (-1.6,0) arc (180:0:.7);
\draw[scalar] (.2,0) arc (180:0:.7);
\draw[fill=white] (-.2,0) circle (.1);
\draw[fill=white] (.2,0) circle (.1);
\end{tikzpicture}}}
\newcommand{\twoptUfour}{\scalebox{\scale}{\begin{tikzpicture}[baseline={([yshift=-3.5ex]current bounding box.center)}]
\draw[wilson] (-2,0) -- (2,0);
\draw[scalar] (-1.6,0) arc (150:30:1.85);
\draw[scalar] (-.2,0) arc (210:-20:.23);
\draw[fill=white] (-.2,0) circle (.1);
\draw[fill=white] (1.6,0) circle (.1);
\end{tikzpicture}}}
\newcommand{\twoptUfive}{\scalebox{\scale}{\begin{tikzpicture}[baseline={([yshift=-2ex]current bounding box.center)}]
\draw[wilson] (-2,0) -- (2,0);
\draw[scalar] (-1.6,0) arc (180:0:.7);
\draw[scalar] (.2,0) arc (180:0:.7);
\draw[fill=white] (-1.6,0) circle (.1);
\draw[fill=white] (1.6,0) circle (.1);
\end{tikzpicture}}}
\newcommand{\twoptUsix}{\scalebox{\scale}{\begin{tikzpicture}[baseline={([yshift=-2ex]current bounding box.center)}]
\draw[wilson] (-2,0) -- (2,0);
\draw[scalar] (-1.6,0) arc (180:0:.7);
\draw[scalar] (.2,0) arc (180:0:.7);
\draw[fill=white] (-1.6,0) circle (.1);
\draw[fill=white] (.2,0) circle (.1);
\end{tikzpicture}}}
\newcommand{\twoptUseven}{\scalebox{\scale}{\begin{tikzpicture}[baseline={([yshift=-3.5ex]current bounding box.center)}]
\draw[wilson] (-2,0) -- (2,0);
\draw[scalar] (-1.6,0) arc (150:30:1.85);
\draw[scalar] (-.2,0) arc (210:-20:.23);
\draw[fill=white] (-1.6,0) circle (.1);
\draw[fill=white] (.2,0) circle (.1);
\end{tikzpicture}}}
\newcommand{\twoptUeight}{\scalebox{\scale}{\begin{tikzpicture}[baseline={([yshift=-3.5ex]current bounding box.center)}]
\draw[wilson] (-2,0) -- (2,0);
\draw[scalar] (-1.6,0) arc (150:30:1.85);
\draw[scalar] (-.2,0) arc (210:-20:.23);
\draw[fill=white] (-1.6,0) circle (.1);
\draw[fill=white] (-.2,0) circle (.1);
\end{tikzpicture}}}

\newcommand{\twoptOijLO}{\scalebox{\scale}{\begin{tikzpicture}[baseline={([yshift=-2ex]current bounding box.center)}]
\draw[wilson] (-1.5,0) -- (1.5,0);
\draw[scalar] (-.75,.3) arc (180:0:.75);
\draw[scalar] (-.75,0) arc (170:10:.77);
\draw[scalar] (-.75,0) -- (-.75,.3);
\draw[scalar] (.75,0) -- (.75,.3);
\draw[fill=white] (-.75,0) circle (.1);
\draw[fill=white] (.75,0) circle (.1);
\node[] at (-.75,-.4) {\small $ij$};
\node[] at (.75,-.4) {\small $kl$};
\end{tikzpicture}}}

\newcommand{\fourpttreeone}{\scalebox{\scale}{\begin{tikzpicture}[baseline={([yshift=-0ex]current bounding box.center)}]
\draw[wilson] (-2,0) -- (2,0);
\draw[scalar] (-1.6,0) arc (180:0:.7);
\draw[scalar] (.2,0) arc (180:0:.7);
\draw[fill=white] (-1.6,0) circle (.1);
\draw[fill=white] (-.2,0) circle (.1);
\draw[fill=white] (.2,0) circle (.1);
\draw[fill=white] (1.6,0) circle (.1);
\node[] at (-1.6,-.4) {\small $i$};
\node[] at (-.2,-.4) {\small $j$};
\node[] at (.2,-.4) {\small $k$};
\node[] at (1.6,-.4) {\small $l$};
\end{tikzpicture}}}
\newcommand{\fourpttreetwo}{\scalebox{\scale}{\begin{tikzpicture}[baseline={([yshift=-1.5ex]current bounding box.center)}]
\draw[wilson] (-2,0) -- (2,0);
\draw[scalar] (-1.6,0) arc (150:30:1.85);
\draw[scalar] (-.2,0) arc (210:-20:.23);
\draw[fill=white] (-1.6,0) circle (.1);
\draw[fill=white] (-.2,0) circle (.1);
\draw[fill=white] (.2,0) circle (.1);
\draw[fill=white] (1.6,0) circle (.1);
\node[] at (-1.6,-.4) {\small $i$};
\node[] at (-.2,-.4) {\small $j$};
\node[] at (.2,-.4) {\small $k$};
\node[] at (1.6,-.4) {\small $l$};
\end{tikzpicture}}}

\newcommand{\fourptSEone}{\scalebox{\scale}{\begin{tikzpicture}[baseline={([yshift=-0ex]current bounding box.center)}]
\draw[wilson] (-2,0) -- (2,0);
\draw[scalar] (-1.6,0) arc (180:0:.7);
\draw[scalar] (.2,0) arc (180:0:.7);
\draw[fill=white] (-1.6,0) circle (.1);
\draw[fill=white] (-.2,0) circle (.1);
\draw[fill=white] (.2,0) circle (.1);
\draw[fill=white] (1.6,0) circle (.1);
\draw[fill=SEcolor,SEcolor] (-.9,.7) circle (.15);
\node[] at (-1.6,-.4) {\small $i$};
\node[] at (-.2,-.4) {\small $j$};
\node[] at (.2,-.4) {\small $k$};
\node[] at (1.6,-.4) {\small $l$};
\end{tikzpicture}}}
\newcommand{\fourptSEtwo}{\scalebox{\scale}{\begin{tikzpicture}[baseline={([yshift=-0ex]current bounding box.center)}]
\draw[wilson] (-2,0) -- (2,0);
\draw[scalar] (-1.6,0) arc (180:0:.7);
\draw[scalar] (.2,0) arc (180:0:.7);
\draw[fill=white] (-1.6,0) circle (.1);
\draw[fill=white] (-.2,0) circle (.1);
\draw[fill=white] (.2,0) circle (.1);
\draw[fill=white] (1.6,0) circle (.1);
\draw[fill=SEcolor,SEcolor] (.9,.7) circle (.15);
\node[] at (-1.6,-.4) {\small $i$};
\node[] at (-.2,-.4) {\small $j$};
\node[] at (.2,-.4) {\small $k$};
\node[] at (1.6,-.4) {\small $l$};
\end{tikzpicture}}}
\newcommand{\fourptSEthree}{\scalebox{\scale}{\begin{tikzpicture}[baseline={([yshift=-0ex]current bounding box.center)}]
\draw[wilson] (-2,0) -- (2,0);
\draw[scalar] (-1.6,0) arc (150:30:1.85);
\draw[scalar] (-.2,0) arc (210:-20:.23);
\draw[fill=white] (-1.6,0) circle (.1);
\draw[fill=white] (-.2,0) circle (.1);
\draw[fill=white] (.2,0) circle (.1);
\draw[fill=white] (1.6,0) circle (.1);
\draw[fill=SEcolor,SEcolor] (0,.9) circle (.15);
\node[] at (-1.6,-.4) {\small $i$};
\node[] at (-.2,-.4) {\small $j$};
\node[] at (.2,-.4) {\small $k$};
\node[] at (1.6,-.4) {\small $l$};
\end{tikzpicture}}}
\newcommand{\fourptSEfour}{\scalebox{\scale}{\begin{tikzpicture}[baseline={([yshift=-0ex]current bounding box.center)}]
\draw[wilson] (-2,0) -- (2,0);
\draw[scalar] (-1.6,0) arc (150:30:1.85);
\draw[scalar] (-.2,0) arc (210:-20:.23);
\draw[fill=white] (-1.6,0) circle (.1);
\draw[fill=white] (-.2,0) circle (.1);
\draw[fill=white] (.2,0) circle (.1);
\draw[fill=white] (1.6,0) circle (.1);
\draw[fill=SEcolor,SEcolor] (0,.33) circle (.15);
\node[] at (-1.6,-.4) {\small $i$};
\node[] at (-.2,-.4) {\small $j$};
\node[] at (.2,-.4) {\small $k$};
\node[] at (1.6,-.4) {\small $l$};
\end{tikzpicture}}}
\newcommand{\fourptX}{\scalebox{\scale}{\begin{tikzpicture}[baseline={([yshift=-0ex]current bounding box.center)}]
\draw[wilson] (-2,0) -- (2,0);
\draw[scalar] (-1.6,0) arc (180:0:.9);
\draw[scalar] (-.2,0) arc (180:0:.9);
\draw[fill=white] (-1.6,0) circle (.1);
\draw[fill=white] (-.2,0) circle (.1);
\draw[fill=white] (.2,0) circle (.1);
\draw[fill=white] (1.6,0) circle (.1);
\draw[fill=vertexcolor,vertexcolor] (0,.55) circle (.075);
\node[] at (-1.6,-.4) {\small $i$};
\node[] at (-.2,-.4) {\small $j$};
\node[] at (.2,-.4) {\small $k$};
\node[] at (1.6,-.4) {\small $l$};
\end{tikzpicture}}}
\newcommand{\fourptHone}{\scalebox{\scale}{\begin{tikzpicture}[baseline={([yshift=-1ex]current bounding box.center)}]
\draw[wilson] (-2,0) -- (2,0);
\draw[scalar] (-1.6,0) arc (180:0:.7);
\draw[scalar] (.2,0) arc (180:0:.7);
\draw[gluon] (-.8,.68) arc (145:20:.95);
\draw[fill=white] (-1.6,0) circle (.1);
\draw[fill=white] (-.2,0) circle (.1);
\draw[fill=white] (.2,0) circle (.1);
\draw[fill=white] (1.6,0) circle (.1);
\draw[fill=vertexcolor,vertexcolor] (-.8,.68) circle (.075);
\draw[fill=vertexcolor,vertexcolor] (.8,.68) circle (.075);
\node[] at (-1.6,-.4) {\small $i$};
\node[] at (-.2,-.4) {\small $j$};
\node[] at (.2,-.4) {\small $k$};
\node[] at (1.6,-.4) {\small $l$};
\end{tikzpicture}}}
\newcommand{\fourptHtwo}{\scalebox{\scale}{\begin{tikzpicture}[baseline={([yshift=-1ex]current bounding box.center)}]
\draw[wilson] (-2,0) -- (2,0);
\draw[scalar] (-1.6,0) arc (150:30:1.85);
\draw[scalar] (-.2,0) arc (210:-20:.23);
\draw[gluon] (0,.3) -- (0,.9);
\draw[fill=white] (-1.6,0) circle (.1);
\draw[fill=white] (-.2,0) circle (.1);
\draw[fill=white] (.2,0) circle (.1);
\draw[fill=white] (1.6,0) circle (.1);
\draw[fill=vertexcolor,vertexcolor] (0,.91) circle (.075);
\draw[fill=vertexcolor,vertexcolor] (0,.33) circle (.075);
\node[] at (-1.6,-.4) {\small $i$};
\node[] at (-.2,-.4) {\small $j$};
\node[] at (.2,-.4) {\small $k$};
\node[] at (1.6,-.4) {\small $l$};
\end{tikzpicture}}}
\newcommand{\fourptYone}{\scalebox{\scale}{\begin{tikzpicture}[baseline={([yshift=-0.7ex]current bounding box.center)}]
\draw[wilson] (-2,0) -- (2,0);
\draw[scalar] (-1.6,0) arc (180:0:.7);
\draw[scalar] (.2,0) arc (180:0:.7);
\draw[gluon] (-.9,.7) -- (-.9,1.1);
\draw[fill=white] (-1.6,0) circle (.1);
\draw[fill=white] (-.2,0) circle (.1);
\draw[fill=white] (.2,0) circle (.1);
\draw[fill=white] (1.6,0) circle (.1);
\draw[fill=gluoncolor,gluoncolor] (-1.85,0) circle (.08);
\draw[fill=gluoncolor,gluoncolor] (-.9,0) circle (.08);
\draw[fill=gluoncolor,gluoncolor] (0,0) circle (.08);
\draw[fill=gluoncolor,gluoncolor] (1.85,0) circle (.08);
\draw[fill=vertexcolor,vertexcolor] (-.9,.7) circle (.075);
\node[] at (-1.6,-.4) {\small $i$};
\node[] at (-.2,-.4) {\small $j$};
\node[] at (.2,-.4) {\small $k$};
\node[] at (1.6,-.4) {\small $l$};
\end{tikzpicture}}}
\newcommand{\fourptYtwo}{\scalebox{\scale}{\begin{tikzpicture}[baseline={([yshift=-0.7ex]current bounding box.center)}]
\draw[wilson] (-2,0) -- (2,0);
\draw[scalar] (-1.6,0) arc (180:0:.7);
\draw[scalar] (.2,0) arc (180:0:.7);
\draw[gluon] (.9,.7) -- (.9,1.1);
\draw[fill=white] (-1.6,0) circle (.1);
\draw[fill=white] (-.2,0) circle (.1);
\draw[fill=white] (.2,0) circle (.1);
\draw[fill=white] (1.6,0) circle (.1);
\draw[fill=gluoncolor,gluoncolor] (-1.85,0) circle (.08);
\draw[fill=gluoncolor,gluoncolor] (0,0) circle (.08);
\draw[fill=gluoncolor,gluoncolor] (.9,0) circle (.08);
\draw[fill=gluoncolor,gluoncolor] (1.85,0) circle (.08);
\draw[fill=vertexcolor,vertexcolor] (.9,.7) circle (.075);
\node[] at (-1.6,-.4) {\small $i$};
\node[] at (-.2,-.4) {\small $j$};
\node[] at (.2,-.4) {\small $k$};
\node[] at (1.6,-.4) {\small $l$};
\end{tikzpicture}}}
\newcommand{\fourptYthree}{\scalebox{\scale}{\begin{tikzpicture}[baseline={([yshift=-0.2ex]current bounding box.center)}]
\draw[wilson] (-2,0) -- (2,0);
\draw[scalar] (-1.6,0) arc (150:30:1.85);
\draw[scalar] (-.2,0) arc (210:-20:.23);
\draw[gluon] (0,.9) -- (0,1.3);
\draw[fill=white] (-1.6,0) circle (.1);
\draw[fill=white] (-.2,0) circle (.1);
\draw[fill=white] (.2,0) circle (.1);
\draw[fill=white] (1.6,0) circle (.1);
\draw[fill=gluoncolor,gluoncolor] (-1.85,0) circle (.08);
\draw[fill=gluoncolor,gluoncolor] (-.9,0) circle (.08);
\draw[fill=gluoncolor,gluoncolor] (.9,0) circle (.08);
\draw[fill=gluoncolor,gluoncolor] (1.85,0) circle (.08);
\draw[fill=vertexcolor,vertexcolor] (0,.9) circle (.075);
\node[] at (-1.6,-.4) {\small $i$};
\node[] at (-.2,-.4) {\small $j$};
\node[] at (.2,-.4) {\small $k$};
\node[] at (1.6,-.4) {\small $l$};
\end{tikzpicture}}}
\newcommand{\fourptYfour}{\scalebox{\scale}{\begin{tikzpicture}[baseline={([yshift=-0ex]current bounding box.center)}]
\draw[wilson] (-2,0) -- (2,0);
\draw[scalar] (-1.6,0) arc (150:30:1.85);
\draw[scalar] (-.2,0) arc (210:-20:.23);
\draw[gluon] (0,.33) -- (0,.73);
\draw[fill=white] (-1.6,0) circle (.1);
\draw[fill=white] (-.2,0) circle (.1);
\draw[fill=white] (.2,0) circle (.1);
\draw[fill=white] (1.6,0) circle (.1);
\draw[fill=gluoncolor,gluoncolor] (-.9,0) circle (.08);
\draw[fill=gluoncolor,gluoncolor] (0,0) circle (.08);
\draw[fill=gluoncolor,gluoncolor] (.9,0) circle (.08);
\draw[fill=vertexcolor,vertexcolor] (0,.33) circle (.075);
\node[] at (-1.6,-.4) {\small $i$};
\node[] at (-.2,-.4) {\small $j$};
\node[] at (.2,-.4) {\small $k$};
\node[] at (1.6,-.4) {\small $l$};
\end{tikzpicture}}}

\newcommand{\recursiontree}{\scalebox{\scale}{\begin{tikzpicture}[baseline={([yshift=0ex]current bounding box.center)}]
\draw[wilson] (-2,0) -- (2,0);
\draw[scalar] (-1.5,0) arc (180:0:1);
\draw[scalar] (-.8,.3) arc (180:0:.3);
\draw[scalar] (-.8,0) -- (-.8,.3);
\draw[scalar] (-.2,0) -- (-.2,.3);
\draw[scalar] (.95,.3) arc (180:0:.3);
\draw[scalar] (.95,0) -- (.95,.3);
\draw[scalar] (1.55,0) -- (1.55,.3);
\draw[fill=white] (-1.5,0) circle (.1);
\draw[fill=white] (.5,0) circle (.1);
\node[] at (-1.5,-.4) {\small $1$};
\node[] at (.5,-.4) {\small $2j+2$};
\node[] at (-.5,.25) {\small $t$};
\node[] at (1.25,.25) {\small $t$};
\end{tikzpicture}}}
\newcommand{\treelevelblob}{\scalebox{\scale}{\begin{tikzpicture}[baseline={([yshift=-1.7ex]current bounding box.center)}]
\draw[wilson] (-1,0) -- (0,0);
\draw[scalar] (-.8,.3) arc (180:0:.3);
\draw[scalar] (-.8,0) -- (-.8,.3);
\draw[scalar] (-.2,0) -- (-.2,.3);
\node[] at (-.5,.25) {\small $t$};
\end{tikzpicture}}}
\newcommand{\treelevelblobodd}{\scalebox{\scale}{\begin{tikzpicture}[baseline={([yshift=-1.7ex]current bounding box.center)}]
\draw[wilson] (-1,0) -- (0,0);
\draw[scalar] (-.8,.3) arc (180:0:.3);
\draw[scalar] (-.8,0) -- (-.8,.3);
\draw[scalar] (-.2,0) -- (-.2,.3);
\node[] at (-.5,.25) {\small $t_O$};
\end{tikzpicture}}}

\newcommand{\propagatorS}{\begin{tikzpicture}[baseline={([yshift=-.35ex]current bounding box.center)}]
\draw[scalar] (0,.5) -- (1,.5);
\draw[fill=white] (0,.5) circle (.075);
\draw[fill=white] (1,.5) circle (.075);
\node[] at (0,.1) {\footnotesize \makebox[\widthof{$\smash{\mu,a}$}][c]{$\smash{i,a}$}};
\node[] at (0,.9) {\footnotesize $1$};
\node[] at (1,.1) {\footnotesize \makebox[\widthof{$\smash{\mu,a}$}][c]{$\smash{j,b}$}};
\node[] at (1,.9) {\footnotesize $2$};
\end{tikzpicture}}
\newcommand{\propagatorG}{\begin{tikzpicture}[baseline={([yshift=-.35ex]current bounding box.center)}]
\draw[gluon] (0,.5) -- (1,.5);
\draw[fill=white] (0,.5) circle (.075);
\draw[fill=white] (1,.5) circle (.075);
\node[] at (0,.1) {\footnotesize \makebox[\widthof{$\mu,a$}][c]{$\smash{\mu,a}$}};
\node[] at (0,.9) {\footnotesize $1$};
\node[] at (1,.1) {\footnotesize \makebox[\widthof{$\nu,b$}][c]{$\smash{\nu,b}$}};
\node[] at (1,.9) {\footnotesize $2$};
\end{tikzpicture}}
\newcommand{\propagatorF}{\begin{tikzpicture}[baseline={([yshift=-.35ex]current bounding box.center)}]
\draw[scalar] (0,.5) -- (1,.5);
\draw[fill=white] (0,.5) circle (.075);
\draw[fill=white] (1,.5) circle (.075);
\draw[-Latex] (.62,.5)--(.63,.5);
\node[] at (0,.1) {\footnotesize \makebox[\widthof{$\smash{\mu,a}$}][c]{$\smash{a}$}};
\node[] at (0,.9) {\footnotesize $1$};
\node[] at (1,.1) {\footnotesize \makebox[\widthof{$\smash{\nu,b}$}][c]{$\smash{b}$}};
\node[] at (1,.9) {\footnotesize $2$};
\end{tikzpicture}}
\newcommand{\propagatorGh}{\begin{tikzpicture}[baseline={([yshift=-.35ex]current bounding box.center)}]
\draw[ghost] (0,.5) -- (1,.5);
\draw[fill=white] (0,.5) circle (.075);
\draw[fill=white] (1,.5) circle (.075);
\node[] at (0,.1) {\footnotesize \makebox[\widthof{$\smash{\mu,a}$}][c]{$\smash{a}$}};
\node[] at (0,.9) {\footnotesize $1$};
\node[] at (1,.1) {\footnotesize \makebox[\widthof{$\smash{\nu,b}$}][c]{$\smash{b}$}};
\node[] at (1,.9) {\footnotesize $2$};
\end{tikzpicture}}

\newcommand{\propagatorSSEnotext}{\begin{tikzpicture}[baseline={([yshift=-.35ex]current bounding box.center)}]
\draw[scalar] (0,.5) -- (1,.5);
\draw[fill=white] (0,.5) circle (.075);
\draw[fill=white] (1,.5) circle (.075);
\draw[fill=SEcolor,SEcolor] (.5,.5) circle (.15);
\end{tikzpicture}}
\newcommand{\SSEone}{\begin{tikzpicture}[baseline={([yshift=-.4ex]current bounding box.center)}]
\draw[scalar] (0,.5) -- (2,.5);
\draw[gluon] (.5,.5) arc (180:-20:.5);
\draw[fill=white] (0,.5) circle (.075);
\draw[fill=white] (2,.5) circle (.075);
\draw[fill=vertexcolor,vertexcolor] (.5,.5) circle (.075);
\draw[fill=vertexcolor,vertexcolor] (1.5,.5) circle (.075);
\end{tikzpicture}}
\newcommand{\SSEtwo}{\begin{tikzpicture}[baseline={([yshift=-.5ex]current bounding box.center)}]
\draw[scalar] (0,.5) -- (.5,.5);
\draw[scalar] (1.5,.5) -- (2,.5);
\draw[scalar,postaction={decorate}] (.5,.5) arc (180:0:.5);
\draw[-Latex] (1.1,1)--(1.11,1);
\draw[scalar,postaction={decorate}] (.5,.5) arc (-180:0:.5);
\draw[-Latex] (.9,0)--(.89,0);
\draw[fill=white] (0,.5) circle (.075);
\draw[fill=white] (2,.5) circle (.075);
\draw[fill=vertexcolor,vertexcolor] (.5,.5) circle (.075);
\draw[fill=vertexcolor,vertexcolor] (1.5,.5) circle (.075);
\end{tikzpicture}}
\newcommand{\SSEthree}{\begin{tikzpicture}[baseline={([yshift=-.75ex]current bounding box.center)}]
\draw[scalar] (0,.5) -- (2,.5);
\draw[scalar] (1,1) circle (.5);
\draw[fill=white] (0,.5) circle (.075);
\draw[fill=white] (2,.5) circle (.075);
\draw[fill=vertexcolor,vertexcolor] (1,.5) circle (.075);
\end{tikzpicture}}
\newcommand{\SSEfour}{\begin{tikzpicture}[baseline={([yshift=-1.05ex]current bounding box.center)}]
\draw[scalar] (0,.5) -- (2,.5);
\draw[gluon] (1,1) circle (.5);
\draw[fill=white] (0,.5) circle (.075);
\draw[fill=white] (2,.5) circle (.075);
\draw[fill=vertexcolor,vertexcolor] (1,.5) circle (.075);
\end{tikzpicture}}
\newcommand{\vertexSSG}{\begin{tikzpicture}[baseline={([yshift=-.35ex]current bounding box.center)}]
\draw[gluon] (0,.5) -- (.75,.5);
\draw[scalar] (.75,.5) -- (1.25,1);
\draw[scalar] (.75,.5) -- (1.25,0);
\draw[fill=vertexcolor,vertexcolor] (.75,.5) circle (.075);
\draw[fill=white] (0,.5) circle (.075);
\draw[fill=white] (1.25,1) circle (.075);
\draw[fill=white] (1.25,0) circle (.075);
\node[anchor=west] at (1.3,1) {\footnotesize $i,a$};
\node[] at (1.25,1.3) {\footnotesize $1$};
\node[anchor=west] at (1.3,0) {\footnotesize $j,b$};
\node[] at (1.25,-.3) {\footnotesize $2$};
\node[] at (0,.1) {\footnotesize $\mu,c$};
\node[] at (0,.9) {\footnotesize $3$};
\end{tikzpicture}}
\newcommand{\vertexSSSS}{\begin{tikzpicture}[baseline={([yshift=-.35ex]current bounding box.center)}]
\draw[scalar] (0,0) -- (1.25,1);
\draw[scalar] (0,1) -- (1.25,0);
\draw[fill=vertexcolor,vertexcolor] (.625,.5) circle (.075);
\draw[fill=white] (0,0) circle (.075);
\draw[fill=white] (0,1) circle (.075);
\draw[fill=white] (1.25,1) circle (.075);
\draw[fill=white] (1.25,0) circle (.075);
\node[anchor=west] at (1.3,1) {\footnotesize $i,a$};
\node[] at (1.25,1.3) {\footnotesize $1$};
\node[anchor=west] at (1.3,0) {\footnotesize $j,b$};
\node[] at (1.25,-.3) {\footnotesize $2$};
\node[anchor=east] at (-0.05,1) {\footnotesize $k,c$};
\node[] at (0,1.3) {\footnotesize $3$};
\node[anchor=east] at (-0.05,0) {\footnotesize $l,d$};
\node[] at (0,-.3) {\footnotesize $4$};
\end{tikzpicture}}

\newcommand{\Hinsertone}{\begin{tikzpicture}[baseline={([yshift=-.55ex]current bounding box.center)}]
\draw[scalar] (0,0) -- (1.25,0);
\draw[scalar] (0,1) -- (1.25,1);
\draw[gluon] (.625,0) -- (.625,1);
\draw[fill=vertexcolor,vertexcolor] (.625,0) circle (.075);
\draw[fill=vertexcolor,vertexcolor] (.625,1) circle (.075);
\draw[fill=white] (0,0) circle (.075);
\draw[fill=white] (0,1) circle (.075);
\draw[fill=white] (1.25,1) circle (.075);
\draw[fill=white] (1.25,0) circle (.075);
\node[anchor=west] at (1.3,1) {\footnotesize $i,a$};
\node[] at (1.25,1.3) {\footnotesize $1$};
\node[anchor=west] at (1.3,0) {\footnotesize $j,b$};
\node[] at (1.25,-.3) {\footnotesize $2$};
\node[anchor=east] at (-0.05,1) {\footnotesize $k,c$};
\node[] at (0,1.3) {\footnotesize $3$};
\node[anchor=east] at (-0.05,0) {\footnotesize $l,d$};
\node[] at (0,-.3) {\footnotesize $4$};
\end{tikzpicture}}

\newcommand{\diagYonetwotwo}{\scalebox{\x}{\begin{tikzpicture}[baseline={([yshift=-.55ex]current bounding box.center)}]
\draw[scalar] (0,.5) -- (1,.5);
\draw[scalar] (1.5,.5) circle (.5);
\draw[fill=white] (0,.5) circle (.075);
\draw[fill=white] (2,.5) circle (.075);
\draw[fill=vertexcolor,vertexcolor] (1,.5) circle (.075);
\end{tikzpicture}}}
\newcommand{\diagXoneonetwothree}{\scalebox{\x}{\begin{tikzpicture}[baseline={([yshift=-.55ex]current bounding box.center)}]
\draw[scalar] (.5,1) -- (1,.5);
\draw[scalar] (.5,0) -- (1,.5);
\draw[scalar] (1.5,.5) circle (.5);
\draw[fill=white] (.5,1) circle (.075);
\draw[fill=white] (.5,0) circle (.075);
\draw[fill=white] (2,.5) circle (.075);
\draw[fill=vertexcolor,vertexcolor] (1,.5) circle (.075);
\end{tikzpicture}}}

\newcommand{\recursionAodd}{\scalebox{\scale}{\begin{tikzpicture}[baseline={([yshift=-1ex]current bounding box.center)}]
\draw[wilson] (-2.5,0) -- (2.5,0);
\draw[scalar] (-.75,.3) arc (180:0:.75);
\draw[scalar] (-.75,0) -- (-.75,.3);
\draw[scalar] (.75,0) -- (.75,.3);
\draw[scalar] (-1.8,.3) arc (180:0:.3);
\draw[scalar] (-1.8,0) -- (-1.8,.3);
\draw[scalar] (-1.2,0) -- (-1.2,.3);
\draw[scalar] (-.3,.3) arc (180:0:.3);
\draw[scalar] (-.3,0) -- (-.3,.3);
\draw[scalar] (.3,0) -- (.3,.3);
\draw[scalar] (1.2,.3) arc (180:0:.3);
\draw[scalar] (1.2,0) -- (1.2,.3);
\draw[scalar] (1.8,0) -- (1.8,.3);
\draw[fill=white] (.75,0) circle (.1);
\node[] at (-1.5,.25) {\small $t$};
\node[] at (0,.25) {\small $t$};
\node[] at (1.5,.25) {\small $t$};
\node[] at (-1.25,-.4) {\small $2j$};
\node[] at (.75,-.4) {\small $i$};
\end{tikzpicture}}}

\newcommand{\recursionBodd}{\scalebox{\scale}{\begin{tikzpicture}[baseline={([yshift=-1ex]current bounding box.center)}]
\draw[wilson] (-2.5,0) -- (2.5,0);
\draw[scalar] (-.75,.3) arc (180:0:.75);
\draw[scalar] (-.75,0) -- (-.75,.3);
\draw[scalar] (.75,0) -- (.75,.3);
\draw[scalar] (-1.8,.3) arc (180:0:.3);
\draw[scalar] (-1.8,0) -- (-1.8,.3);
\draw[scalar] (-1.2,0) -- (-1.2,.3);
\draw[scalar] (-.3,.3) arc (180:0:.3);
\draw[scalar] (-.3,0) -- (-.3,.3);
\draw[scalar] (.3,0) -- (.3,.3);
\draw[scalar] (1.2,.3) arc (180:0:.3);
\draw[scalar] (1.2,0) -- (1.2,.3);
\draw[scalar] (1.8,0) -- (1.8,.3);
\draw[fill=white] (-0.75,0) circle (.1);
\node[] at (-1.5,.25) {\small $t$};
\node[] at (0,.25) {\small $t$};
\node[] at (1.5,.25) {\small $t$};
\node[] at (-0.75,-.4) {\small $i$};
\node[] at (.25,-.4) {\small $2j$};
\end{tikzpicture}}}

\newcommand{\recursionCodd}{\scalebox{\scale}{\begin{tikzpicture}[baseline={([yshift=-1ex]current bounding box.center)}]
\draw[wilson] (-2.5,0) -- (2.5,0);
\draw[scalar] (-.75,.3) arc (180:0:.75);
\draw[scalar] (-.75,0) -- (-.75,.3);
\draw[scalar] (.75,0) -- (.75,.3);
\draw[scalar] (-1.8,.3) arc (180:0:.3);
\draw[scalar] (-1.8,0) -- (-1.8,.3);
\draw[scalar] (-1.2,0) -- (-1.2,.3);
\draw[scalar] (-.3,.3) arc (180:0:.3);
\draw[scalar] (-.3,0) -- (-.3,.3);
\draw[scalar] (.3,0) -- (.3,.3);
\draw[scalar] (1.2,.3) arc (180:0:.3);
\draw[scalar] (1.2,0) -- (1.2,.3);
\draw[scalar] (1.8,0) -- (1.8,.3);
\draw[fill=white] (.75,0) circle (.1);
\draw[fill=white] (-0.75,0) circle (.1);
\node[] at (-1.5,.25) {\small $t$};
\node[] at (0,.25) {\small $t_O$};
\node[] at (1.5,.25) {\small $t$};
\node[] at (-0.75,-.4) {\small $i$};
\node[] at (0.75,-.4) {\small $j$};
\end{tikzpicture}}}

\newcommand{\recursionphione}{\scalebox{\scalephi}{\begin{tikzpicture}[baseline={([yshift=-2ex]current bounding box.center)}]
\draw[wilson] (-5,0) -- (2,0);
\draw[scalar] (-.75,0) -- (-.75,.3);
\draw[scalar] (.75,0) -- (.75,.3);
\draw[scalar] (-2.25,.3) arc (180:0:.75);
\draw[scalar] (-3.75,0) -- (-3.75,.3);
\draw[scalar] (-2.25,0) -- (-2.25,.3);
\draw[scalar] (-4.8,.3) arc (180:0:.3);
\draw[scalar] (-4.8,0) -- (-4.8,.3);
\draw[scalar] (-4.2,0) -- (-4.2,.3);
\draw[scalar] (-3.3,.3) arc (180:0:.3);
\draw[scalar] (-3.3,0) -- (-3.3,.3);
\draw[scalar] (-2.7,0) -- (-2.7,.3);
\draw[scalar] (-1.8,.3) arc (180:0:.3);
\draw[scalar] (-1.8,0) -- (-1.8,.3);
\draw[scalar] (-1.2,0) -- (-1.2,.3);
\draw[scalar] (-.3,.3) arc (180:0:.3);
\draw[scalar] (-.3,0) -- (-.3,.3);
\draw[scalar] (.3,0) -- (.3,.3);
\draw[scalar] (1.2,.3) arc (180:0:.3);
\draw[scalar] (1.2,0) -- (1.2,.3);
\draw[scalar] (1.8,0) -- (1.8,.3);
\draw[scalar] (.75,.3) arc(0:180:2.25cm and 1.4cm);
\draw[fill=white] (-3.75,0) circle (.1);
\draw[fill=white] (-2.25,0) circle (.1);
\node[] at (-4.5,.25) {\small $t$};
\node[] at (-3,.25) {\small $t$};
\node[] at (-1.5,.25) {\small $t$};
\node[] at (0,.25) {\small $t$};
\node[] at (1.5,.25) {\small $t$};
\node[] at (-3.75,-.5) {\Large $j$};
\node[] at (-2.25,-.5) {\Large $k$};
\node[] at (-1.25,-.5) {\Large $l$};
\node[] at (.25,-.5) {\Large $m$};
\end{tikzpicture}}}

\newcommand{\recursionphioneb}{\scalebox{\scalephi}{\begin{tikzpicture}[baseline={([yshift=-2ex]current bounding box.center)}]
\draw[wilson] (-5,0) -- (2,0);
\draw[scalar] (-.75,0) -- (-.75,.3);
\draw[scalar] (.75,0) -- (.75,.3);
\draw[scalar] (-2.25,.3) arc (180:0:.75);
\draw[scalar] (-3.75,0) -- (-3.75,.3);
\draw[scalar] (-2.25,0) -- (-2.25,.3);
\draw[scalar] (-4.8,.3) arc (180:0:.3);
\draw[scalar] (-4.8,0) -- (-4.8,.3);
\draw[scalar] (-4.2,0) -- (-4.2,.3);
\draw[scalar] (-3.3,.3) arc (180:0:.3);
\draw[scalar] (-3.3,0) -- (-3.3,.3);
\draw[scalar] (-2.7,0) -- (-2.7,.3);
\draw[scalar] (-1.8,.3) arc (180:0:.3);
\draw[scalar] (-1.8,0) -- (-1.8,.3);
\draw[scalar] (-1.2,0) -- (-1.2,.3);
\draw[scalar] (1.2,.3) arc (180:0:.3);
\draw[scalar] (1.2,0) -- (1.2,.3);
\draw[scalar] (1.8,0) -- (1.8,.3);
\draw[scalar] (.75,.3) arc(0:180:2.25cm and 1.4cm);
\draw[fill=white] (-3.75,0) circle (.1);
\draw[fill=white] (-2.25,0) circle (.1);
\node[] at (-4.5,.25) {\small $t$};
\node[] at (-3,.25) {\small $t$};
\node[] at (-1.5,.25) {\small $t$};
\node[] at (1.5,.25) {\small $t$};
\node[] at (-3.75,-.5) {\Large $j$};
\node[] at (-2.25,-.5) {\Large $k$};
\node[] at (-1.25,-.5) {\Large $l$};
\end{tikzpicture}}}

\newcommand{\recursionphitwo}{\scalebox{\scalephi}{\begin{tikzpicture}[baseline={([yshift=-1ex]current bounding box.center)}]
\draw[wilson] (-5,0) -- (2,0);
\draw[scalar] (-.75,0) -- (-.75,.3);
\draw[scalar] (.75,0) -- (.75,.3);
\draw[scalar] (-2.25,.3) arc (180:0:.75);
\draw[scalar] (-3.75,0) -- (-3.75,.3);
\draw[scalar] (-2.25,0) -- (-2.25,.3);
\draw[scalar] (-4.8,.3) arc (180:0:.3);
\draw[scalar] (-4.8,0) -- (-4.8,.3);
\draw[scalar] (-4.2,0) -- (-4.2,.3);
\draw[scalar] (-3.3,.3) arc (180:0:.3);
\draw[scalar] (-3.3,0) -- (-3.3,.3);
\draw[scalar] (-2.7,0) -- (-2.7,.3);
\draw[scalar] (-1.8,.3) arc (180:0:.3);
\draw[scalar] (-1.8,0) -- (-1.8,.3);
\draw[scalar] (-1.2,0) -- (-1.2,.3);
\draw[scalar] (-.3,.3) arc (180:0:.3);
\draw[scalar] (-.3,0) -- (-.3,.3);
\draw[scalar] (.3,0) -- (.3,.3);
\draw[scalar] (1.2,.3) arc (180:0:.3);
\draw[scalar] (1.2,0) -- (1.2,.3);
\draw[scalar] (1.8,0) -- (1.8,.3);
\draw[scalar] (.75,.3) arc(0:180:2.25cm and 1.4cm);
\draw[fill=white] (-3.75,0) circle (.1);
\draw[fill=white] (-0.75,0) circle (.1);
\node[] at (-4.5,.25) {\small $t$};
\node[] at (-3,.25) {\small $t$};
\node[] at (-1.5,.25) {\small $t$};
\node[] at (0,.25) {\small $t$};
\node[] at (1.5,.25) {\small $t$};
\node[] at (-3.75,-.5) {\Large $j$};
\node[] at (-2.75, -.5) {\Large $m$};
\node[] at (-0.75,-.5) {\Large $k$};
\node[] at (.25,-.5) {\Large $l$};
\end{tikzpicture}}}

\newcommand{\recursionphithree}{\scalebox{\scalephi}{\begin{tikzpicture}[baseline={([yshift=-1ex]current bounding box.center)}]
\draw[wilson] (-5,0) -- (2,0);
\draw[scalar] (-.75,0) -- (-.75,.3);
\draw[scalar] (.75,0) -- (.75,.3);
\draw[scalar] (-2.25,.3) arc (180:0:.75);
\draw[scalar] (-3.75,0) -- (-3.75,.3);
\draw[scalar] (-2.25,0) -- (-2.25,.3);
\draw[scalar] (-4.8,.3) arc (180:0:.3);
\draw[scalar] (-4.8,0) -- (-4.8,.3);
\draw[scalar] (-4.2,0) -- (-4.2,.3);
\draw[scalar] (-3.3,.3) arc (180:0:.3);
\draw[scalar] (-3.3,0) -- (-3.3,.3);
\draw[scalar] (-2.7,0) -- (-2.7,.3);
\draw[scalar] (-1.8,.3) arc (180:0:.3);
\draw[scalar] (-1.8,0) -- (-1.8,.3);
\draw[scalar] (-1.2,0) -- (-1.2,.3);
\draw[scalar] (-.3,.3) arc (180:0:.3);
\draw[scalar] (-.3,0) -- (-.3,.3);
\draw[scalar] (.3,0) -- (.3,.3);
\draw[scalar] (1.2,.3) arc (180:0:.3);
\draw[scalar] (1.2,0) -- (1.2,.3);
\draw[scalar] (1.8,0) -- (1.8,.3);
\draw[scalar] (.75,.3) arc(0:180:2.25cm and 1.4cm);
\draw[fill=white] (.75,0) circle (.1);
\draw[fill=white] (-0.75,0) circle (.1);
\node[] at (-4.5,.25) {\small $t$};
\node[] at (-3,.25) {\small $t$};
\node[] at (-1.5,.25) {\small $t$};
\node[] at (0,.25) {\small $t$};
\node[] at (1.5,.25) {\small $t$};
\node[] at (-4.25,-.5) {\Large $l$};
\node[] at (-2.75, -.5) {\Large $m$};
\node[] at (-0.75,-.5) {\Large $j$};
\node[] at (.75,-.5) {\Large $k$};
\end{tikzpicture}}}

\newcommand{\recursionphithreeb}{\scalebox{\scalephi}{\begin{tikzpicture}[baseline={([yshift=-1ex]current bounding box.center)}]
\draw[wilson] (-5,0) -- (2,0);
\draw[scalar] (-.75,0) -- (-.75,.3);
\draw[scalar] (.75,0) -- (.75,.3);
\draw[scalar] (-2.25,.3) arc (180:0:.75);
\draw[scalar] (-3.75,0) -- (-3.75,.3);
\draw[scalar] (-2.25,0) -- (-2.25,.3);
\draw[scalar] (-4.8,.3) arc (180:0:.3);
\draw[scalar] (-4.8,0) -- (-4.8,.3);
\draw[scalar] (-4.2,0) -- (-4.2,.3);
\draw[scalar] (-1.8,.3) arc (180:0:.3);
\draw[scalar] (-1.8,0) -- (-1.8,.3);
\draw[scalar] (-1.2,0) -- (-1.2,.3);
\draw[scalar] (-.3,.3) arc (180:0:.3);
\draw[scalar] (-.3,0) -- (-.3,.3);
\draw[scalar] (.3,0) -- (.3,.3);
\draw[scalar] (1.2,.3) arc (180:0:.3);
\draw[scalar] (1.2,0) -- (1.2,.3);
\draw[scalar] (1.8,0) -- (1.8,.3);
\draw[scalar] (.75,.3) arc(0:180:2.25cm and 1.4cm);
\draw[fill=white] (.75,0) circle (.1);
\draw[fill=white] (-0.75,0) circle (.1);
\node[] at (-4.5,.25) {\small $t$};
\node[] at (-1.5,.25) {\small $t$};
\node[] at (0,.25) {\small $t$};
\node[] at (1.5,.25) {\small $t$};
\node[] at (-4.25,-.5) {\Large $l$};
\node[] at (-0.75,-.5) {\Large $j$};
\node[] at (.75,-.5) {\Large $k$};
\end{tikzpicture}}}

\newcommand{\recursionphifour}{\scalebox{\scalephi}{\begin{tikzpicture}[baseline={([yshift=-1ex]current bounding box.center)}]
\draw[wilson] (-5,0) -- (2,0);
\draw[scalar] (-.75,0) -- (-.75,.3);
\draw[scalar] (.75,0) -- (.75,.3);
\draw[scalar] (-2.25,.3) arc (180:0:.75);
\draw[scalar] (-3.75,0) -- (-3.75,.3);
\draw[scalar] (-2.25,0) -- (-2.25,.3);
\draw[scalar] (-4.8,.3) arc (180:0:.3);
\draw[scalar] (-4.8,0) -- (-4.8,.3);
\draw[scalar] (-4.2,0) -- (-4.2,.3);
\draw[scalar] (-3.3,.3) arc (180:0:.3);
\draw[scalar] (-3.3,0) -- (-3.3,.3);
\draw[scalar] (-2.7,0) -- (-2.7,.3);
\draw[scalar] (-1.8,.3) arc (180:0:.3);
\draw[scalar] (-1.8,0) -- (-1.8,.3);
\draw[scalar] (-1.2,0) -- (-1.2,.3);
\draw[scalar] (-.3,.3) arc (180:0:.3);
\draw[scalar] (-.3,0) -- (-.3,.3);
\draw[scalar] (.3,0) -- (.3,.3);
\draw[scalar] (1.2,.3) arc (180:0:.3);
\draw[scalar] (1.2,0) -- (1.2,.3);
\draw[scalar] (1.8,0) -- (1.8,.3);
\draw[scalar] (.75,.3) arc(0:180:2.25cm and 1.4cm);
\draw[fill=white] (-2.25,0) circle (.1);
\draw[fill=white] (0.75,0) circle (.1);
\node[] at (-4.5,.25) {\small $t$};
\node[] at (-3,.25) {\small $t$};
\node[] at (-1.5,.25) {\small $t$};
\node[] at (0,.25) {\small $t$};
\node[] at (1.5,.25) {\small $t$};
\node[] at (-4.25,-.5) {\Large $m$};
\node[] at (-2.25, -.5) {\Large $j$};
\node[] at (-1.25,-.5) {\Large $l$};
\node[] at (.75,-.5) {\Large $k$};
\end{tikzpicture}}}

\newcommand{\recursionphifive}{\scalebox{\scalephi}{\begin{tikzpicture}[baseline={([yshift=-0ex]current bounding box.center)}]
\draw[wilson] (-5,0) -- (2,0);
\draw[scalar] (-.75,.3) arc (180:0:.75);
\draw[scalar] (-.75,0) -- (-.75,.3);
\draw[scalar] (.75,0) -- (.75,.3);
\draw[scalar] (-3.75,.3) arc (180:0:.75);
\draw[scalar] (-3.75,0) -- (-3.75,.3);
\draw[scalar] (-2.25,0) -- (-2.25,.3);
\draw[scalar] (-4.8,.3) arc (180:0:.3);
\draw[scalar] (-4.8,0) -- (-4.8,.3);
\draw[scalar] (-4.2,0) -- (-4.2,.3);
\draw[scalar] (-3.3,.3) arc (180:0:.3);
\draw[scalar] (-3.3,0) -- (-3.3,.3);
\draw[scalar] (-2.7,0) -- (-2.7,.3);
\draw[scalar] (-1.8,.3) arc (180:0:.3);
\draw[scalar] (-1.8,0) -- (-1.8,.3);
\draw[scalar] (-1.2,0) -- (-1.2,.3);
\draw[scalar] (-.3,.3) arc (180:0:.3);
\draw[scalar] (-.3,0) -- (-.3,.3);
\draw[scalar] (.3,0) -- (.3,.3);
\draw[scalar] (1.2,.3) arc (180:0:.3);
\draw[scalar] (1.2,0) -- (1.2,.3);
\draw[scalar] (1.8,0) -- (1.8,.3);
\draw[fill=white] (-.75,0) circle (.1);
\draw[fill=white] (-3.75,0) circle (.1);
\node[] at (-4.5,.25) {\small $t$};
\node[] at (-3,.25) {\small $t$};
\node[] at (-1.5,.25) {\small $t$};
\node[] at (0,.25) {\small $t$};
\node[] at (1.5,.25) {\small $t$};
\node[] at (-3.75,-.5) {\Large $j$};
\node[] at (-2.75,-.5) {\Large $l$};
\node[] at (-.75,-.5) {\Large $k$};
\node[] at (.25,-.5) {\Large $m$};
\end{tikzpicture}}}

\newcommand{\recursionphisix}{\scalebox{\scalephi}{\begin{tikzpicture}[baseline={([yshift=-0ex]current bounding box.center)}]
\draw[wilson] (-5,0) -- (2,0);
\draw[scalar] (-.75,.3) arc (180:0:.75);
\draw[scalar] (-.75,0) -- (-.75,.3);
\draw[scalar] (.75,0) -- (.75,.3);
\draw[scalar] (-3.75,.3) arc (180:0:.75);
\draw[scalar] (-3.75,0) -- (-3.75,.3);
\draw[scalar] (-2.25,0) -- (-2.25,.3);
\draw[scalar] (-4.8,.3) arc (180:0:.3);
\draw[scalar] (-4.8,0) -- (-4.8,.3);
\draw[scalar] (-4.2,0) -- (-4.2,.3);
\draw[scalar] (-3.3,.3) arc (180:0:.3);
\draw[scalar] (-3.3,0) -- (-3.3,.3);
\draw[scalar] (-2.7,0) -- (-2.7,.3);
\draw[scalar] (-1.8,.3) arc (180:0:.3);
\draw[scalar] (-1.8,0) -- (-1.8,.3);
\draw[scalar] (-1.2,0) -- (-1.2,.3);
\draw[scalar] (-.3,.3) arc (180:0:.3);
\draw[scalar] (-.3,0) -- (-.3,.3);
\draw[scalar] (.3,0) -- (.3,.3);
\draw[scalar] (1.2,.3) arc (180:0:.3);
\draw[scalar] (1.2,0) -- (1.2,.3);
\draw[scalar] (1.8,0) -- (1.8,.3);
\draw[fill=white] (.75,0) circle (.1);
\draw[fill=white] (-3.75,0) circle (.1);
\node[] at (-4.5,.25) {\small $t$};
\node[] at (-3,.25) {\small $t$};
\node[] at (-1.5,.25) {\small $t$};
\node[] at (0,.25) {\small $t$};
\node[] at (1.5,.25) {\small $t$};
\node[] at (-3.75,-.5) {\Large $j$};
\node[] at (-2.75,-.5) {\Large $l$};
\node[] at (-1.25,-.5) {\Large $m$};
\node[] at (.75,-.5) {\Large $k$};
\end{tikzpicture}}}

\newcommand{\recursionphisixb}{\scalebox{\scalephi}{\begin{tikzpicture}[baseline={([yshift=-0ex]current bounding box.center)}]
\draw[wilson] (-5,0) -- (2,0);
\draw[scalar] (-.75,.3) arc (180:0:.75);
\draw[scalar] (-.75,0) -- (-.75,.3);
\draw[scalar] (.75,0) -- (.75,.3);
\draw[scalar] (-3.75,.3) arc (180:0:.75);
\draw[scalar] (-3.75,0) -- (-3.75,.3);
\draw[scalar] (-2.25,0) -- (-2.25,.3);
\draw[scalar] (-4.8,.3) arc (180:0:.3);
\draw[scalar] (-4.8,0) -- (-4.8,.3);
\draw[scalar] (-4.2,0) -- (-4.2,.3);
\draw[scalar] (-3.3,.3) arc (180:0:.3);
\draw[scalar] (-3.3,0) -- (-3.3,.3);
\draw[scalar] (-2.7,0) -- (-2.7,.3);
\draw[scalar] (-.3,.3) arc (180:0:.3);
\draw[scalar] (-.3,0) -- (-.3,.3);
\draw[scalar] (.3,0) -- (.3,.3);
\draw[scalar] (1.2,.3) arc (180:0:.3);
\draw[scalar] (1.2,0) -- (1.2,.3);
\draw[scalar] (1.8,0) -- (1.8,.3);
\draw[fill=white] (.75,0) circle (.1);
\draw[fill=white] (-3.75,0) circle (.1);
\node[] at (-4.5,.25) {\small $t$};
\node[] at (-3,.25) {\small $t$};
\node[] at (0,.25) {\small $t$};
\node[] at (1.5,.25) {\small $t$};
\node[] at (-3.75,-.5) {\Large $j$};
\node[] at (-2.75,-.5) {\Large $l$};
\node[] at (.75,-.5) {\Large $k$};
\end{tikzpicture}}}

\newcommand{\recursionphiseven}{\scalebox{\scalephi}{\begin{tikzpicture}[baseline={([yshift=-0ex]current bounding box.center)}]
\draw[wilson] (-5,0) -- (2,0);
\draw[scalar] (-.75,.3) arc (180:0:.75);
\draw[scalar] (-.75,0) -- (-.75,.3);
\draw[scalar] (.75,0) -- (.75,.3);
\draw[scalar] (-3.75,.3) arc (180:0:.75);
\draw[scalar] (-3.75,0) -- (-3.75,.3);
\draw[scalar] (-2.25,0) -- (-2.25,.3);
\draw[scalar] (-4.8,.3) arc (180:0:.3);
\draw[scalar] (-4.8,0) -- (-4.8,.3);
\draw[scalar] (-4.2,0) -- (-4.2,.3);
\draw[scalar] (-3.3,.3) arc (180:0:.3);
\draw[scalar] (-3.3,0) -- (-3.3,.3);
\draw[scalar] (-2.7,0) -- (-2.7,.3);
\draw[scalar] (-1.8,.3) arc (180:0:.3);
\draw[scalar] (-1.8,0) -- (-1.8,.3);
\draw[scalar] (-1.2,0) -- (-1.2,.3);
\draw[scalar] (-.3,.3) arc (180:0:.3);
\draw[scalar] (-.3,0) -- (-.3,.3);
\draw[scalar] (.3,0) -- (.3,.3);
\draw[scalar] (1.2,.3) arc (180:0:.3);
\draw[scalar] (1.2,0) -- (1.2,.3);
\draw[scalar] (1.8,0) -- (1.8,.3);
\draw[fill=white] (-.75,0) circle (.1);
\draw[fill=white] (-2.25,0) circle (.1);
\node[] at (-4.5,.25) {\small $t$};
\node[] at (-3,.25) {\small $t$};
\node[] at (-1.5,.25) {\small $t$};
\node[] at (0,.25) {\small $t$};
\node[] at (1.5,.25) {\small $t$};
\node[] at (-2.25,-.5) {\Large $j$};
\node[] at (-4.25,-.5) {\Large $l$};
\node[] at (.35,-.5) {\Large $m$};
\node[] at (-.75,-.5) {\Large $k$};
\end{tikzpicture}}}

\newcommand{\recursionphieight}{\scalebox{\scalephi}{\begin{tikzpicture}[baseline={([yshift=-0ex]current bounding box.center)}]
\draw[wilson] (-5,0) -- (2,0);
\draw[scalar] (-.75,.3) arc (180:0:.75);
\draw[scalar] (-.75,0) -- (-.75,.3);
\draw[scalar] (.75,0) -- (.75,.3);
\draw[scalar] (-3.75,.3) arc (180:0:.75);
\draw[scalar] (-3.75,0) -- (-3.75,.3);
\draw[scalar] (-2.25,0) -- (-2.25,.3);
\draw[scalar] (-4.8,.3) arc (180:0:.3);
\draw[scalar] (-4.8,0) -- (-4.8,.3);
\draw[scalar] (-4.2,0) -- (-4.2,.3);
\draw[scalar] (-3.3,.3) arc (180:0:.3);
\draw[scalar] (-3.3,0) -- (-3.3,.3);
\draw[scalar] (-2.7,0) -- (-2.7,.3);
\draw[scalar] (-1.8,.3) arc (180:0:.3);
\draw[scalar] (-1.8,0) -- (-1.8,.3);
\draw[scalar] (-1.2,0) -- (-1.2,.3);
\draw[scalar] (-.3,.3) arc (180:0:.3);
\draw[scalar] (-.3,0) -- (-.3,.3);
\draw[scalar] (.3,0) -- (.3,.3);
\draw[scalar] (1.2,.3) arc (180:0:.3);
\draw[scalar] (1.2,0) -- (1.2,.3);
\draw[scalar] (1.8,0) -- (1.8,.3);
\draw[fill=white] (.75,0) circle (.1);
\draw[fill=white] (-2.25,0) circle (.1);
\node[] at (-4.5,.25) {\small $t$};
\node[] at (-3,.25) {\small $t$};
\node[] at (-1.5,.25) {\small $t$};
\node[] at (0,.25) {\small $t$};
\node[] at (1.5,.25) {\small $t$};
\node[] at (-4.25,-.5) {\Large $l$};
\node[] at (-2.25,-.5) {\Large $j$};
\node[] at (-1.25,-.5) {\Large $m$};
\node[] at (.75,-.5) {\Large $k$};
\end{tikzpicture}}}

\newcommand{\recursionphinine}{\scalebox{\scalephi}{\begin{tikzpicture}[baseline={([yshift=-0ex]current bounding box.center)}]
\draw[wilson] (-5,0) -- (2,0);
\draw[scalar] (-.75,.3) arc (180:0:.75);
\draw[scalar] (-.75,0) -- (-.75,.3);
\draw[scalar] (.75,0) -- (.75,.3);
\draw[scalar] (-3.75,.3) arc (180:0:.75);
\draw[scalar] (-3.75,0) -- (-3.75,.3);
\draw[scalar] (-2.25,0) -- (-2.25,.3);
\draw[scalar] (-4.8,.3) arc (180:0:.3);
\draw[scalar] (-4.8,0) -- (-4.8,.3);
\draw[scalar] (-4.2,0) -- (-4.2,.3);
\draw[scalar] (-3.3,.3) arc (180:0:.3);
\draw[scalar] (-3.3,0) -- (-3.3,.3);
\draw[scalar] (-2.7,0) -- (-2.7,.3);
\draw[scalar] (-1.8,.3) arc (180:0:.3);
\draw[scalar] (-1.8,0) -- (-1.8,.3);
\draw[scalar] (-1.2,0) -- (-1.2,.3);
\draw[scalar] (-.3,.3) arc (180:0:.3);
\draw[scalar] (-.3,0) -- (-.3,.3);
\draw[scalar] (.3,0) -- (.3,.3);
\draw[scalar] (1.2,.3) arc (180:0:.3);
\draw[scalar] (1.2,0) -- (1.2,.3);
\draw[scalar] (1.8,0) -- (1.8,.3);
\draw[fill=white] (-.75,0) circle (.1);
\draw[fill=white] (-2.25,0) circle (.1);
\draw[fill=white] (-3.75,0) circle (.1);
\node[] at (-4.5,.25) {\small $t$};
\node[] at (-3,.25) {\small $t_O$};
\node[] at (-1.5,.25) {\small $t$};
\node[] at (0,.25) {\small $t$};
\node[] at (1.5,.25) {\small $t$};
\node[] at (-3.75,-.5) {\Large $j$};
\node[] at (-2.25,-.5) {\Large $k$};
\node[] at (-.75,-.5) {\Large $l$};
\node[] at (.25,-.5) {\Large $m$};
\end{tikzpicture}}}

\newcommand{\recursionphiten}{\scalebox{\scalephi}{\begin{tikzpicture}[baseline={([yshift=-0ex]current bounding box.center)}]
\draw[wilson] (-5,0) -- (2,0);
\draw[scalar] (-.75,.3) arc (180:0:.75);
\draw[scalar] (-.75,0) -- (-.75,.3);
\draw[scalar] (.75,0) -- (.75,.3);
\draw[scalar] (-3.75,.3) arc (180:0:.75);
\draw[scalar] (-3.75,0) -- (-3.75,.3);
\draw[scalar] (-2.25,0) -- (-2.25,.3);
\draw[scalar] (-4.8,.3) arc (180:0:.3);
\draw[scalar] (-4.8,0) -- (-4.8,.3);
\draw[scalar] (-4.2,0) -- (-4.2,.3);
\draw[scalar] (-3.3,.3) arc (180:0:.3);
\draw[scalar] (-3.3,0) -- (-3.3,.3);
\draw[scalar] (-2.7,0) -- (-2.7,.3);
\draw[scalar] (-1.8,.3) arc (180:0:.3);
\draw[scalar] (-1.8,0) -- (-1.8,.3);
\draw[scalar] (-1.2,0) -- (-1.2,.3);
\draw[scalar] (-.3,.3) arc (180:0:.3);
\draw[scalar] (-.3,0) -- (-.3,.3);
\draw[scalar] (.3,0) -- (.3,.3);
\draw[scalar] (1.2,.3) arc (180:0:.3);
\draw[scalar] (1.2,0) -- (1.2,.3);
\draw[scalar] (1.8,0) -- (1.8,.3);
\draw[fill=white] (.75,0) circle (.1);
\draw[fill=white] (-2.25,0) circle (.1);
\draw[fill=white] (-3.75,0) circle (.1);
\node[] at (-4.5,.25) {\small $t$};
\node[] at (-3,.25) {\small $t_O$};
\node[] at (-1.5,.25) {\small $t$};
\node[] at (0,.25) {\small $t$};
\node[] at (1.5,.25) {\small $t$};
\node[] at (-3.75,-.5) {\Large $j$};
\node[] at (-2.25,-.5) {\Large $k$};
\node[] at (.75,-.5) {\Large $l$};
\node[] at (-1.25,-.5) {\Large $m$};
\end{tikzpicture}}}

\newcommand{\recursionphieleven}{\scalebox{\scalephi}{\begin{tikzpicture}[baseline={([yshift=-0ex]current bounding box.center)}]
\draw[wilson] (-5,0) -- (2,0);
\draw[scalar] (-.75,.3) arc (180:0:.75);
\draw[scalar] (-.75,0) -- (-.75,.3);
\draw[scalar] (.75,0) -- (.75,.3);
\draw[scalar] (-3.75,.3) arc (180:0:.75);
\draw[scalar] (-3.75,0) -- (-3.75,.3);
\draw[scalar] (-2.25,0) -- (-2.25,.3);
\draw[scalar] (-4.8,.3) arc (180:0:.3);
\draw[scalar] (-4.8,0) -- (-4.8,.3);
\draw[scalar] (-4.2,0) -- (-4.2,.3);
\draw[scalar] (-3.3,.3) arc (180:0:.3);
\draw[scalar] (-3.3,0) -- (-3.3,.3);
\draw[scalar] (-2.7,0) -- (-2.7,.3);
\draw[scalar] (-1.8,.3) arc (180:0:.3);
\draw[scalar] (-1.8,0) -- (-1.8,.3);
\draw[scalar] (-1.2,0) -- (-1.2,.3);
\draw[scalar] (-.3,.3) arc (180:0:.3);
\draw[scalar] (-.3,0) -- (-.3,.3);
\draw[scalar] (.3,0) -- (.3,.3);
\draw[scalar] (1.2,.3) arc (180:0:.3);
\draw[scalar] (1.2,0) -- (1.2,.3);
\draw[scalar] (1.8,0) -- (1.8,.3);
\draw[fill=white] (.75,0) circle (.1);
\draw[fill=white] (-.75,0) circle (.1);
\draw[fill=white] (-3.75,0) circle (.1);
\node[] at (-4.5,.25) {\small $t$};
\node[] at (-3,.25) {\small $t$};
\node[] at (-1.5,.25) {\small $t$};
\node[] at (0,.25) {\small $t_O$};
\node[] at (1.5,.25) {\small $t$};
\node[] at (-3.75,-.5) {\Large $l$};
\node[] at (-.75,-.5) {\Large $j$};
\node[] at (.75,-.5) {\Large $k$};
\node[] at (-2.75,-.5) {\Large $m$};
\end{tikzpicture}}}

\newcommand{\recursionphitwelve}{\scalebox{\scalephi}{\begin{tikzpicture}[baseline={([yshift=-0ex]current bounding box.center)}]
\draw[wilson] (-5,0) -- (2,0);
\draw[scalar] (-.75,.3) arc (180:0:.75);
\draw[scalar] (-.75,0) -- (-.75,.3);
\draw[scalar] (.75,0) -- (.75,.3);
\draw[scalar] (-3.75,.3) arc (180:0:.75);
\draw[scalar] (-3.75,0) -- (-3.75,.3);
\draw[scalar] (-2.25,0) -- (-2.25,.3);
\draw[scalar] (-4.8,.3) arc (180:0:.3);
\draw[scalar] (-4.8,0) -- (-4.8,.3);
\draw[scalar] (-4.2,0) -- (-4.2,.3);
\draw[scalar] (-3.3,.3) arc (180:0:.3);
\draw[scalar] (-3.3,0) -- (-3.3,.3);
\draw[scalar] (-2.7,0) -- (-2.7,.3);
\draw[scalar] (-1.8,.3) arc (180:0:.3);
\draw[scalar] (-1.8,0) -- (-1.8,.3);
\draw[scalar] (-1.2,0) -- (-1.2,.3);
\draw[scalar] (-.3,.3) arc (180:0:.3);
\draw[scalar] (-.3,0) -- (-.3,.3);
\draw[scalar] (.3,0) -- (.3,.3);
\draw[scalar] (1.2,.3) arc (180:0:.3);
\draw[scalar] (1.2,0) -- (1.2,.3);
\draw[scalar] (1.8,0) -- (1.8,.3);
\draw[fill=white] (.75,0) circle (.1);
\draw[fill=white] (-.75,0) circle (.1);
\draw[fill=white] (-2.25,0) circle (.1);
\node[] at (-4.5,.25) {\small $t$};
\node[] at (-3,.25) {\small $t$};
\node[] at (-1.5,.25) {\small $t$};
\node[] at (0,.25) {\small $t_O$};
\node[] at (1.5,.25) {\small $t$};
\node[] at (-2.25,-.5) {\Large $l$};
\node[] at (-.75,-.5) {\Large $j$};
\node[] at (.75,-.5) {\Large $k$};
\node[] at (-4.25,-.5) {\Large $m$};
\end{tikzpicture}}}

\newcommand{\recursionphithirteen}{\scalebox{\scalephi}{\begin{tikzpicture}[baseline={([yshift=-1ex]current bounding box.center)}]
\draw[wilson] (-5,0) -- (2,0);
\draw[scalar] (-.75,0) -- (-.75,.3);
\draw[scalar] (.75,0) -- (.75,.3);
\draw[scalar] (-2.25,.3) arc (180:0:.75);
\draw[scalar] (-3.75,0) -- (-3.75,.3);
\draw[scalar] (-2.25,0) -- (-2.25,.3);
\draw[scalar] (-4.8,.3) arc (180:0:.3);
\draw[scalar] (-4.8,0) -- (-4.8,.3);
\draw[scalar] (-4.2,0) -- (-4.2,.3);
\draw[scalar] (-3.3,.3) arc (180:0:.3);
\draw[scalar] (-3.3,0) -- (-3.3,.3);
\draw[scalar] (-2.7,0) -- (-2.7,.3);
\draw[scalar] (-1.8,.3) arc (180:0:.3);
\draw[scalar] (-1.8,0) -- (-1.8,.3);
\draw[scalar] (-1.2,0) -- (-1.2,.3);
\draw[scalar] (-.3,.3) arc (180:0:.3);
\draw[scalar] (-.3,0) -- (-.3,.3);
\draw[scalar] (.3,0) -- (.3,.3);
\draw[scalar] (1.2,.3) arc (180:0:.3);
\draw[scalar] (1.2,0) -- (1.2,.3);
\draw[scalar] (1.8,0) -- (1.8,.3);
\draw[scalar] (.75,.3) arc(0:180:2.25cm and 1.4cm);
\draw[fill=white] (-2.25,0) circle (.1);
\draw[fill=white] (-0.75,0) circle (.1);
\draw[fill=white] (0.75,0) circle (.1);
\node[] at (-4.5,.25) {\small $t$};
\node[] at (-3,.25) {\small $t$};
\node[] at (-1.5,.25) {\small $t_O$};
\node[] at (0,.25) {\small $t$};
\node[] at (1.5,.25) {\small $t$};
\node[] at (-2.25,-.5) {\Large $j$};
\node[] at (-4.25, -.5) {\Large $m$};
\node[] at (-0.75,-.5) {\Large $k$};
\node[] at (.75,-.5) {\Large $l$};
\end{tikzpicture}}}

\newcommand{\recursionphifourteen}{\scalebox{\scalephi}{\begin{tikzpicture}[baseline={([yshift=-1ex]current bounding box.center)}]
\draw[wilson] (-5,0) -- (2,0);
\draw[scalar] (-.75,0) -- (-.75,.3);
\draw[scalar] (.75,0) -- (.75,.3);
\draw[scalar] (-2.25,.3) arc (180:0:.75);
\draw[scalar] (-3.75,0) -- (-3.75,.3);
\draw[scalar] (-2.25,0) -- (-2.25,.3);
\draw[scalar] (-4.8,.3) arc (180:0:.3);
\draw[scalar] (-4.8,0) -- (-4.8,.3);
\draw[scalar] (-4.2,0) -- (-4.2,.3);
\draw[scalar] (-3.3,.3) arc (180:0:.3);
\draw[scalar] (-3.3,0) -- (-3.3,.3);
\draw[scalar] (-2.7,0) -- (-2.7,.3);
\draw[scalar] (-1.8,.3) arc (180:0:.3);
\draw[scalar] (-1.8,0) -- (-1.8,.3);
\draw[scalar] (-1.2,0) -- (-1.2,.3);
\draw[scalar] (-.3,.3) arc (180:0:.3);
\draw[scalar] (-.3,0) -- (-.3,.3);
\draw[scalar] (.3,0) -- (.3,.3);
\draw[scalar] (1.2,.3) arc (180:0:.3);
\draw[scalar] (1.2,0) -- (1.2,.3);
\draw[scalar] (1.8,0) -- (1.8,.3);
\draw[scalar] (.75,.3) arc(0:180:2.25cm and 1.4cm);
\draw[fill=white] (-3.75,0) circle (.1);
\draw[fill=white] (-0.75,0) circle (.1);
\draw[fill=white] (-2.25,0) circle (.1);
\node[] at (-4.5,.25) {\small $t$};
\node[] at (-3,.25) {\small $t$};
\node[] at (-1.5,.25) {\small $t_O$};
\node[] at (0,.25) {\small $t$};
\node[] at (1.5,.25) {\small $t$};
\node[] at (-3.75,-.5) {\Large $l$};
\node[] at (0.25, -.5) {\Large $m$};
\node[] at (-0.75,-.5) {\Large $k$};
\node[] at (-2.25,-.5) {\Large $j$};
\end{tikzpicture}}}

\newcommand{\recursionphififthteen}{\scalebox{\scalephi}{\begin{tikzpicture}[baseline={([yshift=-0ex]current bounding box.center)}]
\draw[wilson] (-5,0) -- (2,0);
\draw[scalar] (-.75,.3) arc (180:0:.75);
\draw[scalar] (-.75,0) -- (-.75,.3);
\draw[scalar] (.75,0) -- (.75,.3);
\draw[scalar] (-3.75,.3) arc (180:0:.75);
\draw[scalar] (-3.75,0) -- (-3.75,.3);
\draw[scalar] (-2.25,0) -- (-2.25,.3);
\draw[scalar] (-4.8,.3) arc (180:0:.3);
\draw[scalar] (-4.8,0) -- (-4.8,.3);
\draw[scalar] (-4.2,0) -- (-4.2,.3);
\draw[scalar] (-3.3,.3) arc (180:0:.3);
\draw[scalar] (-3.3,0) -- (-3.3,.3);
\draw[scalar] (-2.7,0) -- (-2.7,.3);
\draw[scalar] (-1.8,.3) arc (180:0:.3);
\draw[scalar] (-1.8,0) -- (-1.8,.3);
\draw[scalar] (-1.2,0) -- (-1.2,.3);
\draw[scalar] (-.3,.3) arc (180:0:.3);
\draw[scalar] (-.3,0) -- (-.3,.3);
\draw[scalar] (.3,0) -- (.3,.3);
\draw[scalar] (1.2,.3) arc (180:0:.3);
\draw[scalar] (1.2,0) -- (1.2,.3);
\draw[scalar] (1.8,0) -- (1.8,.3);
\draw[fill=white] (-3.75,0) circle (.1);
\draw[fill=white] (.75,0) circle (.1);
\draw[fill=white] (-.75,0) circle (.1);
\draw[fill=white] (-2.25,0) circle (.1);
\node[] at (-4.5,.25) {\small $t$};
\node[] at (-3,.25) {\small $t_O$};
\node[] at (-1.5,.25) {\small $t$};
\node[] at (0,.25) {\small $t_O$};
\node[] at (1.5,.25) {\small $t$};
\node[] at (-2.25,-.5) {\Large $k$};
\node[] at (-.75,-.5) {\Large $l$};
\node[] at (.75,-.5) {\Large $m$};
\node[] at (-3.75,-.5) {\Large $j$};
\end{tikzpicture}}}

\newcommand{\Uoneint}{\scalebox{\scaleU}{\begin{tikzpicture}[baseline={([yshift=-0ex]current bounding box.center)}]
\draw[wilson] (-3.05,0) -- (0.5,0);
\draw[scalar] (-1.5,.3) arc (180:0:.75);
\draw[scalar] (-1.5,0) -- (-1.5,.3);
\draw[scalar] (.0,0) -- (.0,.3);
\draw[fill=white] (0,0) circle (.1);
\draw[scalar] (-1.05,.3) arc (180:0:.3);
\draw[scalar] (-1.05,0) -- (-1.05,.3);
\draw[scalar] (-.45,0) -- (-.45,.3);
\draw[scalar] (-2.55,.3) arc (180:0:.3);
\draw[scalar] (-2.55,0) -- (-2.55,.3);
\draw[scalar] (-1.95,0) -- (-1.95,.3);
\node[] at (-1.95,-.47) {$i$};
\node[] at (-1.05,-.5) {$j$};
\node[] at (-1.5,-.5) {$n$};
\node[] at (.0,-.5) {$a$};
\node[] at (-0.75,.25) {\small $t$};
\node[] at (-2.25,.25) {\small $t$};
\end{tikzpicture}}}

\newcommand{\Utwoint}{\scalebox{\scaleU}{\begin{tikzpicture}[baseline={([yshift=-0ex]current bounding box.center)}]
\draw[wilson] (-4.25,0) -- (0.5,0);
\draw[scalar] (-1.5,.3) arc (180:0:.75);
\draw[scalar] (-1.5,0) -- (-1.5,.3);
\draw[scalar] (.0,0) -- (.0,.3);
\draw[scalar] (-3.75,.3) arc (180:0:.75);
\draw[scalar] (-3.75,0) -- (-3.75,.3);
\draw[scalar] (-2.25,0) -- (-2.25,.3);
\draw[fill=white] (0,0) circle (.1);
\draw[fill=white] (-3.75,0) circle (.1);
\draw[fill=blue] (-0.75,0) circle (.1);
\draw[fill=blue] (-3,0) circle (.1);
\node[] at (-2.25,-.5) {$n$};
\node[] at (-1.5,-.5) {$m$};
\node[] at (.0,-.45) {$b$};
\node[] at (-3.75,-.5) {$a$};
\node[] at (-3,-.47) {$i$};
\node[] at (-0.75,-.5) {$j$};
\end{tikzpicture}}}

\newcommand{\Utwointtwo}{\scalebox{\scaleU}{\begin{tikzpicture}[baseline={([yshift=-0ex]current bounding box.center)}]
\draw[wilson] (-4.25,0) -- (0.5,0);
\draw[scalar] (-0.75,0) arc(0:180:0.8cm and 0.5cm);
\draw[scalar] (0,0) arc(0:180:1.5cm and 1.1cm);
\draw[fill=white] (0,0) circle (.1);
\draw[fill=white] (-0.75,0) circle (.1);
\draw[fill=blue] (-1.5,0) circle (.1);
\draw[fill=blue] (-3.75,0) circle (.1);
\node[] at (-3,-.5) {$n$};
\node[] at (-0.75,-.5) {$a$};
\node[] at (.0,-.45) {$b$};
\node[] at (-2.25,-.5) {$m$};
\node[] at (-1.5,-.5) {$j$};
\node[] at (-3.75,-.47) {$i$};
\end{tikzpicture}}}

\newcommand{\Utwointthree}{\scalebox{\scaleU}{\begin{tikzpicture}[baseline={([yshift=-0ex]current bounding box.center)}]
\draw[wilson] (-4.25,0) -- (0.5,0);
\draw[scalar] (-1.5,0) arc(0:180:0.8cm and 0.5cm);
\draw[scalar] (-0.75,0) arc(0:180:1.5cm and 1.1cm);
\draw[fill=white] (-3.1,0) circle (.1);
\draw[fill=white] (-3.75,0) circle (.1);
\draw[fill=blue] (0,0) circle (.1);
\draw[fill=blue] (-2.25,0) circle (.1);
\node[] at (-1.5,-.5) {$n$};
\node[] at (-3.75,-.5) {$a$};
\node[] at (-3.1,-.45) {$b$};
\node[] at (-0.75,-.5) {$m$};
\node[] at (0,-.5) {$j$};
\node[] at (-2.25,-.47) {$i$};
\end{tikzpicture}}}

\newcommand{\Utwo}{\scalebox{\scale}{\begin{tikzpicture}[baseline={([yshift=-0ex]current bounding box.center)}]
\draw[wilson] (-4.5,0) -- (0,0);
\draw[scalar] (-1.5,0) arc(0:180:0.8cm and 0.5cm);
\draw[scalar] (-0.75,0) arc(0:180:1.5cm and 1.1cm);
\draw[fill=white] (-3.1,0) circle (.1);
\draw[fill=white] (-3.75,0) circle (.1);
\end{tikzpicture}}}

\newcommand{\combblocks}{\scalebox{\scaleCC}{\begin{tikzpicture}[baseline={([yshift=-0ex]current bounding box.center)}]
\draw[scalar] (-4.5,0) -- (0.5,0);
\draw[scalar] (-3.75,0) -- (-3.75,0.7);
\draw[scalar] (-2.75,0) -- (-2.75,0.7);
\draw[scalar] (-1.25,0) -- (-1.25,0.7);
\draw[scalar] (-0.25,0) -- (-0.25,0.7);
\node[] at (-3.2,.2) {\tiny $\Delta_1$};
\node[] at (-.7,.2) {\tiny $\Delta_{n-3}$};
\node[] at (-4.7,.1) {\tiny $\phi^{I_1}$};
\node[] at (-3.75,0.9) {\tiny $\phi^{I_2}$};
\node[] at (-2.75,0.9) {\tiny $\phi^{I_{3}}$};
\node[] at (-1.25,0.9) {\tiny $\phi^{I_{n-2}}$};
\node[] at (-0.25,0.9) {\tiny $\phi^{I_{n-1}}$};
\node[] at (0.8,.1) {\tiny $\phi^{I_n}$};
\node[] at (-2,.5) {$\ldots$};
\draw[fill=teal] (-3.75,0) circle (.05);
\draw[fill=teal] (-2.75,0) circle (.05);
\draw[fill=teal] (-1.25,0) circle (.05);
\draw[fill=teal] (-0.25,0) circle (.05);
\end{tikzpicture}}}

\newcommand{\snowblocks}{\scalebox{\scaleCC}{\begin{tikzpicture}[baseline={([yshift=-0ex]current bounding box.center)}]
\draw[scalar] (0,0) -- (0,.7);
\draw[scalar] (0,.7) -- (.7,1.15);
\draw[scalar] (0,.7) -- (-.7,1.15);

\draw[scalar] (0,0) -- (-.7,-.45);
\draw[scalar] (-.7,-.45) -- (-1.4,0);
\draw[scalar] (-.7,-.45) -- (-.7,-1.15);

\draw[scalar] (0,0) -- (.7,-.45);
\draw[scalar] (.7,-.45) -- (1.4,0);
\draw[scalar] (.7,-.45) -- (.7,-1.15);

\node[] at (.9,1.3) {\tiny $\phi^{I_1}$};
\node[] at (-.8,1.3) {\tiny $\phi^{I_2}$};
\node[] at (-1.5,0.1) {\tiny $\phi^{I_3}$};
\node[] at (-.7,-1.35) {\tiny $\phi^{I_4}$};
\node[] at (.7,-1.35) {\tiny $\phi^{I_5}$};
\node[] at (1.6,0.1) {\tiny $\phi^{I_6}$};

\node[] at (-.43,-.05) {\tiny $\Delta_1$};
\node[] at (.25,-.35) {\tiny $\Delta_2$};
\node[] at (.2,.4) {\tiny $\Delta_3$};

\draw[fill=teal] (0,0) circle (.05);
\draw[fill=teal] (0,.7) circle (.05);
\draw[fill=teal] (-.7,-.45) circle (.05);
\draw[fill=teal] (.7,-.45) circle (.05);
\end{tikzpicture}}}

\newcommand{\fourcombblocks}{\scalebox{\scaleCB}{\begin{tikzpicture}[baseline={([yshift=-0ex]current bounding box.center)}]
\draw[scalar] (-4.5,0) -- (-2,0);
\draw[scalar] (-3.75,0) -- (-3.75,0.7);
\draw[scalar] (-2.75,0) -- (-2.75,0.7);
\node[] at (-3.25,.2) {\tiny $\Delta$};
\node[] at (-4.7,.1) {\tiny $\phi^{I_1}$};
\node[] at (-3.75,0.9) {\tiny $\phi^{I_2}$};
\node[] at (-2.75,0.9) {\tiny $\phi^{I_3}$};
\node[] at (-1.7,.1) {\tiny $\phi^{I_4}$};
\node[] at (-3.75,-0.2) {\textcolor{teal}{\tiny 
$C_{\scalebox{.5}{$\scriptscriptstyle \phi^{I_1} \phi^{I_2} \Op_{\Delta}$}}$}};
\node[] at (-2.75,-0.2) {\textcolor{teal}{\tiny 
$C_{\scalebox{.5}{$\scriptscriptstyle \Op_{\Delta} \phi^{I_3} \phi^{I_4}$}}$ }};
\draw[fill=teal] (-3.75,0) circle (.05);
\draw[fill=teal] (-2.75,0) circle (.05);
\end{tikzpicture}}}

\newcommand{\fivecombblocks}{\scalebox{\scaleCB}{\begin{tikzpicture}[baseline={([yshift=-0ex]current bounding box.center)}]
\draw[scalar] (-4.5,0) -- (-1,0);
\draw[scalar] (-3.75,0) -- (-3.75,0.7);
\draw[scalar] (-2.75,0) -- (-2.75,0.7);
\draw[scalar] (-1.75,0) -- (-1.75,0.7);
\node[] at (-3.25,.2) {\tiny $\Delta_1$};
\node[] at (-2.25,.2) {\tiny $\Delta_2$};
\node[] at (-4.7,.1) {\tiny $\phi^{I_1}$};
\node[] at (-3.75,0.9) {\tiny $\phi^{I_2}$};
\node[] at (-2.75,0.9) {\tiny $\phi^{I_3}$};
\node[] at (-1.75 ,0.9) {\tiny $\phi^{I_4}$};
\node[] at (-0.7,.1) {\tiny $\phi^{I_5}$};
\node[] at (-3.75,-0.2) {\textcolor{teal}{\tiny $C_{ \scalebox{.5}{$\scriptscriptstyle \phi^{I_1} \phi^{I_2} \Op_{\Delta_1} $}}$ }};
\node[] at (-2.75,-0.2) {\textcolor{teal}{\tiny $C_{ \scalebox{.5}{$\scriptscriptstyle \Op_{\Delta_1} \phi^{I_3} \Op_{\Delta_2} $}}$ }};
\node[] at (-1.75,-0.2) {\textcolor{teal}{\tiny $C_{ \scalebox{.5}{$\scriptscriptstyle \Op_{\Delta_2} \phi^{I_4} \phi^{I_5}$}}$ }};
\draw[fill=teal] (-3.75,0) circle (.05);
\draw[fill=teal] (-2.75,0) circle (.05);
\draw[fill=teal] (-1.75,0) circle (.05);
\end{tikzpicture}}}

\newcommand{\sixcombblocks}{\scalebox{\scaleCB}{\begin{tikzpicture}[baseline={([yshift=-0ex]current bounding box.center)}]
\draw[scalar] (-4.5,0) -- (0,0);
\draw[scalar] (-3.75,0) -- (-3.75,0.7);
\draw[scalar] (-2.75,0) -- (-2.75,0.7);
\draw[scalar] (-1.75,0) -- (-1.75,0.7);
\draw[scalar] (-.75,0) -- (-.75,0.7);
\node[] at (-3.25,.2) {\tiny $\Delta_1$};
\node[] at (-2.25,.2) {\tiny $\Delta_2$};
\node[] at (-1.25,.2) {\tiny $\Delta_3$};
\node[] at (-4.7,.1) {\tiny $\phi^{I_1}$};
\node[] at (-3.75,0.9) {\tiny $\phi^{I_2}$};
\node[] at (-2.75,0.9) {\tiny $\phi^{I_3}$};
\node[] at (-1.75 ,0.9) {\tiny $\phi^{I_4}$};
\node[] at (-.75 ,0.9) {\tiny $\phi^{I_5}$};
\node[] at (.3,.1) {\tiny $\phi^{I_6}$};
\node[] at (-3.75,-0.2) {\textcolor{teal}{\tiny $C_{ \scalebox{.5}{$\scriptscriptstyle \phi^{I_1} \phi^{I_2} \Op_{\Delta_1} $}}$ }};
\node[] at (-2.75,-0.2) {\textcolor{teal}{\tiny $C_{ \scalebox{.5}{$\scriptscriptstyle \Op_{\Delta_1} \phi^{I_3} \Op_{\Delta_2} $}}$ }};
\node[] at (-1.75,-0.2) {\textcolor{teal}{\tiny $C_{ \scalebox{.5}{$\scriptscriptstyle \Op_{\Delta_2} \phi^{I_4} \Op_{\Delta_3} $}}$ }};
\node[] at (-.75,-0.2) {\textcolor{teal}{\tiny $C_{ \scalebox{.5}{$\scriptscriptstyle \Op_{\Delta_3} \phi^{I_5} \phi^{I_6}$}}$ }};
\draw[fill=teal] (-3.75,0) circle (.05);
\draw[fill=teal] (-2.75,0) circle (.05);
\draw[fill=teal] (-1.75,0) circle (.05);
\draw[fill=teal] (-.75,0) circle (.05);
\end{tikzpicture}}}
\usepackage{simplewick}
\usepackage{dsfont}
\usepackage{float}
\usepackage{pgfplots}
\pgfplotsset{compat=1.14}

\let\oldbfseries=\bfseries
\let\oldmdseries=\mdseries
\let\oldnormalfont=\normalfont
\renewcommand{\bfseries}{\oldbfseries\boldmath}
\renewcommand{\mdseries}{\oldmdseries\unboldmath}
\renewcommand{\normalfont}{\oldnormalfont\unboldmath}

\makeatletter
\newlength{\apb@width}
\newcommand{\autoparbox}[2][c]{\settowidth{\apb@width}{#2}\parbox[#1]{\apb@width}{#2}}

\makeatother

\DeclareMathOperator{\tr}{tr}

\def\Am{{\mathcal{A}}}

\def\Km{{\mathcal{K}}}

\def\Nm{{\mathcal{N}}}

\def\Pm{{\mathcal{P}}}

\newcommand{\beq}{\begin{equation}}
\newcommand{\eeq}{\end{equation}}

\definecolor{nicegreen}{rgb}{0.1,0.6,0.1}

\newmuskip\pFqskip
\mathchardef\pFcomma=\mathcode`,

\makeatletter
\renewcommand*\env@matrix[1][\arraystretch]{%
  \edef\arraystretch{#1}%
  \hskip -\arraycolsep
  \let\@ifnextchar\new@ifnextchar
  \array{*\c@MaxMatrixCols c}}
\makeatother

\graphicspath{{./figures/}}

\title{\center{Multipoint correlators on the supersymmetric Wilson \\ line defect CFT II: Unprotected operators}}

\author[1]{Julien Barrat,}
\author[2]{Pedro Liendo,}
\author[1]{Giulia Peveri.}

\affiliation[1]{Institut f\"ur Physik und IRIS Adlershof, Humboldt-Universit{\"a}t zu Berlin, Zum Gro{\ss}en Windkanal 2, 12489 Berlin, Germany}
\affiliation[2]{Theory Group, DESY Hamburg, Notkestra{\ss}e 85, D-22607 Hamburg, Germany}

\emailAdd{julien.barrat@hu-berlin.de, pedro.liendo@desy.de, giulia.peveri@physik.hu-berlin.de}

\preprint{HU-EP-22/35-RTG}

\abstract{We continue our study of multipoint correlators of scalar fields on the $1d$ defect CFT generated by inserting operators along the Maldacena-Wilson line in $\Nm = 4$ SYM. We present a weak-coupling recursion relation that captures correlators at next-to-leading order involving an arbitrary number of the elementary scalar fields $\phi^i$ and $\phi^6$, the latter being unprotected. We can then build correlators of composite operators by pinching the scalar fields together. As a demonstration of our method, we give explicit results for correlators containing up to six points. We also expand some selected correlators using recently obtained conformal blocks in the comb and snowflake channel, and check that the extracted low-lying CFT data is consistent with explicit computations.}

\begin{document} 

\setcounter{tocdepth}{2}
\maketitle
\setcounter{page}{1}

\section{Introduction}
\label{sec:intro}

Extended objects or \textit{defects} are important observables in quantum field theory. They enrich the dynamics of a system by probing new physics, and are relevant in particle physics and condensed matter. A \textit{conformal defect} is an extended object in conformal field theory (CFT) that preserves a fraction of the original conformal algebra. Conformal defects in higher dimensional CFTs have received significant attention in recent years, due in part to the revival of the conformal bootstrap program and its natural extension to conformal defects.

In the presence of a defect, several configurations of correlation functions of local operators are possible. One can consider the local operators to either be \textit{bulk} operators, i.e. they live \textit{outside} of the defect, or to be excitations localized on the defect itself. In this paper, we focus on the latter and refer to this setup as correlation functions of \textit{defect} operators. Such correlators are described by a (non-local) CFT in a lower-dimensional space, where the usual constraints coming from conformal invariance apply. In particular, line defects form an interesting setup, as they are the most natural way of implementing $1d$ conformal symmetry, and their dynamics are expected to be simpler than their higher-dimensional counterparts. This makes the study of $1d$ defect CFTs an ideal laboratory where new techniques can be tested. Indeed, modern developments such as analytic functionals were first formulated in $1d$ \cite{Mazac:2016qev,Mazac:2018mdx,Mazac:2018ycv}, before being generalized to higher dimensional CFTs \cite{Mazac:2019shk}.

In this work we focus on a well-known $1d$ model: the supersymmetric Wilson line in $4d$ $\Nm$=4 Super Yang-Mills (SYM), defined as
\begin{equation}
\Wm_C := \frac{1}{N} \text{tr}\, \Pm \exp \int_C d\tau\, (\dot{x}_\mu A^\mu (x) + 
\sqrt{\dot x^{\mu}\, \dot x_{\mu}}\, \theta^I \phi^I (x))\,,
\label{eq:WC}
\end{equation}
where $\theta^{I}$ ($I=1,\ldots, 6$) is a $SO(6)$ vector parametrizing a path on $S^{5}$.  
The coupling is such that from a $10d$ perspective the path $C$ is light-like:  $\dot x^{M}=\{
\dot x^{\mu}, \theta^{i} \sqrt{\dot x^{\mu}\, \dot x_{\mu}}\}$ with $\dot x^{M} \dot x_{M}=0$. This operator is locally half-BPS and conformally invariant, and the supersymmetry enhances its total symmetry to a powerful superconformal algebra. For special geometries of the path $C$, such as the infinite straight line, the expectation value of this operator was obtained at all orders by summing up Feynman diagrams \cite{Erickson:2000af,Drukker:2000rr}; a result which was later confirmed rigorously using supersymmetric localization \cite{Pestun:2007rz,Pestun:2009nn}.  The $1d$ CFT can be generated by inserting local excitations along the line, and the fact that the resulting theory is conformal, supersymmetric and embedded in $\Nm=4$ SYM places it at a crossroads where several modern techniques converge, like perturbation theory \cite{Cooke:2017qgm,Kiryu:2018phb,Barrat:2021tpn}, integrability \cite{Correa:2018fgz,Grabner:2020nis,Cavaglia:2021bnz,Cavaglia:2022qpg,Cavaglia:2022yvv}, localization \cite{Giombi:2009ds, Giombi:2012ep, Billo:2018oog,Giombi:2018qox,Giombi:2018hsx,Beccaria:2020ykg}, holography \cite{Buchbinder:2012vr, Giombi:2017cqn} and the conformal bootstrap \cite{Liendo:2018ukf,Barrat:2021yvp,Ferrero:2021bsb}. In this work we compute multipoint correlators using standard perturbative techniques. One of our motivations is the multipoint conformal bootstrap in general, and in particular, recent developments in the study of multipoint correlators in CFT \cite{Fortin:2019zkm,Fortin:2020yjz,Fortin:2020bfq,Fortin:2020zxw,Buric:2020dyz,Buric:2021ywo,Buric:2021ttm,Buric:2021kgy,Bercini:2020msp,Antunes:2021kmm}. Although multipoint correlators are often technically challenging, $1d$ CFTs offer the advantage that the number of cross-ratios is reduced and the kinematics simplifies.

Our goal is to obtain efficient recursion relations that reproduce general $n$-point correlators built out of the single-trace fundamental scalars $\phi^I(\tau)$ at next-to-leading order in the couping constant $\lambda:= g^2 N$. This work is then a natural continuation of our previous paper \cite{Barrat:2021tpn}, where we focused on (single-trace) protected operators formed by inserting $\phi^i(\tau)$ ($i=1,\ldots, 5$). In this work we extend that analysis to include \textit{unprotected} operators. The most studied example is the fundamental field $\phi^6$ itself, which is the only scalar of length $L=1$ that couples to the Wilson line for the choice $\theta = (0,0,0,0,0,1)$. In many ways, this operator can be seen as the $1d$ analog of the famous Konishi operator in $4d$ $\Nm=4$ SYM: it is the lowest-dimensional unprotected operator at weak coupling, and it is not degenerate. Perhaps the main difference is that it flows to a ``two-particle" state at strong coupling (i.e. $\Delta = 2$), whereas the Konishi operator decouples in this limit. As a result, the conformal dimension of $\phi^6$ has been determined up to five loops at weak coupling \cite{Grabner:2020nis} and up to four loops at strong coupling \cite{Ferrero:2021bsb}.

The structure of the paper is as follows. In section \ref{sec:preliminaries} we review the construction of the $1d$ defect CFT defined along the Maldacena-Wilson line and give the perturbative rules of $\Nm = 4$ SYM. In section \ref{sec:correlators} we present the recursion relations for computing correlation functions of single-trace scalar operators up to next-to-leading order. We present applications of these formulae in section \ref{sec:applications}, including the computation of anomalous dimensions and higher-point correlators of protected and unprotected operators. Section \ref{sec:blockexpansion} is dedicated to the expansion of the correlators $\vvev{\phi^6 \ldots \phi^6}$ in conformal blocks, both in the comb and snowflake channels, and to the analysis of the corresponding CFT data. Finally we review our results in section \ref{sec:conclusions} and discuss possible future directions.
\section{Preliminaries}
\label{sec:preliminaries}

In this section we introduce the $1d$ defect CFT defined by inserting operators along the Wilson line. We review how correlation functions can be constructed and give the perturbative rules of $\Nm = 4$ SYM.

\subsection{The $1d$ defect CFT}
\label{subsec:defectCFT}

Our focus in this paper is the Maldacena-Wilson line operator in $4d$ $\Nm = 4$ SYM, i.e. the extended operator defined in \eqref{eq:WC} and for which the path is a straight line:
\begin{equation}
\Wl := \frac{1}{N} \tr \mathcal{P} \exp \int_{-\infty}^{\infty} d\tau\, \bigl( i 
\dot{x}^\mu \tensor{A}{_{\smash{\mu}}}(\tau) + | \dot{x}|\, \tensor{\phi}{^6}(\tau) \bigr) \,.
\label{eq:wilsonline}
\end{equation}
Here we have chosen the scalar $\phi^6$ to be the one coupling to the line by setting $\theta = (0,0,0,0,0,1)$ in \eqref{eq:WC}. Note that we have Wick-rotated to Euclidean space and defined the path such that the line extends in the Euclidean time direction, i.e. $\dot{x}_\mu = (0,0,0,1)$ and $|\dot{x}|=1$. The straight Maldacena-Wilson line is a half-BPS operator, and its expectation value is just
\begin{equation}
\vev{\Wl} = 1\,.
\end{equation}

One can consider this extended operator as a \textit{defect}. In Minkowski space, it corresponds to a point-like impurity in the $3d$ space, which evolves in time.  As a consequence, if we consider the defect to be part of the vacuum, the conformal symmetry of $\Nm = 4$ SYM is broken from $SO(4,2)$ to $SO(1,2) \times SO(3)$.  If we restrict ourselves to operators inserted \textit{on} the line (as explained in more detail in the next subsection), then the symmetry group $SO(1,2)$ corresponds to a $1d$ CFT for which the representations carry the quantum number $\Delta$, which is the scaling dimension of the operators. On the other hand, the subgroup $SO(3)$ refers to rotations orthogonal to the defect, which in this $1d$ picture is an internal symmetry with quantum number $s$ (spin).

Because of the presence of the scalar field $\phi^6$ in \eqref{eq:wilsonline}, the defect also breaks the $R$-symmetry from $SO(6)_R$ to $SO(5)_R$. This choice entails that $SO(5)_R$ corresponds to the five scalars $\phi^i\, (i =1, \ldots, 5)$ which do not couple to the line. In this setup, the full superconformal algebra $\mathfrak{psu}(2,2|4)$ of $\Nm=4$ SYM breaks into the $\Nm=8$ superconformal quantum mechanics algebra $\mathfrak{osp}(4^*|4)$.

In this work we consider correlation functions of operators in the scalar sector, which involve only the six fundamental scalar fields $\phi^I (\tau)$ ($I=1, \ldots, 6$) of the bulk theory. Operators are constructed by effectively inserting them inside the trace of the Wilson line and for this reason we refer to them as \textit{insertions}. Moreover we consider only \textit{single-trace} representations of the algebra. Details about these single-trace insertions can be found in the next subsection, where we formally introduce the correlators.

\subsection{Correlation functions}
\label{subsec:correlators}

\begin{figure}
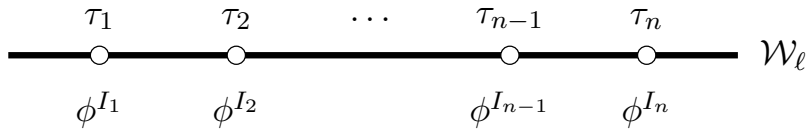

\centering
\scalebox{1.2}{\wilsonline}
\caption{Representation of the correlation functions \eqref{eq:correlators} in the $1d$ defect CFT defined by inserting operators along the Maldacena-Wilson line. At the points $\tau_1\,, \ldots\,, \tau_n$, scalar fields are inserted \textit{inside} the trace of the Wilson line operator.}
\label{fig:insertions}
\end{figure}

The $n$-point correlation functions of the defect single-trace operators are to be understood in the following way:
\begin{equation}
\vev{\phi^{I_1}\, \ldots \phi^{I_n}}_{1d} := \frac{1}{N} \vev{\tr \Pm \left[ \phi^{I_1}\, \ldots \phi^{I_n} \exp \int_{-\infty}^{\infty} d\tau \bigl( i \tensor{\dot{x}}{^{\smash{\mu}}} \tensor{A}{_{\smash{\mu}}} + | \dot{x} |\, \tensor{\phi}{^6} \bigr) \right]}_{4d}\,,
\label{eq:correlators}
\end{equation}
where we suppressed the dependency on $\tau_1, \ldots, \tau_n$ (for the local insertions) and on $\tau$ (for the Wilson line itself) for compactness. Without loss of generality we consider the $\tau$'s to be ordered i.e. $\tau_1 < \tau_2 < \ldots < \tau_n$.
This type of correlators is illustrated in figure \ref{fig:insertions}.
The brackets on the left-hand side refer to correlators in the $1d$ defect theory, while the ones on the right-hand side correspond to correlators in the $4d$ bulk theory. From now on $\vev{\ldots}$ always refers to $1d$ correlators, hence we drop the subscript.

These correlation functions correspond to single-trace operators\footnote{In principle \textit{multi-trace} operators with the same quantum numbers can also be constructed. See footnote 4 in \cite{Barrat:2021tpn} for more detail.}, in the sense that there is only one overall color trace in \eqref{eq:correlators}. This is different from the bulk theory case, where each operator carries its own trace. This property is specific to the defect theory and will be crucial later on for constructing correlators involving operators of higher $R$-charge. This can be done by bringing two operators close to each other, we refer to this limit as \textit{pinching}, and is explained in more detail at the end of this subsection.

In section \ref{sec:applications}, where we present perturbative results, we consider \textit{unit-normalized} correlation functions, which are defined in the following way:
\begin{equation}
\vvev{\phi^{I_1}\, \ldots \phi^{I_n}} := \frac{ \vev{\phi^{I_1}\, \ldots \phi^{I_n}} }{\sqrt{n_{I_1} \ldots n_{I_n}}}\,,
\label{eq:unitnormalized}
\end{equation}
with $n_I$ the normalization constants related to two-point functions. Indeed this definition is chosen such that
\begin{equation}
\vvev{\phi^{I} (\tau_1) \phi^{J} (\tau_2)} = \frac{\delta^{IJ}}{\tau_{12}^2}\,,
\label{eq:twopt}
\end{equation}
with $\tau_{ij} := \tau_i - \tau_j$. Note that the (classical) scaling dimension of the fundamental scalar fields $\phi^I$ is $\Delta = 1$ due to their origin from a $4d$ bulk theory, which explains the form of the propagator in \eqref{eq:twopt},  in spite of the theory being one-dimensional.

The normalization constants can be obtained by computing two-point functions for the correlators \eqref{eq:correlators}. For the protected operators $\phi^i$, conformal symmetry fixes their form to be
\begin{equation}
\vev{\phi^i (\tau_1) \phi^j (\tau_2)} = \frac{n_i}{\tau_{12}^2} \delta^{ij}\,,
\end{equation}
which follows from the fact that the operators $\phi^i$ have exact conformal dimensions $\Delta = 1$. The normalization constant is known to be \cite{Drukker:2011za,Correa:2012at}
\begin{equation}
n_i=2\,\mathbb{B}(\lambda)= \frac{\sqrt{\lambda}}{2\pi^2}\frac{I_2(\sqrt{\lambda})}{I_1(\sqrt{\lambda})}\,,
\label{eq:normalization1}
\end{equation}
where $\mathbb{B}(\lambda)$ is the Bremsstrahlung function with the leading weak-coupling terms being explicitly $\mathbb{B}(\lambda)= \frac{\lambda}{16 \pi^2}-\frac{\lambda^2}{384 \pi^2} +\Op(\lambda^3)$.

\begin{figure}
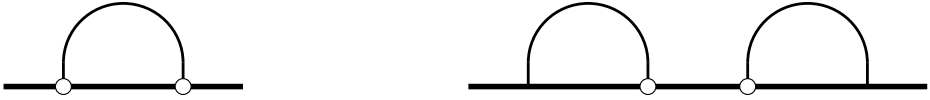

\centering
\begin{subfigure}{.4\textwidth}
  \centering
  \scalebox{1.5}{\twopointpr}
\end{subfigure}%
\begin{subfigure}{.6\textwidth}
  \centering
  \scalebox{1.5}{\twopointnpr}
\end{subfigure}
\caption{Examples of diagrams contributing to the two-point functions of fundamental scalars $\phi^I$. The diagram on the left contributes to both $\vev{\phi^i \phi^j}$ and $\vev{\phi^6 \phi^6}$, while the one on the right contributes only to $\vev{\phi^6 \phi^6}$ and hence is contained in the function $\Lambda(\lambda)$ defined in \eqref{eq:normalization6}. We refer to this type of diagram as \textit{$U$-diagram}. }
\label{fig:Lambda}
\end{figure}

For the unprotected operator $\phi^6$, the non-normalized two-point function reads
\begin{equation}
\vev{\phi^6 (\tau_1) \phi^6 (\tau_2)} = \frac{n_6}{\tau_{12}^{2 \Delta_6}}\,.
\label{eq:2ptphi6}
\end{equation}
Here the normalization constant takes the form:
\begin{equation}
n_6 = 2\,\mathbb{B}(\lambda) +\Lambda(\lambda)\,,
\label{eq:normalization6}
\end{equation}
which can be understood from Feynman diagrams in the following way \cite{Alday:2007he,Correa:2018fgz}: the first term corresponds to diagrams that are common to both $\phi^i$ and $\phi^6$, while the term $\Lambda(\lambda)$ refers to the diagrams unique to $\phi^6$, i.e. the diagrams where the scalar field couples directly to the Wilson line (see figure \ref{fig:Lambda} for an example). Since the tree-level diagrams are the same for $\vev{\phi^i \phi^j}$ and $\vev{\phi^6 \phi^6}$, it is clear that $\Lambda(0) = 0$.

The scaling dimension of $\phi^6$ takes the following form from the weak coupling perspective:
\begin{equation}
\Delta_{\phi^6} = 1 + \sum_{n=1}^\infty \lambda^n \gamma_{\phi^6}^{(n)}\,,
\label{eq:scalingdim6}
\end{equation}
where the anomalous dimensions $\gamma_{\phi^6}^{(n)}$ are known up to order $n=5$ \cite{Grabner:2020nis}. The first correction reads
\begin{equation}
\gamma_{\phi^6}^{(1)} = \frac{1}{4\pi^2}\,.
\label{eq:gamma6}
\end{equation}
As a consequence, the operator $\phi^6$ requires a renormalization procedure. Expanding the two-point function at $\lambda \sim 0$, we have
\begin{equation}
\vev{\phi^6(\tau_1)\phi^6(\tau_2)}= \frac{1}{\tau^2_{12}} \left\lbrace 1+\lambda\gamma_{\phi^6}^{(1)}\, \log\frac{\epsilon^2}{\tau^2_{12}}+\dots \right\rbrace \,,
\label{eq:renormalize1}
\end{equation}
with $\epsilon \to 0$. In order to cancel the divergence, we promote $\phi^6$ to its renormalized version:
\begin{equation}
\phi_R^6(\tau):=\phi^6(\tau) \left\lbrace 1-\lambda\gamma_{\phi^6}^{(1)}\, \log\frac{\epsilon^2}{\mu^2}+\dots \right\rbrace \,,
\end{equation}
where $\mu$ corresponds to some choice of scale. This results in a finite two-point function:
\begin{equation}
\vev{\phi^6_R(\tau_1)\phi^6_R(\tau_2)}= \frac{1}{\tau^2_{12}} \left\lbrace 1+\lambda\gamma_{\phi^6}^{(1)}\, \log\frac{\mu^2}{\tau^2_{12}}+\dots \right\rbrace\,.
\end{equation}
This correlation function is still conformal upon renormalization of the dilatation operator. In the rest of this work we drop the subscript since we always refer to renormalized operators. We also assume normal ordering in the correlators, such that all operators have vanishing one-point functions, as required by conformal symmetry.

Three-point functions of generic operators $\Op_k$ are also kinematically fixed by conformal symmetry and read
\begin{equation}
\vvev{\Op_1 (\tau_1) \Op_2 (\tau_2) \Op_3 (\tau_3)} = \frac{C_{\Op_1 \Op_2 \Op_3}}{\tau_{12}^{\Delta_{123}} \tau_{23}^{\Delta_{231}} \tau_{31}^{\Delta_{312}}}\,,
\label{eq:threept}
\end{equation}
with $\Delta_{ijk} := \Delta_i + \Delta_j - \Delta_k$. For protected operators, an appropriate tensor with $R$-symmetry indices
must be inserted in \eqref{eq:threept} (see e.g. equation (2.9) in \cite{Barrat:2021tpn}), and it follows directly that three-point functions with an odd number of fundamental fields $\phi^i$ vanish since all $R$-symmetry indices must be contracted.

For higher $n$-point functions, conformal symmetry is not strong enough to fix the kinematical form of the correlators.  Nevertheless, it constrains them to be functions of $n-3$ cross-ratios $\chi_i$. For convenience, we now restrict ourselves to correlators of the operators $\phi^I$ introduced above, and we use the following factorized form: 
\begin{equation}
\vvev{\phi^{I_1}\, \ldots \phi^{I_n}}= \mathcal{K} (\tau_1, \Delta_{\phi^{I_1}}; \dots\,; \tau_n, \Delta_{\phi^{I_n}}) \,\mathcal{A}^{I_1 \dots I_n}(\chi_1\,, \ldots\,, \chi_{n-3})\, ,
\label{eq:factorisedcorr}
\end{equation}
where $\chi_i$ are the spacetime cross-ratios. The prefactor reads
\begin{equation}
\mathcal{K} (\tau_1, \Delta_{\phi^{I_1}}; \dots\,; \tau_n, \Delta_{\phi^{I_n}}) = \left(\frac{\tau_{32}}{\tau_{21} \tau_{31}}\right)^{\Delta_{1}}\left(\frac{\tau_{n-1, n-2}}{\tau_{n, n-2} \tau_{n, n-1}}\right)^{\Delta_{n}} \prod_{i=1}^{n-2}\left(\frac{\tau_{i+2, i}}{\tau_{i+1, i} \tau_{i+2, i+1}}\right)^{\Delta_{i+1}}\,,
\label{eq:prefactor}
\end{equation}
with $\tau_{ij}:=\tau_i-\tau_j$ as usual, while we refer to $\Am^{I_1 \ldots I_n}$ as the \textit{reduced correlator}.  The spacetime cross-ratios are defined as
\begin{equation}
\chi_i:= \frac{\tau_{i,i+1}\tau_{i+2,i+3}}{\tau_{i,i+2}\tau_{i+1,i+3}}\,,
\label{eq:crossratio}
\end{equation}
and they are positive-definite. The prefactor as well as the cross-ratios are adopted from \cite{Rosenhaus:2018zqn}\footnote{Note that in \cite{Rosenhaus:2018zqn} the points are ordered as $\tau_1 > \tau_2 > \ldots > \tau_n$}, where they emerge naturally in the derivation of the conformal blocks in the comb channel. These expressions generalize straightforwardly to operators of higher lengths.

To conclude this section, we review the notion of pinching already discussed in \cite{Barrat:2021tpn}. For a given correlation function, one can bring two operators or more together in order to produce single-trace operators with a higher length. For example,
\begin{equation}
\vvev{\phi^{I_1} (\tau_1)\, \ldots\, \underbrace{\phi^{I_{n-1}} (\tau_{n-1}) \phi^{I_n} (\tau_n)}_{\text{two operators of length $1$}}} \overset{\tau_n \to \tau_{n-1}}{\longrightarrow} \vvev{\phi^{I_1} (\tau_1)\, \ldots\, \underbrace{\phi^{I_{n-1}} (\tau_{n-1}) \phi^{I_n} (\tau_{n-1})}_{\text{one operator of length $2$}}}\,.
\end{equation}
This pinching technique allows to construct \textit{any} single-trace scalar operator made of fundamental scalar fields from correlation functions involving operators of length $L=1$.
Note that this is \textit{not} the case in the bulk theory, where the pinching of two single-trace operators produces a double-trace operator, since each operator carries its own trace.

\subsection{Bulk action and propagators}
\label{subsec:action}

As stated above, although the correlators satisfy the axioms of a $1d$ CFT, we perform the computations using the $4d$ action of $\Nm = 4$ SYM. The latter is given by
\begin{align}
S &= \frac{1}{g^2} \int d^4 x\ \text{Tr} \left\lbrace \frac{1}{2} \tensor{F}{_{\mu\nu}} \tensor{F}{^{\mu\nu}} + \tensor{D}{_\mu} \tensor{\phi}{_i} \tensor{D}{^\mu} \tensor{\phi}{^i} - \frac{1}{2} [ \tensor{\phi}{_i} , \tensor{\phi}{_j} ] [ \tensor{\phi}{^i} , \tensor{\phi}{^j} ] \notag \right.\\
& \left. \qquad \qquad \qquad \qquad \qquad \qquad + i \bar{\psi} \tensor{\gamma}{^\mu} \tensor{D}{_\mu} \psi + \bar{\psi} \tensor{\Gamma}{^i} [ \tensor{\phi}{_i} , \psi ] + \tensor{\partial}{_\mu} \bar{c} \tensor{D}{^\mu} c + \xi \left( \tensor{\partial}{_\mu} \tensor{A}{^\mu} \right)^2 \right\rbrace\;,
\label{eq:action}
\end{align}
where we include the ghosts and the gauge fixing. Our conventions are collected in appendix \ref{sec:insertion}. The resulting propagators in Feynman gauge ($\xi = 1$) take the following form in position space:
\begin{subequations}
\begin{align}
\text{Scalars:} \qquad 
& \propagatorS = g^2 \tensor{\delta}{_{ij}} \tensor{\delta}{^{ab}} I_{12}\;, \label{subeq:propagatorS} \\
\text{Gluons:} \qquad 
& \propagatorG = g^2 \tensor{\delta}{_{\mu\nu}} \tensor{\delta}{^{ab}} I_{12}\;, \label{subeq:propagatorG} \\
\text{Gluinos:} \qquad 
& \propagatorF = i g^2 \tensor{\delta}{^{ab}} \slashed{\partial}_{\Delta} I_{12}\;, \label{subeq:propagatorF} \\
\text{Ghosts:} \qquad 
& \propagatorGh = g^2 \tensor{\delta}{^{ab}} I_{12}\;, \label{subeq:propagatorGh}
\end{align}
\label{eq:propagators}%
\end{subequations}
where we have defined for brevity the $4d$ propagator
\begin{equation}
I_{ij} := \frac{1}{(2\pi)^2 x_{ij}^2}\;,
\label{eq:I12}
\end{equation}
with $x^{\mu}_{ij} := x^{\mu}_i - x^{\mu}_j$ and
\begin{equation}
\slashed{\partial}_{\Delta} := \gamma \cdot \frac{\partial}{\partial \Delta}\;, \qquad \qquad \Delta^{\mu} := x_{12}^{\mu}\;,
\end{equation}
with $\gamma_\mu$ the Dirac matrices. The Feynman rules are easy to obtain, and a set of convenient insertion rules can be found in appendix \ref{sec:insertion}.
\section{Correlation functions of fundamental scalar insertions}
\label{sec:correlators}
\begingroup
\allowdisplaybreaks

In this section we derive a series of recursion relations for correlators involving an arbitrary number of insertions of the fundamental scalar fields $\phi^I$ ($I=1\,, \ldots\,, 6$). The case where all the operators are protected (i.e. $I=1\,, \ldots\,, 5$) was already worked out in \cite{Barrat:2021tpn}. In order to implement the remaining field $\phi^6$, we have to distinguish between two cases: the formulae depend on whether an \textit{even} or \textit{odd} number of $\phi^6$ operators is inserted on the Wilson line.

Similarly to \cite{Barrat:2021tpn}, in this section we use for compactness the following shorthand notation for the correlators:
\begin{equation}
A^{I_1 \ldots I_n} := \vev{\phi^{I_1} (\tau_1) \ldots \phi^{I_n} (\tau_n)}\,.
\label{eq:defA}
\end{equation}
Note that this differs from (3.1) in \cite{Barrat:2021tpn} by the fact that we keep the $R$-symmetry indices open. We consider correlation functions that are \textit{not} unit-normalized, since this is the natural normalization to work with when doing perturbative computations. However the results presented in the subsequent sections are going to be unit-normalized. Note that the recursion relation presented in this section is implemented in the ancillary \textsc{Mathematica} notebook and ready to use.

As stated in \cite{Barrat:2021tpn}, if a correlator contains an odd number of protected scalars $\phi^i$ in \eqref{eq:defA}, then it vanishes because of the $R$-symmetry indices. Therefore in the following we consider the number of $\phi^i$ to always be even.

\subsection{Even case}
\label{subsec:evencase}

We start our analysis by studying the case where an \textit{even} number of unprotected scalars $\phi^6$ is included in the correlator. This provides a generalization of the equations (3.3) and (3.5) of \cite{Barrat:2021tpn}.

\subsubsection{Leading order}
\label{subsubsec:evenLO}

We begin with a formula for the leading order.
In this case,
computing correlation functions with an even number of $\phi^6$ operators is the same as computing correlation functions of only protected operators $\phi^i$,  and thus the recursion relation is the same as equation (3.3) in \cite{Barrat:2021tpn},  with the difference that we now keep the $R$-symmetry indices open:
\begin{equation}
A_\text{\tiny{LO}}^{I_1 \ldots I_n}  = \sum_{j=0}^{\frac{n}{2}-1} A_\text{\tiny{LO}}^{I_1 I_{2j+2}} A_\text{\tiny{LO}}^{I_2 \ldots I_{2j+1}} A_\text{\tiny{LO}}^{I_{2j+3} \ldots I_n}\,.
\label{eq:recursioneventree}
\end{equation}
This can be represented diagrammatically as
\begin{equation}
A_\text{\tiny{LO}}^{I_1 \ldots I_n} = \sum_{j=0}^{\frac{n}{2}-1} \recursiontree\,,
\end{equation}
where \treelevelblob\, stands for the leading-order correlation functions $A_\text{\tiny{LO}}$ of appropriate lengths. Arbitrary correlation functions of scalar fields can then be obtained by selecting $R$-symmetry indices, as long as the number of $\phi^6$ is kept even.

In the expression above, the starting values for the recursion are given by the vacuum expectation value and by the two-point functions:
\begin{equation} 
A_\text{\tiny{LO}} = 1\,, \qquad A_\text{\tiny{LO}}^{I_1 I_2} = \frac{\lambda}{8 \pi^2} \frac{\delta^{I_1 I_2}}{\tau_{12}^2}\,,
\label{eq:evenstart}
\end{equation}
the second one corresponding to the leading order of equation \eqref{eq:normalization1} at weak coupling.

\subsubsection{Next-to-leading order}
\label{subsubsec:evenNLO}

We now turn our attention to the next-to-leading order, where the recursion relation becomes more involved.
For arbitrary operators (still with an even number of $\phi^6$), it consists of the diagrams appearing in $\vev{\phi^{i_1} \ldots \phi^{i_n}}$, which must be complemented with the so-called \textit{$U$-diagrams}, i.e. diagrams accounting for the coupling to $\phi^6$ present in the definition of the Maldacena-Wilson line in \eqref{eq:wilsonline}.
An example of such diagrams can be found on the right in figure \ref{fig:Lambda}.
Explicitly, we have
\begin{equation}
A_\text{\tiny{NLO}}^{I_1 \ldots I_n} = \left. A_\text{\tiny{NLO}}^{I_1 \ldots I_n} \right|_\text{old} + \left. A_\text{\tiny{NLO}}^{I_1 \ldots I_n} \right|_\text{new}\,,
\label{eq:oldnew}
\end{equation}
where $\left. A_\text{\tiny{NLO}}^{I_1 \ldots I_n} \right|_\text{old}$ refers to equation (3.5) in \cite{Barrat:2021tpn}\footnote{Note that the notation is slightly different here due to the fact that we keep the indices open. This change is easy to implement in (3.5) of \cite{Barrat:2021tpn}, by removing the null-vectors $u$'s and keeping the $R$-symmetry indices of the fundamental fields open.}, while the second term can be determined by considering all the possible $U$-contractions:
\begin{align}
\left. A_\text{\tiny{NLO}}^{I_1 \ldots I_n} \right|_\text{new} =& \sum_{j=1}^{n-1} \sum_{k=j+1}^n \Biggl( \sum_{l=k}^{n-2} \sum_{m=l+2}^n \recursionphione + \sum_{l=k}^n \sum_{m=j}^{k-1} \recursionphitwo \notag \\
& \qquad\qquad\quad + \sum_{l=0}^{j-3} \sum_{m=l+2}^{j-1} \recursionphithree + \sum_{l=j}^{k-1} \sum_{m=0}^{j-1} \recursionphifour  \notag \\
& \qquad\qquad\quad + \sum_{l=j}^{k-1} \sum_{m=k}^n \recursionphifive  + \sum_{l=j}^{k-3} \sum_{m=l+2}^{k-1} \recursionphisix \notag \\
& \qquad\qquad\quad + \sum_{l=0}^{j-1} \sum_{m=k}^n \recursionphiseven + \sum_{l=0}^{j-1} \sum_{m=j}^{k-1} \recursionphieight
\Biggr) \notag \\
& + \sum_{j=1}^{n-3} \sum_{k=j+2}^{n-1} \Biggl( \sum_{l=k+1}^n \sum_{m=l}^n \recursionphinine + \sum_{l=k+1}^n \sum_{m=k}^{l-1} \recursionphiten \Biggr) \notag \\
& + \sum_{j=2}^{n-2} \sum_{k=j+2}^n \Biggl( \sum_{l=1}^{j-1} \sum_{m=l}^{j-1} \recursionphieleven + \sum_{l=1}^{j-1} \sum_{m=0}^{l-1} \recursionphitwelve \Biggr) \notag \\
& + \sum_{j=1}^{n-3} \sum_{k=j+2}^{n-1} \sum_{l=k+1}^n \sum_{m=0}^{j-1} \recursionphithirteen +  \sum_{j=3}^{n-2} \sum_{k=j+2}^{n} \sum_{l=1}^{j-1} \sum_{m=k}^{n} \recursionphifourteen \notag \\
& + \sum_{j=1}^{n-5} \sum_{k=j+2}^{n-3} \sum_{l=k+1}^{n-2} \sum_{m=l+2}^n \recursionphififthteen\,,
\label{eq:recursionevenNLO}
\end{align}
where every sum should be considered as going in steps of $2$. In the recursion, we find \treelevelblobodd\, to indicate that a leading-order contribution of appropriate \textit{odd} length has to be inserted there. These contributions are derived in the next subsection and are given in equation \eqref{eq:recursionoddtree}.

This recursive expression is lengthy but easy to understand: it corresponds to summing over all the possible $U$-diagrams. When propagators end on the Wilson line without a dot, it means that the integration limit of the $U$-integral goes from the previous propagator to the next. More concretely:
\begin{equation}
\Uoneint\, :=  \int_{\tau_i}^{\tau_j} d\tau_n\, I_{an}\,,
\label{eq:Uoneint}
\end{equation}
where we have not included the leading-order insertions \treelevelblob\, on the right-hand side for the sake of clarity.

The explicit form of this diagrammatic expression is particularly long, therefore we give it in appendix \ref{sec:recursionapp}. It is important to note that two types of $U$-integrals appear in that expression, which are the ones defined in \eqref{eq:Uint} and \eqref{eq:Utwo}.
This formula is implemented in the ancillary notebook and can readily be used for producing arbitrary correlators composed of fundamental scalar fields $\phi^I$ ($I=1\,, \ldots\,, 6$). The technical details related to the recursion relation and to the $U$-integrals can be found in appendices \ref{sec:recursionapp} and \ref{subsubsec:Uintegrals}.

In later sections, we refer to the terms $\left. A_\text{\tiny{NLO}}^{I_1 \ldots I_n} \right|_\text{old}$ appearing in equation \eqref{eq:oldnew} as \textit{building blocks}, since these terms appear in all the correlation functions involving fundamental scalar fields. On the other hand, the second term $\left. A_\text{\tiny{NLO}}^{I_1 \ldots I_n} \right|_\text{new}$ is only relevant when some of the $R$-symmetry indices are set to $I_k = 6$.

To conclude, note that $\left. A_\text{\tiny{NLO}}^{I_1 \ldots I_n} \right|_\text{old}$ contains a recursive term (see equation (3.19) in \cite{Barrat:2021tpn}). As it is obvious from thinking in terms of Feynman diagrams, the \textit{full} expression $A_\text{\tiny{NLO}}^{I_1 \ldots I_n}$ on the left-hand side of \eqref{eq:oldnew} should be used as input for this recursive term.

\subsection{Odd case}
\label{subsec:oddcase}

We now consider the case where an \textit{odd} number of $\phi^6$ appears in the correlators, while the number of protected scalars $\phi^i$ is still kept even. We restrict our analysis to the leading order, since a coupling to the Wilson line already appears here and hence it corresponds to the interacting theory.

Diagrammatically the formula reads
\begin{align}
A_{\text{\tiny{LO}}}^{I_1 \ldots I_n} (1,\ldots , n) =& \sum^{n}_{i=1} \left( \sum_{j=0}^{\frac{i-1}{2}}  \recursionAodd + \sum^{\frac{n}{2}}_{j=\frac{i}{2}} \recursionBodd \right) \notag \\
&+\sum^{n-1}_{i=1} 	\sum_{j=i+2}^{n} \recursionCodd \,,
\label{eq:recursionoddtree}
\end{align}
where the sum in the second line goes in steps of $2$, and where \treelevelblobodd\, are the leading-order correlation functions with an \textit{odd} number of points of the appropriate length.
Again it is fairly easy to understand the formula: these three terms ensure that all the possible $U$-diagrams are represented, either when the propagator of equation \eqref{eq:Uoneint} closes over leading-order contractions (the first two terms) or when the $U$-integral is contained \textit{inside} a leading-order propagator (the third one).

The diagrammatic expression given above can be expressed formally as
\begin{align}
(\ref{eq:recursionoddtree})&= \sum_{i=1}^n \Biggl( \sum_{j=0}^{\frac{i-1}{2}} \frac{\lambda}{8\pi^2} \delta^{i6} U_{i;2j(2j+1)} A^{I_1,\dots,I_{2j}}_{\text{\tiny{LO}}} A^{I_{2j+1},\dots,I_{i-1}}_{\text{\tiny{LO}}} A^{I_{i+1},\dots,I_{n}}_{\text{\tiny{LO}}} \notag \\
&\quad\quad\,\,+\sum_{j=\frac{i}{2}}^{\frac{n}{2}} \frac{\lambda}{8\pi^2} \delta^{i6} U_{i;2j(2j+1)} A^{I_1,\dots,I_{i-1}}_{\text{\tiny{LO}}} A^{I_{i+1},\dots,I_{2j}}_{\text{\tiny{LO}}} A^{I_{2j+1},\dots,I_{n}}_{\text{\tiny{LO}}} \Biggr) \notag \\
&\,\,+ \sum_{i=1}^{n-1} \sum^n_{j=i+2} A^{I_i I_j}_{\text{\tiny{LO}}} A^{I_{1},\dots,I_{i-1}}_{\text{\tiny{LO}}} A^{I_{j+1},\dots,I_{n}}_{\text{\tiny{LO}}} A^{I_{i+1},\dots,I_{j-1}}_{\text{\tiny{LO}}}\,,
\end{align}
\label{eq:recursionoddtreeexp}
again with the sum in the last line going in steps of $2$. The starting values of the recursion are the same as in \eqref{eq:evenstart}, and this formula is also fully implemented in the ancillary notebook.

In this section, we have presented recursion relations that allow to compute arbitrary correlators of fundamental scalar fields $\phi^I$, both at leading and next-to-leading orders for the even case and at leading order for the odd case. We now consider concrete examples of correlators that can be computed using these expressions.

\endgroup
\section{Applications}
\label{sec:applications}
\begingroup
\allowdisplaybreaks

In this section, we gather explicit results for correlators that can be derived using the recursion relations presented in the previous section. More precisely, we compute examples of $n$-point functions that include the unprotected scalar field $\phi^6$, as well as composite operators, for $n=2\,, \ldots\,, 6$, complementing the results of \cite{Barrat:2021tpn}. For operators of lengths $L=1,2$, we obtain the normalization constants and scaling dimensions up to next-to-leading order, and compare the results to the literature when possible.

\subsection{Two-point functions and anomalous dimensions}
\label{subsec:twopoint}

We start by computing two-point functions both for protected and unprotected operators of lengths $L=1,2$. We obtain normalization constants as well as anomalous dimensions, which for the latter can be compared to the existing literature, while to the best of our knowledge the normalization constants are new results. The method presented here can be extended straightforwardly to operators of higher length consisting of the elementary scalar fields $\phi^I$.

\subsubsection{Operators of length $L=1$}
\label{subsubsec:twoptone}

As explained in section \ref{subsec:defectCFT}, there are two distinct operators of length $L=1$, which are the half-BPS operators $\phi^i$ and the unprotected scalar field $\phi^6$. As a sanity check, we compute here their two-point functions in order to compare them to equations \eqref{eq:normalization1}, \eqref{eq:normalization6} and \eqref{eq:gamma6}.

We start with the two-point function $\vev{\phi^i \phi^j}$. At leading order, the recursion relation given in equation \eqref{eq:recursioneventree} trivially produces the following diagram:
\begin{equation}
\vev{\phi^i(\tau_1) \phi^j(\tau_2)}_{\text{LO}} = \twopttree = \frac{\delta^{ij}}{\tau_{12}^2} \frac{\lambda}{8 \pi^2}\,,
\label{eq:twoptidiagLO}
\end{equation}
where the explicit result on the right-hand side simply comes from the starting value given in \eqref{eq:evenstart}.

At next-to-leading order, we use equation \eqref{eq:oldnew} in order to generate the following diagrams:
\begin{equation}
\vev{\phi^i(\tau_1) \phi^j(\tau_2)}_{\text{NLO}} = \twoptSE\, + \twoptY\,,
\label{eq:twoptidiags}
\end{equation}
where the red dots in the second diagram indicate the places where the gluon line should be connected. This corresponds to integrals along the Wilson line, similarly to the case of the $U$-diagrams and as explained in \cite{Barrat:2021tpn} around equation (3.12). The two diagrams are individually divergent and refer to the starting values given in equations (3.10) and (3.14) of \cite{Barrat:2021tpn}. The divergences cancel and the two-point function at next-to-leading order reads
\begin{equation}
\vev{\phi^i(\tau_1) \phi^j(\tau_2)}_{\text{NLO}} = - \frac{\delta^{ij}}{\tau_{12}^2} \frac{\lambda^2}{192 \pi^2}\,,
\label{eq:twoptiNLO}
\end{equation}
in perfect agreement with \eqref{eq:normalization1}.

Analogously we can compute the two-point function of $\phi^6$. At leading order we find that it coincides with $\vev{\phi^i \phi^j}$:
\begin{align}
\vev{\phi^6(\tau_1) \phi^6(\tau_2)}_{\text{LO}} = \left. \vev{\phi^i(\tau_1) \phi^j(\tau_2)}_{\text{LO}} \right|_{i=j} =\, \twopttree\, = \frac{1}{\tau_{12}^2} \frac{\lambda}{8 \pi^2}\,,
\end{align}
and so the function $\Lambda(\lambda)$ defined in \eqref{eq:normalization6} indeed satisfies $\Lambda(0) = 0$.

However, at next-to-leading order we observe that new diagrams contribute:
\begin{align}
\vev{\phi^6(\tau_1) \phi^6(\tau_2)}_{\text{NLO}} =& \left. \vev{\phi^i(\tau_1) \phi^j(\tau_2)}_{\text{NLO}} \right|_{i=j} \notag \\
& +\, \twoptUone\, +\, \twoptUtwo\, +\, \twoptUthree \notag \\
& +\, \twoptUfour\, +\, \twoptUfive\, +\, \twoptUsix \notag \\
& +\, \twoptUseven\, +\, \twoptUeight\, + \Op(\lambda^3)\,.
\end{align}
The new diagrams are $U$-integrals, which are the ones contributing to the function $\Lambda(\lambda)$. Using the integrals given in appendix \ref{subsubsec:Uintegrals} and following the renormalization procedure described in section \ref{subsec:correlators}, we find the following result for the leading and next-to-leading orders combined:
\begin{equation}
\vev{\phi^6(\tau_1) \phi^6(\tau_2)} = \frac{1}{\tau_{12}^2} \frac{\lambda}{8 \pi^2} \left(1 - \frac{\lambda}{24} \frac{6+\pi^2}{\pi^2} + \Op(\lambda^2) \right) \left( 1 - \frac{\lambda}{4\pi^2} \log \tau_{12}^2  + \Op(\lambda^2) \right)\,.
\end{equation}
This factorized form is useful for reading off the normalization coefficient as well as the anomalous dimension, since it can be compared to \eqref{eq:2ptphi6}. The first-order correction to the scaling dimension reads
\begin{equation}
\gamma^{(1)}_{\phi^6} = \frac{1}{4\pi^2}\,,
\label{eq:gamma6two}
\end{equation}
in perfect agreement with \eqref{eq:gamma6}, while the normalization constant is
\begin{equation}
n_6 = \frac{\lambda}{8 \pi^2} \left(1 - \frac{\lambda}{24}  \frac{6+\pi^2}{\pi^2} + \Op(\lambda^2) \right)\,,
\label{eq:norm6}
\end{equation}
which to the best of our knowledge has not been given explicitly in the literature yet.
Comparing this result to \eqref{eq:normalization6}, we determine
\begin{equation}
\Lambda(\lambda) = - \frac{\lambda^2}{32 \pi^4} + \Op(\lambda^3)\,.
\end{equation}

\subsubsection{Operators of length $L=2$}
\label{subsubsec:twopttwo}

We now move our attention to the operators of length $L=2$. Orthogonal eigenstates of the dilatation operator at next-to-leading order have been constructed in \cite{Correa:2018fgz}:
\begin{align}
\begin{split}
O^{ij}_S :=&\ \phi^i \phi^j + \phi^j \phi^i - \frac{2}{5} \delta^{ij} \phi^k \phi^k\,, \\
O^{ij}_A :=&\, \phi^i \phi^j  - \phi^j  \phi^i \,, \\
O^{i}_A :=&\, \phi^6 \phi^i - \phi^i \phi^6 \,, \\
O^{i}_S :=&\, \phi^6 \phi^i + \phi^i \phi^6 \,, \\
O_{\pm} :=&\, \phi^i \phi^i \pm \sqrt{5}\, \phi^6 \phi^6\,.
\end{split}
\label{eq:opL2}
\end{align}
In this case, the operator $O^{ij}_S$ is protected while the other ones are not. In the following, we compute the two-point functions of all these operators up to next-to-leading order.

For the protected operator $O^{ij}_S$, inserting the definition given in \eqref{eq:opL2} results in
\begin{align}
\vev{O^{ij}_S (\tau_1) O^{kl}_S (\tau_2)} =&\, \vev{\phi^i_1 \phi^j_1 \phi^k_2 \phi^l_2} + \vev{\phi^i_1 \phi^j_1 \phi^l_2 \phi^k_2} - \frac{2}{5} \delta^{kl} \vev{\phi^i_1 \phi^j_1 \phi^m_2 \phi^m_2} \notag \\
&+ \vev{\phi^j_1 \phi^i_1 \phi^k_2 \phi^l_2} + \vev{\phi^j_1 \phi^i_1 \phi^l_2 \phi^k_2} - \frac{2}{5} \delta^{kl} \vev{\phi^j_1 \phi^i_1 \phi^m_2 \phi^m_2} \notag \\
&- \frac{2}{5} \delta^{ij} \vev{\phi^m_1 \phi^m_1 \phi^k_2 \phi^l_2} - \frac{2}{5} \delta^{ij} \vev{\phi^m_1 \phi^m_1 \phi^l_2 \phi^k_2} + \frac{4}{25} \delta^{ij} \delta^{kl} \vev{\phi^m_1 \phi^m_1 \phi^n_2 \phi^n_2}\,,
\label{eq:fullexpOS}
\end{align}
where we defined $\phi^i_1 := \phi^i (\tau_1)$ for compactness. Each term can be seen as the \textit{pinching limit} of a four-point function of the fundamental protected scalars $\phi^i$, e.g.
\begin{equation}
\vev{\phi^i_1 \phi^j_1 \phi^k_3 \phi^l_3} = \lim\limits_{(2,4) \to (1,3)} \vev{\phi^i_1 \phi^j_2 \phi^k_3 \phi^l_4}\,.
\end{equation}
The recursion relations \eqref{eq:recursioneventree} and \eqref{eq:oldnew} can be used to efficiently compute each of these terms up to next-to-leading order.

We now illustrate with an example at leading order how the pinching of the recursion relation works. In the planar limit, the four-point function consists of the following two diagrams:
\begin{equation}
\vev{\phi^i_1 \phi^j_2 \phi^k_3 \phi^l_4}_\text{LO} = \fourpttreeone\, +\, \fourpttreetwo\,.
\label{eq:fourptLO}
\end{equation}
In order to generate the first term of \eqref{eq:fullexpOS}, $\vev{\phi^i_1 \phi^j_1 \phi^k_2 \phi^l_2}$, we need to pinch $(\tau_2, \tau_4) \to (\tau_1, \tau_3)$ and then relabel $\tau_3$ to $\tau_2$. Only the second diagram in \eqref{eq:fourptLO} survives this pinching, since the first one results into self-contractions, and we have
\begin{equation}
\vev{\phi^i_1 \phi^j_1 \phi^k_2 \phi^l_2}  =\, \twoptOijLO\,  = \delta^{il} \delta^{jk} \frac{\lambda^2}{64 \pi^4 \tau_{12}^4}\,.
\end{equation}
We can repeat the same procedure for the other terms at leading order and for the next-to-leading order\footnote{See section \ref{subsubsec:fourpt-buildingblocks} and in particular equation \eqref{eq:fourptNLO} for more detail on the four-point function at next-to-leading order.}. We finally obtain
\begin{equation}
\vev{O^{ij}_S (\tau_1) O^{kl}_S (\tau_2)} = 2 \left( \delta^{ik} \delta^{jl} + \delta^{il} \delta^{jk} - \frac{2}{5} \delta^{ij} \delta^{kl} \right) \frac{\lambda^2}{64 \pi^4 \tau_{12}^4} \left(1 - \frac{\lambda}{24} + \Op(\lambda^2) \right)\,.
\end{equation}
In this case there is no factor corresponding to the correction to the scaling dimension since this operator is half-BPS and hence protected ($\Delta = 2$). The normalization constant is
\begin{equation}
n_{O^{ij}_S} = \frac{\lambda^2}{64 \pi^4} \left(1 - \frac{\lambda}{24} + \Op(\lambda^2) \right)\,,
\end{equation}
which agrees with (4.3) in \cite{Barrat:2021tpn} after identifying $(u_1 \cdot u_2)^2 \to 2 \left( \delta^{ik} \delta^{jl} + \delta^{il} \delta^{jk} - \frac{2}{5} \delta^{ij} \delta^{kl} \right)$.\

One can proceed in a similar way for the other unprotected operators in order to read their normalization constants as well as the anomalous dimensions. For example, the two-point function of $\Op_A^{ij}$ can be obtained in the following way:
\begin{equation}
\vev{O_A^{ij} (\tau_1) O_A^{kl} (\tau_2)} =\, \vev{\phi^i_1 \phi^j_1 \phi^k_2 \phi^l_2} - \vev{\phi^i_1 \phi^j_1 \phi^l_2 \phi^k_2} - \vev{\phi^j_1 \phi^i_1 \phi^k_2 \phi^l_2} + \vev{\phi^j_1 \phi^i_1 \phi^l_2 \phi^k_2}\,. 
\end{equation}
Note that there are only correlators of protected operators of length $L=1$ on the right-hand side, but that the pinching operation generates logarithmic divergences that can be related to the anomalous dimension of the operator, as explained in equation \eqref{eq:renormalize1} and below. We find that the normalization constant is
\begin{equation}
n_{O_A^{ij}} = - \frac{\lambda^2}{32 \pi^4} \left(1 - \frac{\lambda}{24} + \Op(\lambda^2) \right)\,,
\end{equation}
while the anomalous dimension turns out to be
\begin{equation}
\gamma^{(1)}_{O_A^{ij}} = \frac{1}{4\pi^2}\,,
\end{equation}
in perfect agreement with \cite{Correa:2018fgz}.

All the other operators can be treated the same way, even when they involve $\phi^6$. For $O^i_A$ we find the normalization constant to be
\begin{equation}
n_{O^i_A} = - \frac{\lambda^2}{32 \pi^4} \left(1 - \frac{\lambda}{24} \frac{6 + \pi^2}{\pi^2} + \Op(\lambda^2) \right)\,,
\end{equation}
while the anomalous dimension reads
\begin{equation}
\gamma^{(1)}_{O^i_A} = \frac{3}{8\pi^2}\,.
\label{eq:anomalousOA}
\end{equation}

Similarly, for $O^i_S$ the normalization constant turns out to be
\begin{equation}
n_{O^i_S} = \frac{\lambda^2}{32 \pi^4} \left(1 - \frac{\lambda}{24} + \Op(\lambda^2) \right)\,,
\end{equation}
and the anomalous dimension is
\begin{equation}
\gamma^{(1)}_{O^i_S} = \frac{1}{8\pi^2}\,.
\end{equation}

Finally, for the last operator $O_{\pm}$ we find
\begin{equation}
n_{O_{\pm}} = \frac{5\lambda^2}{32 \pi^4} \left(1 - \frac{\lambda}{24 \pi^2} \left( \pi^2 - \frac{9}{2} (1 \pm \sqrt{5}) \right) + \Op(\lambda^2) \right)\,,
\end{equation}
and
\begin{equation}
\gamma^{(1)}_{O_{\pm}} = \frac{5\pm\sqrt{5}}{16\pi^2}\,.
\end{equation}

All the anomalous dimensions listed above perfectly match the results of \cite{Correa:2018fgz} for the supersymmetric case $\zeta = 1$.

\subsection{Three-point functions}
\label{subsec:threepoint}

In this section, we compute selected three-point functions using the recursion relations given in section \ref{sec:correlators}. We focus our attention on correlators involving the two operators of length $L=1$,  $\phi^i$ and $\phi^6$, but in the ancillary \textsc{Mathematica} notebook we provide examples of three-point functions involving unprotected operators of length $L=2$ as well.

Note that from now and for the rest of this work, we are going to consider unit-normalized correlation functions, following the definition given in \eqref{eq:unitnormalized} and using the results of subsection \ref{subsec:twopoint}.

\subsubsection{$\vvev{\phi^i \phi^j \phi^6}$}
\label{subsubsec:ij6}

We start by computing the three-point function involving two protected operators $\phi^i$ together with the only unprotected operator of length $L=1$, $\phi^6$.
Using the recursion relation for an odd number of $\phi^6$ operators given in \eqref{eq:recursionoddtree}, we find the following result:
\begin{equation}
\vvev{\phi^i \phi^j \phi^6} = \frac{\vev{\phi^i \phi^j \phi^6}}{n_i \sqrt{n_6}} = \frac{\delta^{ij}}{\tau_{12} \tau_{23} \tau_{31}} \left( - \frac{\sqrt{\lambda}}{2\sqrt{2} \pi} + \ldots \right)\,,
\end{equation}
which yields, by comparison to \eqref{eq:threept}, the OPE coefficient
\begin{equation}
C_{\phi^i \phi^j \phi^6} = - \frac{\sqrt{\lambda}}{2\sqrt{2} \pi} + \Op(\lambda^{3/2})\,.
\label{eq:ij6}
\end{equation}

\subsubsection{$\vvev{\phi^6 \phi^6 \phi^6}$}
\label{subsubsec:666}

The same computation can easily be performed for three unprotected operators $\phi^6$. In this case we obtain
\begin{equation}
C_{\phi^6 \phi^6 \phi^6} = - \frac{3\sqrt{\lambda}}{2\sqrt{2} \pi} + \Op(\lambda^{3/2})\,.
\label{eq:3ptphi6}
\end{equation}
These results are going to be used as consistency checks for the correlation functions that we expand in conformal blocks in section \ref{sec:blockexpansion}.

\subsection{Four-point functions}
\label{subsec:fourpoint}

We move now our attention to the four-point functions that can be computed using the recursion relations of section \ref{sec:correlators}. These are the first correlators that have a non-trivial kinematic dependence.

We consider here three examples, two of which involve the fundamental scalars $\phi^I$ only and one involving a composite operator of length $L=2$. For both types, more correlators can be found in the ancillary notebook.

\subsubsection{Building blocks and $\vvev{\phi^i \phi^j \phi^k \phi^l}$}
\label{subsubsec:fourpt-buildingblocks}

As defined in \eqref{eq:oldnew} and explained in the text,
the $R$-symmetry channels of the correlator $\vvev{\phi^i \phi^j \phi^k \phi^l}$ ($i,j,k,l = 1\,, \ldots\,, 5$) can be used as building blocks for other correlators involving an even number of unprotected operators.
This correlator has been computed in \cite{Kiryu:2018phb} and can also be generated using the recursion relation of \cite{Barrat:2021tpn}. 
We repeat this computation here in order to show how the recursion relations work in this case.

In our convention, the reduced correlator can be extracted from the full correlator following
\begin{equation}
\vvev{\phi^i (\tau_1) \phi^j (\tau_2) \phi^k (\tau_3) \phi^l (\tau_4)} = \frac{1}{\tau_{12}^2 \tau_{34}^2} \Am^{ijkl} (\chi)\,,
\label{eq:corr-ijkl}
\end{equation}
where the conformal prefactor is obtained following \eqref{eq:prefactor}.
The spacetime cross-ratio is defined as
\begin{equation}
\chi := \frac{\tau_{12} \tau_{34}}{\tau_{13} \tau_{24}}\,,
\end{equation}
which satisfies $0 < \chi < 1$, with the ordering of the spacetime points $\tau_1 < \tau_2 < \tau_3 < \tau_4$.

The reduced correlator can be expanded in three $R$-symmetry channels:
\begin{equation}
\Am^{ijkl} (\chi) = \delta^{ij} \delta^{kl} F_0 (\chi) + \delta^{ik} \delta^{jl} \chi^2 F_1 (\chi) + \delta^{il} \delta^{jk} \frac{\chi^2}{(1-\chi)^2} F_2 (\chi)\,,
\label{eq:Aijkl-channels}
\end{equation}
where the prefactors have been chosen such that they correspond to the same channels as in equation (4.8) in \cite{Barrat:2021tpn}, even if here we keep the $R$-symmetry indices open and the correlator is unit-normalized.

These channels (which we will call \textit{building blocks} from now on) obey the following perturbative expansion:
\begin{equation}
F_j (\chi) = \sum_{k=0}^{\infty} \lambda^{k} F^{(k)}_j (\chi)\,.
\label{eq:perturbative}
\end{equation}
At leading order the recursion relation given in equation \eqref{eq:recursioneventree} produces the diagrams given in \eqref{eq:fourptLO}. Unit-normalizing the correlator by following \eqref{eq:unitnormalized} and using the normalization constant computed in \eqref{eq:normalization1} results in the following channels:
\begin{equation}
F_0^{(0)} (\chi) = F_2^{(0)} (\chi) = 1 \,, \quad F_1^{(0)} (\chi) = 0\,.
\label{eq:buildingblocksLO}
\end{equation}

At next-to-leading order, the recursion relation generates the following diagrams:
\begin{align}
\vev{\phi^i_1 \phi^j_2 \phi^k_3 \phi^l_4}_\text{NLO} = & \phantom{+}\,\,\, \,\fourptX\, +\, \fourptHone\, +\, \fourptHtwo \notag \\
& +\, \fourptSEone\, +\, \fourptSEtwo\, +\, \fourptSEthree\, \notag \\
& +\, \fourptYone\,+\, \fourptYtwo\, +\, \fourptYthree \notag \\
& +\, \fourptSEfour\, +\, \fourptYfour\,,
\label{eq:fourptNLO}
\end{align}
where we have used the notation $\phi^i_1 := \phi^i (\tau_1)$ on the left-hand side for compactness. This computation was first performed in \cite{Kiryu:2018phb}, and then repeated in \cite{Barrat:2021tpn} with the use of the recursion relation. The unit-normalized $R$-symmetry channels read
\begin{subequations}
\label{eq:buildingblocksNLO}
\begin{align}
F_0^{(1)} (\chi) =& \frac{1}{8\pi^2} \left( 2 L_R (\chi) + \frac{\ell (\chi,1)}{1-\chi} \right) \,, \\
F_1^{(1)} (\chi) =& - \frac{1}{8 \pi^2} \frac{\ell (\chi,1)}{\chi(1-\chi)} \,, \\
F_2^{(1)} (\chi) =& - \frac{1}{8\pi^2} \left( 2 L_R (\chi) - \frac{\ell (\chi,1)}{\chi} - \frac{\pi^2}{3}  \right) \,.
\end{align}
\end{subequations}
Note that we have used the Rogers dilogarithm, defined as
\begin{equation}
L_R(\chi) := \text{Li}_2 (\chi) + \frac{1}{2} \log (\chi) \log(1-\chi)\,,
\label{eq:Rogers}
\end{equation}
and satisfying the following properties:
\begin{subequations}
\begin{align}
&L_R (x) + L_R (1-x) = \frac{\pi^2}{6}\,, \\
&L_R (x) + L_R(y) = L_R (xy) + L_R \left( \frac{x(1-y)}{1-xy} \right) + L_R \left( \frac{y(1-x)}{1-xy} \right)\,.
\end{align}
\end{subequations}
We also use the following two-variable function introduced in \cite{Barrat:2021tpn}:
\begin{equation}
\ell(\chi_1, \chi_2) := \chi_1 \log \chi_1 - \chi_2 \log \chi_2 + (\chi_2 - \chi_1) \log (\chi_2 - \chi_1)\,.
\label{eq:ell}
\end{equation}
Note that the function $\ell(\chi,1)$ is manifestly crossing-symmetric, i.e.
\begin{equation}
\ell(\chi,1) = \ell(1-\chi,1)\,,
\end{equation}
and it is related to a special limit of the Bloch-Wigner function $D(\chi, \bar{\chi})$ in the following sense:
\begin{equation}
\ell(\chi, 1) = \chi(1-\chi) \lim\limits_{\bar{\chi} \to \chi} \frac{D(\chi, \bar{\chi})}{2(\bar{\chi} - \chi)}\,,
\end{equation}
with
\begin{equation}
D(\chi, \bar{\chi}) = 2 \text{Li}_2 (\chi) - 2 \text{Li}_2 (\bar{\chi}) + \log \chi \bar{\chi} \log \frac{1 - \chi}{1 - \bar{\chi}}\,.
\end{equation}
The function $\ell$, which appears in higher-point functions as well, satisfies the following identities:
\begin{subequations}
\begin{align}
&\ell(\chi_1, \chi_2) + \ell(\chi_2, \chi_1) = i \pi (\chi_1 - \chi_2)\,, \\
&\ell(\chi_1, \chi_2) = \chi_1\, \chi_2\, \ell(\chi_2^{-1}, \chi_1^{-1}) \quad \text{for } 0 < \chi_1 < \chi_2 < 1\,, \\
&\ell(\chi_1, \chi_2) + \ell(1-\chi_2, 1-\chi_1) = \ell(\chi_1,1) - \ell(\chi_2,1)\,.
\end{align}
\end{subequations}

\subsubsection{$\vvev{\phi^6 \phi^6 \phi^6 \phi^6}$}
\label{subsubsec:6666}

We now look at the four-point function of unprotected fundamental fields $\phi^6$. The reduced correlator can be read from
\begin{equation}
\vvev{\phi^6 (\tau_1) \phi^6 (\tau_2) \phi^6 (\tau_3) \phi^6 (\tau_4)} = \frac{1} {\tau_{12}^{2\smash{\Delta_{\phi^6}}} \tau_{34}^{2\smash{\Delta_{\phi^6}}}} \Am^{6666} (\chi)\,.
\label{eq:corr-6666}
\end{equation}
Similarly to \eqref{eq:perturbative}, the reduced correlator obeys the following perturbative expansion:
\begin{equation}
\Am^{6666} (\chi) = \sum_{k=0}^{\infty} \lambda^{k} \Am_{6666}^{(k)} (\chi)\,.
\label{eq:Am6666-pert}
\end{equation}
Note that, as opposed to the case $\vvev{\phi^i \phi^j \phi^k \phi^l}$ presented above, this correlator consists of a \textit{single} $R$-symmetry channel.

At leading order, using (\ref{eq:recursioneventree}) and unit-normalizing we find that it agrees with $\Am_{ijkl}^{(0)}$ when all the indices are set equal
\begin{align}
\Am_{6666}^{(0)} (\chi) &= \left. \Am_{ijkl}^{(0)} (\chi) \right|_{i=j=k=l} = \frac{1 - 2 \chi(1-\chi)}{(1-\chi)^2}\,.
\label{eq:4ptphi6LO}
\end{align}

At next-to-leading order, the conformal prefactor in \eqref{eq:corr-6666} produces logs when expanded around $\lambda \sim 0$ because of the anomalous dimension of $\phi^6$:
\begin{equation}
\frac{1}{\tau_{12}^{2\smash{\Delta_{\phi^6}}} \tau_{34}^{2\smash{\Delta_{\phi^6}}}} = \frac{1}{\tau_{12}^2 \tau_{34}^2} \left( 1 - \lambda\, \gamma^{(1)}_{\phi^6} \log \tau_{12}^2 \tau_{34}^2 + \Op(\lambda^2)\right)\,.
\label{eq:log6666}
\end{equation}
The log term must be taken into account in order to isolate the reduced correlator $\Am_{6666}^{(1)}$ at next-to-leading order. Moreover, as discussed in section \ref{subsec:evencase}, the correlator can be expressed as a sum of building blocks and $U$-diagrams. Applying \eqref{eq:oldnew} and computing the integrals with the help of appendix \ref{sec:integrals} results in the following elegant result:
\begin{align}
\Am_{6666}^{(1)} (\chi) =& \left. \Am_{ijkl}^{(1)} (\chi) \right|_{i=j=k=l}  \notag \\
& + \frac{\lambda}{12 (1-\chi)^2} \left( 1-2\chi(1-\chi) \phantom{\frac{3}{\pi^2}} \right. \notag \\
& \phantom{+ \frac{\lambda}{12 (1-\chi)^2} \left( \right.} \left. + \frac{3}{\pi^2} ( 3\chi(1-\chi) + \chi^2 \log \chi - (1-\chi(2-3\chi)) \log(1-\chi) ) \right)\,.
\label{eq:4ptphi6NLO}
\end{align}
The first line corresponds to the building blocks defined in equations \eqref{eq:Aijkl-channels} and \eqref{eq:buildingblocksNLO}, while the second and third ones are the results of computing the $U$-diagrams.

\subsubsection{$\vvev{\phi^i \phi^j \Op_A^k \Op_A^l}$}
\label{subsubsec:ijOAOA}

We now use our algorithm to compute a four-point function involving the composite operator $\Op_A^i$ introduced in \eqref{eq:opL2}, namely $\vvev{\phi^i \phi^j \Op_A^k \Op_A^l}$. The correlator takes the following form:
\begin{equation}
\vvev{\phi^i (\tau_1) \phi^j (\tau_2) \Op_A^k (\tau_3) \Op_A^l (\tau_4)} = \frac{1} {\tau_{12}^2 \tau_{34}^{2\smash{\Delta_{\Op_A}}}} \widehat{\Am}^{ijkl} (\chi)\,.
\label{eq:corr-ijAkAl}
\end{equation}
Here we use the notation $\widehat{\Am}^{ijkl}$ for the reduced correlator in order to distinguish it from the $\Am^{ijkl}$ used in \eqref{eq:corr-ijkl}.

This reduced correlator consists of \textit{three} $R$-symmetry channels, for which we use the following notation:
\begin{equation}
\widehat{\Am}^{ijkl} (\chi) = \delta^{ij} \delta^{kl} G_0 (\chi) + \delta^{ik} \delta^{jl} \chi^2\, G_1 (\chi) +\delta^{il} \delta^{jk} \frac{\chi^2}{(1-\chi)^2} G_2 (\chi)\,.
\end{equation}

At leading order, the channels are related to the building blocks defined in section \ref{subsubsec:fourpt-buildingblocks} in the following way:
\begin{subequations}
\begin{align}
G_0^{(0)} (\chi) &= F_0^{(0)} (\chi) = 1\,, \\
G_1^{(0)} (\chi) &= 0\,, \\
G_2^{(0)} (\chi) &= \frac{1}{2} F_2^{(0)} (\chi) = \frac{1}{2}\,.
\end{align}
\end{subequations}

At next-to-leading order the anomalous dimension computed in \eqref{eq:anomalousOA} must be taken into account in the same fashion as explained around equation \eqref{eq:log6666}. Applying the recursion relation \eqref{eq:oldnew} for six-point functions, pinching to four-point and unit-normalizing, we find the following elegant results in terms of building blocks:
\begin{subequations}
\begin{align}
G_0^{(1)} (\chi) &= F_0^{(1)} (\chi) - \frac{\lambda}{12} F_0^{(0)} (\chi) - \frac{\lambda}{16\pi^2} \left( \frac{8-9\chi}{1-\chi} + \log(1-\chi) \right)\,, \\
G_1^{(1)} (\chi) &= \frac{1}{2} F_1^{(1)} (\chi)\,, \\
G_2^{(1)} (\chi) &= \frac{1}{2} F_2^{(1)} (\chi) - \frac{\lambda}{48} F_2^{(0)} (\chi) - \frac{\lambda}{16\pi^2} (4 + \log(1-\chi))\,.
\end{align}
\end{subequations}

Several other four-point functions can be found in the ancillary notebook. We conclude this section by also reminding that \textit{all} four-point functions of single-trace scalar operators made of fundamental scalar fields and of arbitrary length $L$ can be obtained up to next-to-leading order by using \eqref{eq:recursioneventree} and \eqref{eq:oldnew} (for an even case of $\phi^6$ insertions) or \eqref{eq:recursionoddtree} (for the odd case), and by pinching the operators in the desired way. 

\subsection{Five-point functions}
\label{subsec:fivepoint}

We now consider the case of five-point functions. When the number of operators is odd, there are no building blocks as for the even case, and the recursion relation provides the leading order of the correlators only.

Five-point functions depend on \textit{two} independent cross-ratios:
\begin{equation}
\chi_1 := \frac{\tau_{12}\tau_{34}}{\tau_{13}\tau_{24}}\,, \qquad \chi_2:=\frac{\tau_{23}\tau_{45}}{\tau_{24}\tau_{35}} \,,
\end{equation}
which are defined following \eqref{eq:crossratio}.

\subsubsection{$\vvev{\phi^i \phi^j \phi^6 \phi^6 \phi^6}$}
\label{subsubsec:ij666}

In this subsection, we compute the correlator of two protected operators $\phi^i$ and three unprotected ones $\phi^6$ at leading order. The correlator can be expressed as
\begin{equation}
\vvev{\phi^i (\tau_1) \phi^j (\tau_2) \phi^6 (\tau_3) \phi^6 (\tau_4) \phi^6 (\tau_5)} =\frac{1}{\tau_{12}^2} \left(\frac{\tau_{42}}{\tau_{32}^{\phantom{}}\tau_{43}^{\phantom{}}\tau_{45}^2}\right)^{\smash{\Delta_{\phi^6}}} \Am^{ij666} (\chi_1, \chi_2)\,,
\end{equation}
following equation (\ref{eq:prefactor}) for the prefactor.

The reduced correlator obtained using the recursion relation (\ref{eq:recursionoddtree}) consists of a \textit{single} $R$-symmetry channel corresponding to the contraction between the indices $i$ and $j$, and it obeys the following perturbative expansion:
\begin{equation}
\Am^{ij666} (\chi_1,\chi_2) = \sum_{k=1}^{\infty} \lambda^{k/2} \Am_{ij666}^{(k)} (\chi_1, \chi_2)\,.
\label{eq:Am1166-pert}
\end{equation}
As it was the case for three-point functions, the leading order is $\Op(\sqrt{\lambda})$, and we obtain
\begin{align}
\Am_{ij666}^{(1)} (\chi_1, \chi_2) = - \frac{1}{2\sqrt{2} \pi} \frac{\chi_1}{1-\chi_1-\chi_2} \left(  \frac{3\chi_2 (1-\chi_2)}{\chi_1} + \frac{1}{1-\chi_2} - \chi_1 - 5 \chi_2 \right)\,.
\label{eq:5ptiphi6}
\end{align}

\subsubsection{$\vvev{\phi^6 \phi^6 \phi^6 \phi^6 \phi^6}$}
\label{subsubsec:66666}

We now want to compute the correlator of five unprotected scalars $\phi^6$ at leading order. It can be factorized in
\begin{equation}
\vvev{\phi^6 (\tau_1) \phi^6 (\tau_2) \phi^6 (\tau_3) \phi^6 (\tau_4) \phi^6 (\tau_5)} =\left( \frac{\tau_{42}}{\tau_{12}^{2}\tau_{32}^{\phantom{}}\tau_{43}^{\phantom{}}\tau_{45}^{2}}\right)^{\smash{\Delta_{\phi^6}}} \Am^{66666} (\chi_1, \chi_2)\,,
\end{equation}
with the prefactor following \eqref{eq:prefactor}.

The reduced correlator obtained using the recursion relation \eqref{eq:recursionoddtree} obeys the following perturbative expansion:
\begin{equation}
\Am^{66666} (\chi_1,\chi_2) = \sum_{k=1}^{\infty} \lambda^{k/2} \Am_{66666}^{(k)} (\chi_1, \chi_2)\,.
\label{eq:Am1166-pert}
\end{equation}
The leading order is again $\Op(\sqrt{\lambda})$ and we obtain the following expression:
\begin{align}
\Am_{66666}^{(1)} (\chi_1,\chi_2) = - \frac{3}{2\sqrt{2} \pi} &  \left( \frac{\chi_1 ( 2 (1-\chi_1) - \chi_1^2)}{1-\chi_1} + \frac{\chi_2 (1-\chi_2) - 1}{1-\chi_2} \right. \notag \\
& \quad \left. + \frac{\chi_1^2 (1-\chi_1)^2}{(1-\chi_1 - \chi_2)^2} + \frac{1 -2\chi_1 (1-\chi_1) (1+\chi_1)}{1-\chi_1 - \chi_2} \right)\,.
\label{eq:5ptphi6}
\end{align}

\subsection{Six-point functions}
\label{subsec:sixpoint}

We now turn our attention to six-point functions of operators of length $L=1$, involving both protected and unprotected scalars, using the recursion relations given in section \ref{subsec:evencase}. As before, the results can be extended to more complicated correlators by combining the formulae for length $L=1$ operators and the pinching technique.

\subsubsection{Building blocks and $\vvev{\phi^i \phi^j \phi^k \phi^l \phi^m \phi^n}$}
\label{subsubsec:sixpt-buildingblocks}

We start by analyzing the six-point function of protected operators and collecting the building blocks useful to express other correlators, including the ones involving the unprotected scalar $\phi^6$. As usual we define a reduced correlator through
\begin{align}
\vvev{\phi^i (\tau_1) \phi^j (\tau_2) \phi^k (\tau_3) \phi^l (\tau_4) \phi^m (\tau_5) \phi^n (\tau_6)} &= \frac{\tau_{24}\tau_{35}}{\tau_{12}^2\tau_{23} \tau_{34}^2\tau_{45} \tau_{56}^2} \Am^{ijklmn} (\chi_1\,, \chi_2\,, \chi_3) \notag \\
&= \frac{1}{\tau_{12}^2 \tau_{34}^2 \tau_{56}^2} \frac{1}{\chi_2} \Am^{ijklmn} (\chi_1\,, \chi_2\,, \chi_3)\,,
\label{eq:6point}
\end{align}
where the notation on the second line turns out to be more convenient for expressing the correlator in terms of $R$-symmetry channels. We define \textit{three} independent spacetime cross-ratios from (\ref{eq:crossratio}):
\begin{equation}
\chi_1 := \frac{\tau_{12}\tau_{34}}{\tau_{13}\tau_{24}}\,, \qquad \chi_2:=\frac{\tau_{23}\tau_{45}}{\tau_{24}\tau_{35}}\,, \qquad \chi_3:= \frac{\tau_{34}\tau_{56}}{\tau_{35}\tau_{46}}\,.
\end{equation}
The reduced correlator consists of \textit{fifteen} $R$-symmetry channels, which we choose to define as follows\footnote{Note that the definition of these channels differs from the convention followed in \cite{Barrat:2021tpn}. This choice is due to the fact that another set of cross-ratios is being used in \eqref{eq:6point}, as well as a different conformal prefactor.}:
\begin{align}
\frac{1}{\chi_2} \Am^{ijklmn} =& \phantom{+ } \delta^{ij} \delta^{kl} \delta^{mn} F_0
+ \delta^{ik} \delta^{jl} \delta^{mn} \chi_1^2 F_1
+ \delta^{il} \delta^{jk} \delta^{mn} \frac{\chi_1^2}{(1-\chi_1)^2} F_{2}
+ \delta^{ij} \delta^{km} \delta^{ln} \chi_3^2 F_3 \notag \\
&+ \delta^{ij} \delta^{kn} \delta^{lm} \frac{\chi_3^2}{(1-\chi_3)^2} F_4
+ \delta^{ik} \delta^{jm} \delta^{ln} \frac{\chi_1^2 \chi_3^2}{(1 - \chi_2)^2} F_5
+ \delta^{im} \delta^{jk} \delta^{lm} \frac{\chi_1^2 \chi_3^2}{(1-\chi_1 - \chi_2)^2} F_6 \notag \\
&+ \delta^{ik} \delta^{jn} \delta^{lm} \frac{\chi_1^2 \chi_3^2}{(1 - \chi_2 - \chi_3)^2} F_7
+ \delta^{il} \delta^{jm} \delta^{kn} \frac{\chi_1^2 \chi_2^2 \chi_3^2}{(1-\chi_1)^2 (1-\chi_2)^2 (1-\chi_3)^2} F_{8} \notag \\
&+ \delta^{il} \delta^{jn} \delta^{km} \frac{\chi_1^2 \chi_2^2 \chi_3^2}{(1-\chi_1)^2 (1-\chi_2-\chi_3)^2} F_{9}
+ \delta^{im} \delta^{jl} \delta^{km} \frac{\chi_1^2 \chi_2^2 \chi_3^2}{(1-\chi_3)^2 (1 - \chi_1 - \chi_2)^2} F_{10} \notag \\
&+ \delta^{im} \delta^{jn} \delta^{kl} \frac{\chi_1^2 \chi_2^2 \chi_3^2}{(1 - \chi_1 - \chi_2)^2 (1 - \chi_2 - \chi_3)^2} F_{11} \notag \\
&+ \delta^{in} \delta^{jl} \delta^{km} \frac{\chi_1^2 \chi_2^2 \chi_3^2}{(1-\chi_1-\chi_2-\chi_3 + \chi_1 \chi_3)^2} F_{12} \notag \\
&+ \delta^{in} \delta^{jk} \delta^{lm} \frac{\chi_1^2 \chi_3^2}{(1-\chi_1-\chi_2-\chi_3 + \chi_1 \chi_3)^2} F_{13} \notag \\
&+ \delta^{in} \delta^{jm} \delta^{kl} \frac{\chi_1^2 \chi_2^2 \chi_3^2}{(1 - \chi_2)^2 (1 - \chi_1 - \chi_2 - \chi_3 + \chi_1 \chi_3)^2} F_{14}\,,
\label{eq:6Rchannels}
\end{align}
where we suppressed the dependency on the spacetime cross-ratios, i.e. $F_j := F_j(\chi_1, \chi_2, \chi_3)$.

As usual, these channels (the \textit{building blocks}) have the following perturbative expansion:
\begin{equation}
F_j (\chi_1,\chi_2,\chi_3) = \sum_{k=0}^{\infty} \lambda^{k} F^{(k)}_j (\chi_1,\chi_2,\chi_3)\,.
\end{equation}

The computation up to next-to-leading order was already performed in \cite{Barrat:2021tpn} without unit-normalizing the correlator. If we do so, at leading order it is easy to determine the building blocks and they read
\begin{equation}
F_0^{(0)} = F_{2}^{(0)} = F_4^{(0)} = F_{13}^{(0)} = F_{14}^{(0)}= 1\,,  \qquad F_j^{(0)} = 0\,.
\label{eq:6ptbuildingblocksLO}
\end{equation}
At next-to-leading order the expressions are cumbersome and we gathered them in an ancillary notebook. As an example we give here the highest $R$-symmetry channel:
\begin{align}
8\pi^2 F_0^{(1)} =& \bar{L}_R \left( \frac{1}{\eta_1} \right) + \bar{L}_R \left( \frac{1-\eta_2}{\eta_{32}} \right) + \bar{L}_R \left( - \frac{\eta_2}{\eta_{32}} \right) + \bar{L}_R \left( \frac{\eta_{31}}{\eta_{32}} \right) \notag \\
& + 2 \left( L_R \left( - \frac{\eta_{21}}{\eta_1} \right)+ L_R \left( - \frac{\eta_{31}}{\eta_1} \right) \right) + \frac{\ell (\eta_1, \eta_2)}{\eta_{21}} - \left( \frac{\eta_2}{\eta_3 \eta_{21}} + \frac{i \pi}{\eta_{31}} \right) \ell (\eta_1, \eta_3) \notag \\
& + \left( \frac{1}{1-\eta_3} + \frac{\eta_1}{\eta_3 \eta_{21}} + \frac{i\pi}{\eta_{32}} \right) \ell(\eta_2,\eta_3) + \frac{\ell(\eta_1,1)}{1-\eta_1} - \left( \frac{1}{1-\eta_3} + \frac{i\pi}{1-\eta_2} \right) \ell(\eta_2,1) \notag \\
& + \frac{\ell(\eta_3,1)}{1-\eta_3} + \frac{i\pi \eta_3}{\eta_{31}} \log \eta_1 - \frac{i\pi (\eta_3 (1-\eta_2) - \eta_2 \eta_{32})}{(1-\eta_2) \eta_{32}} \log \eta_2 \notag \\
& - \frac{i\pi (2 \eta_1 \eta_{32} - \eta_3 (\eta_{31} + \eta_{32}))}{\eta_{31} \eta_{32}} \log \eta_3\,,
\label{eq:6ptF0NLO}
\end{align}
where we have defined the following help variables:
\begin{equation}
\eta_1 := \frac{\chi_1 \chi_2 \chi_3}{(1-\chi_1 - \chi_2)(1-\chi_2 - \chi_3)}\,, \quad \eta_2 := \frac{\chi_2 \chi_3}{(1-\chi_1 - \chi_2)(1- \chi_3)}\,, \quad \eta_3 := \frac{(1-\chi_1) \chi_3}{1-\chi_1 - \chi_2}\,,
\label{eq:newcrossratios}
\end{equation}
with $\eta_{ij} := \eta_i - \eta_j$. Note that with these definitions we have the ordering $0< \eta_1 < \eta_2 < \eta_3 <1$\footnote{In fact these cross-ratios correspond to the ones defined in equation (2.12) of \cite{Barrat:2021tpn}.}. We have used the functions $L_R(\chi)$ and $\ell(\chi_1,\chi_2)$ defined respectively in \eqref{eq:Rogers} and \eqref{eq:ell}, while we introduced for compactness the new function
\begin{equation}
\bar{L}_R (\chi) :=  L_R (1-\chi) - L_R (\chi)\,.
\end{equation}

\subsubsection{$\vvev{\phi^6 \phi^6 \phi^6 \phi^6 \phi^6 \phi^6}$}
\label{subsubsec:666666}

To conclude this section, we give another example of six-point function, namely the case where all the operators are the unprotected elementary scalar $\phi^6$. The reduced correlator is defined through
\begin{align}
\vvev{\phi^6 (\tau_1) \phi^6 (\tau_2) \phi^6 (\tau_3) \phi^6 (\tau_4) \phi^6 (\tau_5) \phi^6 (\tau_6)} &= \left( \frac{\tau_{24}\tau_{35}}{\tau_{12}^2\tau_{23} \tau_{34}^2\tau_{45} \tau_{56}^2} \right)^{\smash{\Delta_{\phi^6}}} \Am^{666666} (\chi_1\,, \chi_2\,, \chi_3) \notag \\
&=  \frac{1}{\tau_{12}^{2\smash{\Delta_{\phi^6}}}\tau_{34}^{2\smash{\Delta_{\phi^6}}}\tau_{56}^{2\smash{\Delta_{\phi^6}}}} \frac{1}{\chi_2^{\smash{\Delta_{\phi^6}}}} \Am^{666666} (\chi_1\,, \chi_2\,, \chi_3)\,.
\end{align}

At leading order, the correlator $\vvev{\phi^6 \phi^6 \phi^6 \phi^6 \phi^6 \phi^6}$ coincides with $\vvev{\phi^i \phi^j \phi^k \phi^l \phi^m \phi^n}$ with $i=j=k=l=m=n$, i.e.
\begin{align}
\frac{1}{\chi_2^{\smash{\Delta_{\phi^6}}}} \Am^{(0)}_{666666} &= \frac{1}{\chi_2} \left. \Am^{(0)}_{ijklmn} \right|_{i=j=k=l=m=n} \notag \\
& = 1 + \frac{\eta_1^2}{(1-\eta_1)^2} + \frac{\eta_{23}^2}{(1-\eta_3)^2} + \frac{\eta_1^2 \eta_{23}^2 (1-2 \eta_3 (1-\eta_3))}{\eta_3^2 (1-\eta_3)^2 \eta_{12}^2} \,,
  \label{eq:666666LO}
\end{align}
where in the second equality we have used again the cross-ratios defined in \eqref{eq:newcrossratios} for compactness.

The next order includes $U$-diagrams as well as the next-to-leading order building blocks $F^{(1)}_j$ and it is significantly more involved. These additional terms, as well as the full correlator, can be found in the ancillary notebook.

Several other examples of six-point functions can be found in the ancillary notebook, and it is straightforward to extend these computations to correlators involving composite operators made of fundamental scalar fields.

\endgroup
\section{Expansions in conformal blocks and checks}
\label{sec:blockexpansion}
\begingroup
\allowdisplaybreaks

In this section we expand some of our correlators in conformal blocks, as a consistency check and to extract CFT data.
Conformal blocks allow us to extract the CFT data, which consist of the scaling dimensions and the OPE coefficients of the operators present in the spectrum. 
For $n \geq 6$, different OPE limits lead to decompositions with different topologies, and therefore there exist multiple $n$-point blocks. We will focus on the so-called \textit{comb} channel for $n=4,5,6$, while for the case of six-point functions we also investigate the \textit{snowflake} channel for two correlators.

In the following, we perform several tests for the results presented in the previous section, by comparing the simplest OPE coefficients (always involving either $\phi^i$ or $\phi^6$) to the results derived in section \ref{subsec:threepoint}. Moreover, we present closed forms for the OPE coefficients at leading order for different correlation functions $\vvev{\phi^6 \ldots \phi^6}$. We should point that the analysis of this section is completely bosonic. A full superconformal analysis requires knowledge of the corresponding superconformal blocks, which are only known for some selected four-point functions.

\subsection{Comb channel}
\label{subsec:comb}

In this section we discuss how to expand the correlators that we obtained in section \ref{sec:applications} in the comb channel. 
This channel consists of taking one by one the OPE of an external operator with an internal operator, as represented in figure \ref{fig:combch}\footnote{The exception being of course the two extremities, where we have to take the OPE of two external operators.}.
Four-point blocks in $d=1$ have been known for a long time \cite{Ferrara:1974ny}, but only recently this work was extended to higher-point functions \cite{Rosenhaus:2018zqn}. Five-point point blocks were also derived for generic dimension $d$ in \cite{Goncalves:2019znr}.
From now on, we are going to specialize our analysis to the case where all the external operators are identical scalar fields of length $L=1$.

\begin{figure}[H]
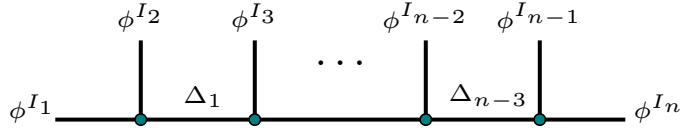

\centering
\combblocks
\caption{Representation of the comb channel for $n$-point correlation functions. The vertices correspond to bosonic OPE coefficients, which can be interpreted as three-point functions in the bosonic theory. For $n$ external operators, there are $n-3$ operators being exchanged.}
\label{fig:combch}
\end{figure}

For a given $R$-symmetry channel, the reduced correlator of such $n$-point functions can be expanded in blocks in the following way:
\begin{align}
\mathcal{A}^{I_1 \dots I_n}
=& \sum_{\Op_1,\dots,\Op_n} C_{\phi^{I_1} \phi^{I_2} \Op_1} C_{\Op_1 \phi^{I_3} \Op_2} \dots C_{\Op_{n-4} \phi^{I_{n-2}} \Op_{n-3}} C_{\Op_{n-3} \smash{\phi^{I_{n-1}}} \smash{\phi^{I_n}}} 
g_{\Delta_1,\dots,\Delta_{n-3}} (\chi_1, \dots, \chi_{n-3})\,,
\label{eq:expansionblocks}
\end{align}
where the $\Delta_k$'s refer to the scaling dimensions of the exchanged operators, and $C_{\Op_1 \Op_2 \Op_3}$ are three-point functions defined in \eqref{eq:threept}. In the case where all the scalar fields are protected, we consider the highest-weight $R$-symmetry channel $F_0$. If the operators are all unprotected, then there is only one $R$-symmetry channel, which is labelled $\Am^{6 \ldots 6}$.

The functions $g_{\Delta_1, \ldots, \Delta_{n-3}}$ correspond to the comb conformal blocks derived in \cite{Rosenhaus:2018zqn} and for identical external operators $\phi$, they are defined as
\begin{align}
&g_{\Delta_1,\dots,\Delta_{n-3}} (\chi_1, \dots, \chi_{n-3}) := \prod_{k=1}^{n-3}\, \chi_k^{\Delta_k} \notag \\
& \qquad \times F_{K}\left[\begin{array}{c}
\left.  \Delta_1, \Delta_1+\Delta_2-\Delta_{\phi}, \ldots, \Delta_{n-4}+\Delta_{n-3}-\Delta_{\phi},\Delta_{n-3}\right.\\
 2 \Delta_1, \ldots, 2 \Delta_{n-3}\end{array} ; \chi_{1}, \ldots, \chi_{n-3} \right]\,,
\label{eq:combblocks}
\end{align} 
where the function $F_K$ is a multivariable hypergeometric function defined by the following expansion:
\small
\begin{align}
& F_{K} \Bigg[\begin{array}{c}a_{1}, b_{1}, \ldots, b_{k-1}, a_{2} \\ c_{1}, \ldots, c_{k}\end{array} ; \, x_1, \ldots, x_{k}\Bigg]  \notag \\
& \qquad \qquad = \sum_{n_{1}, \ldots, n_{k}=0}^{\infty} \frac{\left(a_{1}\right)_{n_{1}}\left(b_{1}\right)_{n_{1}+n_{2}}\left(b_{2}\right)_{n_{2}+n_{3}} \cdots\left(b_{k-1}\right)_{n_{k-1}+n_{k}}\left(a_{2}\right)_{n_{k}}}{\left(c_{1}\right)_{n_{1}} \cdots\left(c_{k}\right)_{n_{k}}} \frac{x_{1}^{n_{1}}}{n_{1} !} \cdots \frac{x_{k}^{n_{k}}}{n_{k} !}\,.
\end{align}
\normalsize
Here $(a)_{n}=\Gamma(a+n) / \Gamma(a)$ refers to the Pochhammer symbol.

\subsubsection{Four-point functions}
\label{subsubsec:fourpointdata}

We start our analysis of the block expansions with four-point functions. Our goal here is to expand the correlators $\vvev{\phi^i \phi^j \phi^k \phi^l}$ and $\vvev{\phi^6 \phi^6 \phi^6 \phi^6}$ following \eqref{eq:expansionblocks} and to check whether the CFT data for the exchanged operator $\Op_\Delta$ with the lowest $\Delta$ agrees with the results computed in section \ref{subsec:threepoint}.

\begin{figure}[H]
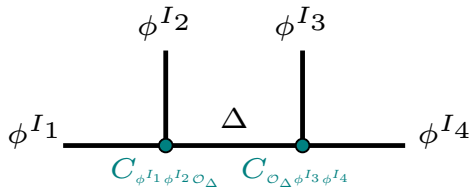

\centering
\fourcombblocks
\caption{Representation of four-point functions in the comb channel. In this case, one operator labelled $\Op_\Delta$ is being exchanged and the OPE coefficients consist of three-point functions squared, when all the external operators are identical.}
\label{fig:fourcombch}
\end{figure}

\paragraph{$\vvev{\phi^i \phi^j \phi^k \phi^l}$.} It can be seen from
figure \ref{fig:fourcombch}
that the four-point function of protected operators $\vvev{\phi^i \phi^j \phi^k \phi^l}$ should contain the three-point function $\vvev{\phi^i \phi^j \phi^6}$, which was computed in equation \eqref{eq:ij6} at leading order, when the exchanged operator is $\Op_\Delta = \phi^6$. 

%
This coefficient can be easily extracted from
\begin{align}
F_0 (\chi) &= 1 + C_{\phi^i \phi^j \phi^6} C_{\phi^6 \phi^k \phi^l} g_{\Delta=1}(\chi) + \ldots \notag \\
&= 1 + C_{\phi^i \phi^j \phi^6} C_{\phi^6 \phi^k \phi^l} \chi + \ldots\,,
\label{eq:expblock}
\end{align}
where the first term corresponds to the exchange of the identity operator $\mathds{1}$ and is $1$ due to the unit-normalization of the two-point function. According to equation \eqref{eq:ij6}, we have
\begin{equation}
C_{\phi^i \phi^j \phi^6} C_{\phi^6 \phi^k \phi^l} = \frac{\lambda}{8 \pi^2} + \Op(\lambda^2)\,.
\label{eq:tocheck}
\end{equation}

We now expand the correlator at $\chi \sim 0$ for the leading and next-to-leading orders, and compare the order $\Op(\chi)$ to \eqref{eq:expblock} and \eqref{eq:tocheck}. From \eqref{eq:buildingblocksLO}, we see that
\begin{equation}
F^{(0)}_0 (\chi) = 1\,,
\end{equation}
and this implies that $C_{\phi^i \phi^j \phi^6} C_{\phi^6 \phi^k \phi^l}$ vanishes at $\Op(\lambda^0)$ as predicted by \eqref{eq:tocheck}. For the next order, we expand \eqref{eq:buildingblocksNLO} to find that
\begin{equation}
F^{(1)}_0 (\chi) = \frac{1}{8\pi^2} \chi + \ldots\,,
\end{equation}
which is in perfect agreement with \eqref{eq:expblock} and \eqref{eq:tocheck}.

\paragraph{$\vvev{\phi^6 \phi^6 \phi^6 \phi^6}$.} We focus now our attention on the four-point function of \textit{unprotected} operators $\phi^6$. In this case, it is clear from $R$-charge conservation that the only operator with (bare) scaling dimension $\Delta = 1$ that can appear in the exchange is the unprotected scalar $\phi^6$ itself, and thus the correlator $\vvev{\phi^6 \phi^6 \phi^6 \phi^6}$ is expected to contain the three-point function $\vvev{\phi^6 \phi^6 \phi^6}$ in its expansion. This coefficient was computed in \eqref{eq:3ptphi6} and can be compared to the four-point function obtained in section \ref{subsubsec:6666}.
%
Expanding the correlator in blocks following \eqref{eq:expansionblocks}, we find
\begin{align}
\mathcal{A}^{6666} (\chi) &= 1 + C_{\phi^6 \phi^6 \phi^6}^2 g_{1} (\chi) + \ldots \notag \\
&= 1 + C_{\phi^6 \phi^6 \phi^6}^2 \chi + \ldots\,,
\end{align}
which we compare to the results listed in section \ref{subsubsec:6666}.

From \eqref{eq:4ptphi6LO}, we find that the correlator at leading order can be expanded as
\begin{equation}
\mathcal{A}^{(0)}_{6666} (\chi) = 1+\chi^2+\ldots \,,
\end{equation}
and thus we observe that $C_{\phi^6 \phi^6 \phi^6}^2$ vanishes at $\Op(\lambda^0)$ as expected from \eqref{eq:3ptphi6}, since there is no term of order $\Op(\chi)$. Again the first term corresponds to the exchange of the identity operator $\mathds{1}$ and it is $1$ due to the unit-normalization of the two-point function.

We are also able to derive a closed-form expression for the OPE coefficients with arbitrary $\Delta$:
\begin{equation}
C_{\phi^6 \phi^6 \Op_\Delta} C_{\Op_\Delta \phi^6 \phi^6} \rvert_{\Op(\lambda^0)} = \frac{4 \sqrt{\pi}\, (\Delta-1)\, \Gamma(\Delta+1)}{4^{\Delta}\,\Gamma(\Delta-\frac{1}{2})}\,.
\end{equation}
%
We expect that there exist several operators corresponding to the bare scaling dimensions $\Delta>1$, and thus that these coefficients are in fact \textit{averages} of three-point functions.

We now move to the next order and we expand the expression of the correlator in (\ref{eq:4ptphi6NLO})
to find
\begin{equation}
\mathcal{A}^{(1)}_{6666} (\chi) = \frac{9}{8 \pi^2} \chi + \ldots \,.
\end{equation}
This coefficient should be compared to equation \eqref{eq:3ptphi6}, which predicts
\begin{equation}
C_{\phi^6 \phi^6 \phi^6}^2 = \frac{9 \lambda}{8 \pi^2} + \Op(\lambda^2)\,,
\end{equation}
and thus we observe a perfect match between the OPE coefficient obtained from the four-point function and the three-point function computed using the recursion relation.

\subsubsection{Five-point functions}
\label{subsubsec:fivepointdata}

We now move to the five-point function $\vvev{\phi^6 \phi^6 \phi^6 \phi^6 \phi^6}$ that has been discussed in section \ref{subsubsec:66666}, and we perform analogous checks as for the four-point functions.
Note that there is no five-point function of protected operators of length $L=1$ only, hence we restrict our attention to the five-point function of unprotected scalars.

\begin{figure}[H]
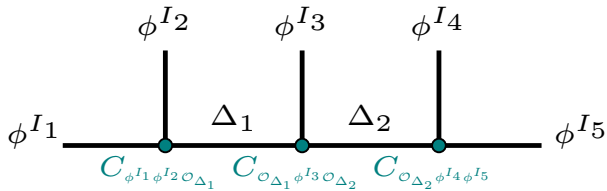

\centering
\fivecombblocks
\caption{Representation of five-point functions in the comb channel. Here there are two operators being exchanged, labelled in the diagram by the scaling dimensions $\Delta_1$ and $\Delta_2$. The OPE coefficients consist of a product of three three-point functions.}
\label{fig:fivecombch}
\end{figure}

The comb channel for five-point functions is represented in figure \ref{fig:fivecombch}. In this case, we are interested in checking the three-point function $\vvev{\phi^6 \phi^6 \phi^6}$, which can be accessed e.g. by setting $\Op_1 = \phi^6$, $\Op_2 = \mathds{1}$.
The expansion of the correlator in blocks up to this term thus reads
\begin{align}
\mathcal{A}^{66666} (\chi_1,\chi_2) &= C_{\phi^6 \phi^6 \phi^6} g_{1,0} (\chi_1,\chi_2) + \ldots \notag \\
&= C_{\phi^6 \phi^6 \phi^6} \chi_1 + \ldots\,,
\end{align}
where the OPE coefficient is just $C_{\phi^6 \phi^6 \phi^6}$ because $C_{\phi^6 \phi^6 \mathds{1}} = 1$. Note that in this case there is no term corresponding to the exchange of two identity operators, since the OPE coefficient $C_{\phi^6 \phi^6 \mathds{1}} C_{\mathds{1} \phi^6 \mathds{1}} C_{\mathds{1} \phi^6 \phi^6}$ vanishes due to the presence of a one-point function in the middle.

Here we only look at the leading order $\Op(\sqrt{\lambda})$ given in equation \eqref{eq:5ptphi6}, which upon expanding at $\chi_1, \chi_2 \sim 0$ reads
\begin{equation}
\Am_{66666}^{(0)} (\chi_1,\chi_2) = \frac{3}{2\sqrt{2} \pi} \chi_1 + \ldots\,,
\end{equation}
and thus we observe a perfect agreement of $C_{\phi^6 \phi^6 \phi^6}$ with equation \eqref{eq:3ptphi6}.

We are able to also derive a closed-form expression for the OPE coefficients at leading order:
\begin{align}
\left. C_{\phi^6 \phi^6 \Op_{\Delta_1}} C_{\Op_{\Delta_1} \phi^6 \Op_{\Delta_2}} C_{\Op_{\Delta_2} \phi^6 \phi^6} \right|_{\Op(\sqrt{\lambda})} =& \frac{12 \sqrt{2} \sqrt{\lambda}}{4^{\Delta_1 + \Delta_2}} \frac{\Gamma(\Delta_1 + \Delta_2)}{\Gamma(\Delta_1 - 1/2) \Gamma(\Delta_2 - 1/2)} \notag \\
& \times (\Delta_1 (\Delta_1 - 1) + \Delta_2 (\Delta_2 - 1) \delta_{\Delta_1,1})
\end{align}
with $\Delta_1 < \Delta_2$.

\subsubsection{Six-point functions}
\label{subsubsec:sixpointdata}

We continue our analysis of the comb channel with the six-point functions of protected fundamental scalars $\vvev{\phi^i \phi^j \phi^k \phi^l \phi^m \phi^n}$ and of unprotected ones $\vvev{\phi^6 \phi^6 \phi^6 \phi^6 \phi^6 \phi^6}$.

\begin{figure}[H]
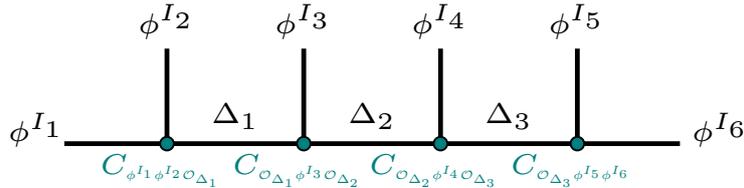

\centering
\sixcombblocks
\caption{Representation of six-point functions in the comb channel. Three operators are being exchanged, and the OPE coefficients consist of products of four three-point functions.}
\label{fig:sixcombch}
\end{figure}

\paragraph{$\vvev{\phi^i \phi^j \phi^k \phi^l \phi^m \phi^n}$.} As in section \ref{subsubsec:fourpointdata}, we can expand in conformal blocks the six-point function of protected operators studied in section \ref{subsubsec:sixpt-buildingblocks}, and compare the three-point function $\vvev{\phi^i \phi^j \phi^6}$ computed in \eqref{eq:ij6} with the prediction obtained from the correlator. The comb channel for this correlator is represented in figure \ref{fig:sixcombch}, and it is easy to see that the lowest coefficient that can be checked corresponds to setting $\Delta_1 = \Delta_2 = 1$, $\Delta_3 = 0$, for which the exchanged operators can only be $\Op_1 = \phi^6$, $\Op_2 = \phi^h\, (h=1,\ldots,5)$, $\Op_3 = \mathds{1}$, due to conservation of the $R$-charge. Noticing that the OPE coefficient vanishes when one $\Delta$ is equal to $1$ and the two other $\Delta$'s are $0$, we can expand the highest-weight channel $F_0$ in blocks in order to compare to that coefficient. We find that
\begin{equation}
F_0 (\chi_1, \chi_2, \chi_3) = 1 + C_{\phi^i \phi^j \phi^6} C_{\phi^6 \phi^k \phi^h} C_{\phi^h \phi^l \mathds{1}} C_{\mathds{1} \phi^m \phi^n} \chi_1 \chi_2 + \ldots\,,
\label{eq:exp6ptF0}
\end{equation}
where we note that $C_{\phi^h \phi^l \mathds{1}} C_{\mathds{1} \phi^m \phi^n} = 1$, due to the unit-normalization of the two-point function. This is also the reason why the leading term is $1$, in perfect analogy with the case of the four-point function. 



We now expand the $R$-symmetry channel $F_0$ of the six-point function studied in section \ref{subsubsec:sixpt-buildingblocks} in order to check whether we find a match for the OPE coefficient mentioned above. At leading order we see from equation \eqref{eq:6ptbuildingblocksLO} that
\begin{equation}
F^{(0)}_0 (\chi_1, \chi_2, \chi_3) = 1\,,
\end{equation}
which matches the expectation that $C_{\phi^i \phi^j \phi^6} C_{\phi^6 \phi^k \phi^h}$ vanishes at order $\Op(\lambda^0)$, as it was the case for the four-point function as well.

At next-to-leading order, expanding equation \eqref{eq:6ptF0NLO} at $\chi_1, \chi_2, \chi_3 \sim 0$ results in
\begin{equation}
F^{(1)}_0 (\chi_1, \chi_2, \chi_3) = \frac{1}{8 \pi^2} \chi_1 \chi_2 + \ldots\,,
\end{equation}
where the coefficient of $\chi_1 \chi_2$ is to be identified with $C_{\phi^i \phi^j \phi^6} C_{\phi^6 \phi^k \phi^h}$ according to \eqref{eq:exp6ptF0}. We observe a perfect match with equation \eqref{eq:tocheck}.

\paragraph{$\vvev{\phi^6 \phi^6 \phi^6 \phi^6 \phi^6 \phi^6}$.} Our last expansion in the comb channel is the six-point function of unprotected operators discussed in subsection \ref{subsubsec:666666}. 
This correlator is expected to contain the three-point function $\vvev{\phi^6 \phi^6 \phi^6}$, which can be checked against \eqref{eq:3ptphi6}. This coefficient can be accessed e.g. by setting as (bare) scaling dimensions $\Delta_1 = \Delta_2 = 1$ and $\Delta_3 = 0$. As for the previous cases, 
the exact correlator can be expanded in blocks and reads
\begin{align}
\mathcal{A}^{666666} (\chi_1,\chi_2,\chi_3) &= 1 + C_{\phi^6 \phi^6 \phi^6}^2 g_{1,1,0} (\chi_1\,, \chi_2\,, \chi_3) + \ldots \notag \\
&= 1 + C_{\phi^6 \phi^6 \phi^6}^2 \chi_1 \chi_2 + \ldots\,,
\end{align}
where the $1$ comes from the exchange of identity operators as always, and the OPE coefficient is just $C_{\phi^6 \phi^6 \phi^6}^2$ because of $C_{\phi^6 \phi^6 \mathds{1}}^2=1$. Other lower combinations such as $\Delta_1 = 1$, $\Delta_2 = \Delta_3 = 0$ vanish because one-point functions are zero in CFT.

We now extract this coefficient at leading and next-to-leading orders and compare it to the direct computation. At leading order, expanding \eqref{eq:666666LO} at $\chi_1\,, \chi_2\,, \chi_3 \sim 0$ gives
\begin{equation}
\Am^{(0)}_{666666} (\chi_1,\chi_2,\chi_3) = 1 + \chi_1^2 + \ldots\,,
\end{equation}
and thus $C_{\phi^6 \phi^6 \phi^6}^2$ vanishes at order $\Op(\lambda^0)$ as predicted by \eqref{eq:3ptphi6}.

We are also able to determine a closed form for the OPE coefficients at leading order:
\begin{align}
\left. C_{\phi^6 \phi^6 \Op_{\Delta_1}} C_{\Op_{\Delta_1} \phi^6 \Op_{\Delta_2}} C_{\Op_{\Delta_2} \phi^6 \Op_{\Delta_3}} C_{\Op_{\Delta_3} \phi^6 \phi^6} \right|_{\Op(\lambda^0)} =& - \frac{64 \pi^{3/2}}{4^{\Delta_1 + \Delta_2 + \Delta_3}} \frac{\Delta_1 (\Delta_1 -1) \Delta_{12}}{(2\Delta_1 - 1)(\Delta_1 + \Delta_2 - 1) } \notag \\
& \times \frac{\Gamma(\Delta_1 + \Delta_2)^2}{\Gamma(\Delta_2) \Gamma(\Delta_1 - 1/2)^2 \Gamma(\Delta_2 - 1/2)} \delta_{\Delta_1, \Delta_3}\,,
\end{align}
with $\Delta_{ij} := \Delta_i - \Delta_j$, valid when $\Delta_1<\Delta_2$.

At next-to-leading order, we expand the correlator given in the ancillary notebook and find
\begin{equation}
\Am^{(1)}_{666666} (\chi_1,\chi_2,\chi_3) = \frac{9}{8\pi^2} \chi_1 \chi_2 + \ldots\,,
\end{equation}
which is in full agreement with \eqref{eq:3ptphi6}.

\subsection{Snowflake channel}
\label{subsec:snowflake}

We now move our attention to the other topology that appears for multipoint functions with $n = 6$, which is called the \textit{snowflake} channel and which is represented diagrammatically in figure \ref{fig:snowch}. 

\begin{figure}[H]
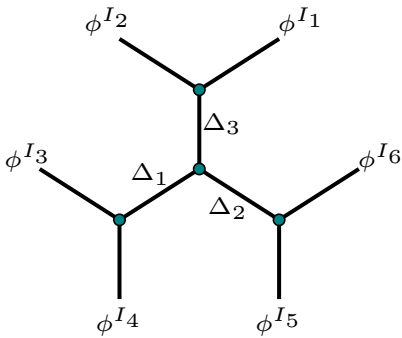

\centering
\snowblocks
\caption{Representation of six-point functions in the snowflake channel. Here the OPEs are taken pairwise between external operators, and lead to the OPE coefficient consisting of products of four three-point functions, represented by the vertices. }
\label{fig:snowch}
\end{figure}

In this case, the OPE limits consist of bringing two neighbouring external operators close to each other pairwise, and this has for consequence that the OPE coefficient in the middle can consist of operators all different from the external ones, as opposed to the comb channel of the previous section where at least one external operator is present in the three-point functions. As above, we specialize our analysis to the case where all the external operators are identical and are of length $L=1$, i.e. correlation functions that involve either the protected fundamental scalars $\phi^i$ or the unprotected one $\phi^6$.

In order to take the proper OPE limits, we have to consider a new set of cross-ratios:
\begin{equation}
z_1=\frac{\tau_{12}\tau_{46}}{\tau_{16}\tau_{24}}\,, \qquad z_2=\frac{\tau_{26}\tau_{34}}{\tau_{23}\tau_{46}}\,, \qquad z_3=\frac{\tau_{24}\tau_{56}}{\tau_{26}\tau_{45}}\,.
\end{equation}
Six-point functions can then as usual be decomposed into conformal prefactor and reduced correlator:
\begin{equation}
\vvev{\phi^{I_1} (\tau_1) \ldots \phi^{I_6} (\tau_6)} = \Km(\tau_1, \Delta_{\smash{\phi^{I_1}}}; \dots\,;\tau_6, \Delta_{\smash{\phi^{I_6}}})\, \Am^{I_1 \ldots I_6  } (z_1\,, z_2\,, z_3)\,.
\end{equation}
For the choice of the conformal prefactor, we also adopt the convention of \cite{Fortin:2022grf}, that we specialize to identical operators:
\begin{equation}
\Km(\tau_1, \Delta_{\phi}; \dots\,;\tau_6, \Delta_{\phi}) = \frac{1}{\tau_{12}^{2 \smash{\Delta_{\phi}}} \tau_{34}^{2 \smash{\Delta_{\phi}}} \tau_{56}^{2 \smash{\Delta_{\phi}}}}\,.
\end{equation}

For a given $R$-symmetry channel, i.e. a fixed choice of indices $I_1, \ldots, I_6$, correlators can be expanded in the following way:
\begin{align}
\Am^{I_1 \ldots I_6} (z_1\,, z_2\,, z_3) =& \sum_{\Op_1\,, \Op_2\,, \Op_3} C_{\phi^{I_1} \phi^{I_2} \Op_1} C_{\phi^{I_3} \phi^{I_4} \Op_2} C_{\phi^{I_5} \phi^{I_6} \Op_3} C_{\Op_1 \Op_2 \Op_3} g_{\Delta_1, \Delta_2, \Delta_3} ( z_1, z_2,z_3 )\,,
\end{align}
where now the function $g_{\Delta_1, \Delta_2, \Delta_3}$ corresponds to the snowflake conformal blocks. For our purposes we write a series expansion of the form
\begin{equation}
\,g_{\Delta_{1}, \Delta_2, \Delta_3}\left(z_1\,, z_2\,,z_3 \right) = z_1 ^{\Delta_1} z_2^{\Delta_2} z_3^{\Delta_3} \sum_{n_1, n_2, n_3} c_{n_1, n_2, n_3} z_1^{n_1} z_2^{n_2} z_3^{n_3} \,,
\end{equation}
where we only need coefficients $c_{n_1, n_2, n_3}$ for low values of $n_1$, $n_2$, $n_3$\footnote{In \cite{Fortin:2020zxw} a different Taylor expansion is used with a closed-form expression for the corresponding coefficients $c_{n_1, n_2, n_3}$.}. It is easy to determine the coefficients up to an overall normalization by applying the Casimir equations on the blocks order by order (see appendix \ref{app:casimir}) and this results in the following expansion of the full correlator:
\begin{equation}
\Am^{I_1 \ldots I_6} (z_1\,, z_2\,, z_3) = 1 + C_{\phi^{I_1} \phi^{I_2} \Op_{\Delta=1}} C_{\phi^{I_3} \phi^{I_4} \Op_{\Delta=1}} C_{\phi^{I_5} \phi^{I_6} \mathds{1}} C_{\Op_{\Delta=1} \Op_{\Delta=1} \mathds{1}} z_1 z_2 + \ldots\,,
\label{eq:snowexp}
\end{equation}
where we have used the fact that terms with two $\Delta$'s set to zero vanish since one-point functions vanish. We note that, as usual, $C_{\phi^{I_5} \phi^{I_6} \mathds{1}} C_{\Op_{\Delta=1} \Op_{\Delta=1} \mathds{1}} = 1$ because of the unit-normalization of two-point functions. We have labeled the exchanged operator with (bare) scaling dimension $\Delta=1$ as $\Op_{\Delta=1}$, but we will see below that in our two cases of interest this operator always turns out to be $\phi^6$.

We now consider the cases where the six-point functions are either $\vvev{\phi^i \phi^j \phi^k \phi^l \phi^m \phi^n}$ or $\vvev{\phi^6 \phi^6 \phi^6 \phi^6 \phi^6 \phi^6}$, and perform checks for the OPE coefficients encountered in \eqref{eq:snowexp}.

\paragraph{$\vvev{\phi^i \phi^j \phi^k \phi^l \phi^m \phi^n}$.} We start by the correlator with the six protected fundamental scalars. As usual let us focus on the highest-weight channel $F_0$, the expansion sketched in \eqref{eq:snowexp} then becomes
\begin{equation}
F_0 (z_1\,, z_2\,, z_3) = 1 + C_{\phi^i \phi^j \phi^6} C_{\phi^6 \phi^k \phi^l} z_1 z_2 + \ldots\,,
\label{eq:snowexp1}
\end{equation}
where the exchanged operator can only be $\phi^6$ because of $R$-charge conservation. At leading order, we have seen in equation \eqref{eq:6ptbuildingblocksLO} that
\begin{equation}
F^{(0)}_0 (z_1\,, z_2\,, z_3) = 1\,,
\end{equation}
and thus $C_{\phi^i \phi^j \phi^6} C_{\phi^6 \phi^k \phi^l} = 0$ at order $\Op(\lambda^0)$, in perfect agreement with \eqref{eq:ij6}.

At next-to-leading order, we can expand \eqref{eq:6ptF0NLO} at $z_1\,, z_2\,, z_3 \sim 0$ to obtain
\begin{equation}
F^{(1)}_0 (z_1\,, z_2\,, z_3) = \frac{1}{8 \pi^2} z_1 z_2 + \ldots\,,
\end{equation}
which perfectly matches the order $\Op(\lambda)$ of \eqref{eq:ij6} squared.

\paragraph{$\vvev{\phi^6 \phi^6 \phi^6 \phi^6 \phi^6 \phi^6}$.} Let us now perform checks on our result for the six-point function of unprotected scalars $\phi^6$ in the snowflake channel. In this case, equation \eqref{eq:snowexp} turns out to be
\begin{equation}
\Am^{666666} (z_1\,, z_2\,, z_3) = 1 + C_{\phi^6 \phi^6 \phi^6}^2 z_1 z_2 + \ldots\,,
\label{eq:snowexp2}
\end{equation}
where again the exchanged operator can only be $\phi^6$ because of conservation of the $R$-charge. At leading order we can expand \eqref{eq:666666LO} and we find
\begin{equation}
\Am^{(0)}_{666666} (z_1\,, z_2\,, z_3) = 1 + z_1^2 z_2^2 + \ldots\,,
\end{equation}
from which we can read that $C_{\phi^6 \phi^6 \phi^6}^2 = 0$ at $\Op(\lambda^0)$, since there is no term of order $\Op(z_1 z_2)$. This is fully consistent with \eqref{eq:3ptphi6}.

At next-to-leading order, we can expand the correlator of the ancillary notebook in order to obtain
\begin{equation}
\Am^{(1)}_{666666} (z_1\,, z_2\,, z_3) = \frac{9}{8 \pi^2} z_1 z_2 + \ldots\,,
\end{equation}
where the prefactor perfectly matches the $C_{\phi^6 \phi^6 \phi^6}^2$ predicted by equation \eqref{eq:3ptphi6} at order $\Op(\lambda)$.

\endgroup

\section{Conclusions}
\label{sec:conclusions}

In this work we investigated correlation functions on the $1d$ defect CFT formed by inserting scalar operators along the half-BPS Wilson line in $4d$ $\Nm =4$ SYM. We derived a recursion relation that allows to compute multipoint correlators $\vev{\phi^{I_1} \ldots \phi^{I_n}}$ made out of an arbitrary number of fundamental scalar fields, up to next-to-leading order for an even number of unprotected operators $\phi^6$, and up to leading order when their number is odd. By pinching operators together it is possible to construct correlation functions of operators of higher length, and in particular our recursion allows us to build correlators containing arbitrary operators made out of the fundamental scalar fields. For operators of length $L=2$, we checked that their anomalous dimensions are properly reproduced, and in addition we generated many correlators of fundamental scalar fields and composite operators up to six-point. This provides an interesting pool of perturbative results, which can be used to test and expand modern analytical bootstrap techniques.\footnote{See for example \cite{Ghosh:2021ruh}, where analytical functionals for $1d$ theories with flavor symmetry were studied, a setup closely related to our supersymmetric Wilson line.} As a check of our results, and to connect with recent work on multipoint conformal blocks, we expanded $\vvev{\phi^{i_1} \ldots \phi^{i_n}}$ ($i_k = 1, \ldots, 5$) and $\vvev{\phi^6 \ldots \phi^6}$ in both the comb and snowflake channels, and extracted low-lying OPE coefficients. Both our correlators and the block literature seem to be consistent with each other.

There are a handful of future directions that can be further explored. In section \ref{sec:blockexpansion} we expanded our correlators in \textit{bosonic} conformal blocks. Although this allowed us to extract the OPE coefficients of the $\phi^6$ operator, our system does have supersymmetry, and it would be desirable to perform a full superblock analysis. A suitable superspace for correlators of half-BPS operators was originally presented in \cite{Liendo:2016ymz}, and more recently, superblocks including long operators were announced in \cite{Ferrero:2021bsb}. A superblock expansion would extract supersymmetric CFT data, which will help disentangle degenerate operators. Morever, multipoint superblocks might pave the way to bootstrapping multipoint correlators at strong coupling, similar to the work done for four-point functions in \cite{Liendo:2018ukf,Ferrero:2021bsb}. One could also repeat our analysis in other superconformal setups, interesting cases include line defects in $4d$ $\Nm=2$ theories \cite{Gimenez-Grau:2019hez} and ABJM \cite{Bianchi:2020hsz}.  

Another natural next step is to include more general operators in our recursion formula, like fields that transform non-trivially under transverse rotations (fermions for example). One way of including more general fundamental fields would be to upgrade our formulae for $\vvev{\phi \ldots \phi}$ to correlation functions of \textit{superfields} $\vvev{\Phi \ldots \Phi}$. A result that might be relevant for this analysis is the superfield formulation for the supersymmetric Wilson line presented in \cite{Beisert:2015jxa}.

Apart from superconformal lines, line defects play a prominent role in condensed matter physics. An interesting setup is the magnetic line defect studied in \cite{Cuomo:2021kfm} (see also \cite{Gimenez-Grau:2022czc}). This line defect can be defined for critical $O(N)$ models in the $\epsilon$-expansion as an exponential of one of the fundamental scalars. Being a model with a Lagrangian formulation makes it well-suited for perturbative calculations. A follow up of this work is to study multipoint correlators on this line defect \cite{Barrat:2023ivo}. Of particular interest are correlators of the displacement operator and the tilt, which measure the breaking of spacetime and flavor symmetry respectively, due to the presence of the line defect. Both these operators can be built out of the fundamental scalars of the $O(N)$ models, and are studied with the techniques used in this paper.\footnote{Incidentally, the fundamental $\phi^i$ scalar in our supersymmetric Wilson line is precisely the tilt operator. The displacement operator on the other hand is a superconformal descendant of the tilt, and cannot be written using the fundamental scalars.}

\acknowledgments

We are particularly grateful to G.~Bliard, V.~Forini, A.~Gimenez-Grau, J. ~Mann, J.~Plefka, L.~Quintavalle, P.~Van Vliet for useful discussions.
PL acknowledges support from the DFG through the Emmy Noether research group ``The Conformal Bootstrap Program'' project number 400570283, and through the German-Israeli Project Cooperation (DIP) grant ``Holography and the Swampland''. JB and GP are funded by the Deutsche Forschungsgemeinschaft (DFG, German Research Foundation) -- Projektnummer 417533893/GRK2575 ``Rethinking Quantum Field Theory''.

\appendix

\section{Insertion rules}
\label{sec:insertion}

In this appendix, we list the insertion rules used for computing the Feynman diagrams of sections \ref{sec:correlators} and \ref{sec:applications}. Those are derived from the action of $\mathcal{N}=4$ SYM in $4d$ Euclidean space, which is given by \eqref{eq:action}. Note that we consider $SU(N)$ as the gauge group and that we work in the large $N$ limit. The generators obey the following commutation relation:
\begin{equation}
[ T^a\,, T^b ] = i f^{abc}\, T^c,
\end{equation}
in which $f^{abc}$ are the structure constants of the $\mathfrak{su}(N)$ Lie algebra. The generators are normalized as
\begin{equation}
\tr T^a \tensor{T}{^b} = \frac{\delta^{ab}}{2}.
\end{equation}
The generators are traceless, i.e.
$\tr \tensor{T}{^a} = 0$. The (contracted) product of structure constants gives $f^{abc} f^{abc} = N (N^2 - 1) \sim N^3$, where the second equality holds in the large $N$ limit.

We start by \textit{two-point} insertions. The only one that we actually need is the self-energy of the scalar propagator at one loop, which is given by the following expression \cite{Erickson:2000af, Plefka:2001bu, Drukker:2008pi}:
\begin{align}
\propagatorSSEnotext\ &=\ \SSEone\ +\ \SSEtwo\ +\ \SSEthree\ +\ \SSEfour \notag \\
&= -2 g^4 N \delta^{ab} \delta_{ij} Y_{112}.
\label{eq:selfenergy}
\end{align}
The integral $Y_{112}$ is given in (\ref{eq:Y112}) and presents a logarithmic divergence.

We also require only one \textit{three-point} insertion, which is the vertex connecting two scalar fields and one gauge field. It is easy to obtain from the action (\ref{eq:action}) and it reads
\begin{equation}
\vertexSSG\ = - g^4 f^{abc} \delta^{ij} \left( \partial_1 - \partial_2 \right)_\mu Y_{123}.
\label{eq:vertexSSG}
\end{equation}
The $Y$-integral is defined in (\ref{subeq:Y123}) and its analytical expression in $1d$ can be found in (\ref{eq:Y123}).

Another relevant vertex is the four-scalars coupling. Similarly to the three-vertex, it is straightforward to read the corresponding Feynman rule from the action and perform the Wick contractions in order to get
\begin{align}
\vertexSSSS &= -g^6 \left\lbrace f^{abe}f^{cde} \left( \delta_{ik}\delta_{jl} - \delta_{il}\delta_{jk} \right) + f^{ace}f^{bde} \left( \delta_{ij}\delta_{kl} - \delta_{il}\delta_{jk} \right) \right. \notag \\[-1.5em]
& \left. \qquad \qquad \qquad + f^{ade}f^{bce} \left( \delta_{ij}\delta_{kl} - \delta_{ik}\delta_{jl} \right) \right\rbrace X_{1234}.
\label{eq:vertexSSSS}
\end{align}
The $X$-integral can be found in (\ref{eq:X1234}) for the $1d$ case.

There is one more sophisticated four-point insertion that we require, which reads
\begin{equation}
\Hinsertone\ = g^6 \left\lbrace \delta_{ik} \delta_{jl} f^{ace} f^{bde} I_{13} I_{24} F_{13,24} + \delta_{il} \delta_{jk} f^{ade} f^{bce} I_{14} I_{23} F_{14,23} \right\rbrace\,.
\label{eq:Hinsertone}
\end{equation}
with $I_{ij}$ the propagator function defined in (\ref{eq:I12}) and $F_{ij,kl}$ as defined in (\ref{subeq:F1324}). An analytical expression for $F_{ij,kl}$ in terms of $X$- and $Y$-integrals is given in (\ref{eq:FXYidentity}).
\section{The recursion relation for an even number of fields}
\label{sec:recursionapp}
\begingroup
\allowdisplaybreaks

In this appendix we give the formal expression for the recursion relation given in equation \eqref{eq:recursionevenNLO} in a diagrammatic way. A close look at \eqref{eq:recursionevenNLO} reveals that there are \textit{two types} of $U$-integrals that one can encounter. These two types are represented in figure \ref{fig:Udiagrams}, and correspond to whether the integration limits are "connected" or not. This distinction is made clear in the definitions of the integrals, which can be found in appendix \ref{subsubsec:Uintegrals}. Equation \eqref{eq:recursionevenNLO} contains both types of integrals, and in order to write an effectively usable formula, we must extract the $U^{(2)}$ contributions. It is easy to check visually which terms contain a $U^{(2)}$:
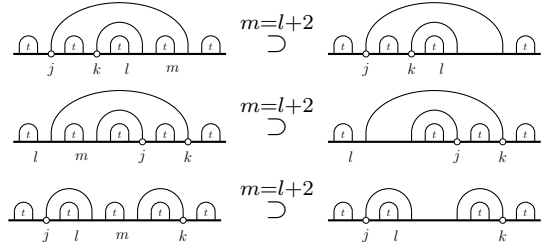
\begin{align*}
\recursionphione\, \overset{m=l+2}{\supset}\,  \recursionphioneb\, \\
\recursionphithree\, \overset{m=l+2}{\supset}\, \recursionphithreeb\, \\
\recursionphisix\ \overset{m=l+2}{\supset}\, \recursionphisixb\,
\end{align*}
\begin{figure}
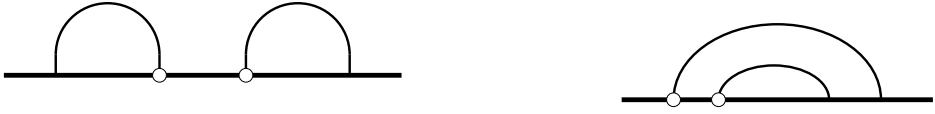

\centering
\begin{subfigure}{.5\textwidth}
  \centering
  \scalebox{1.3}{\twopointnpr}
\end{subfigure}%
\begin{subfigure}{.5\textwidth}
  \centering
  \scalebox{1.3}{\Utwo}
\end{subfigure}
\caption{Illustration of the two types of $U$-integrals that one can encounter in the recursion relation \eqref{eq:recursionevenNLO}. The difference lays in the integration limits, as explained in detail in section \ref{subsubsec:Uintegrals}.}
\label{fig:Udiagrams}
\end{figure}
Hence we only have to change the summation range for these three terms and add them by hand with an explicit mention of $U^{(2)}$. This gives the following expression:
\begin{align}
\eqref{eq:recursionevenNLO} =& \frac{\lambda^2}{4} \sum_{j=1}^{n-1} \sum_{k=j+1}^n \delta^{I_j 6} \delta^{I_k 6} \notag \\
& \times \Biggl( \sum_{l=k}^{n-2} \sum_{m=l+4}^n U_{j;m(m+1)} U_{k;l(l+1)} A_\text{\tiny{LO}}^{I_1 \ldots I_{j-1}} A_\text{\tiny{LO}}^{I_{j+1} \ldots I_{k-1}} A_\text{\tiny{LO}}^{I_{k+1} \ldots I_l} A_\text{\tiny{LO}}^{I_{l+1} \ldots I_m} A_\text{\tiny{LO}}^{I_{m+1} \ldots I_n} \notag \\
& \phantom{\Biggl(} + \sum_{l=k}^n \sum_{m=j}^{k-1} U_{j;l(l+1)} U_{k;m(m+1)} A_\text{\tiny{LO}}^{I_1 \ldots I_{j-1}} A_\text{\tiny{LO}}^{I_{j+1} \ldots I_{m}} A_\text{\tiny{LO}}^{I_{m+1} \ldots k_l} A_\text{\tiny{LO}}^{I_{k+1} \ldots I_l} A_\text{\tiny{LO}}^{I_{l+1} \ldots I_n} \notag \\
& \phantom{\Biggl(} + \sum_{l=0}^{j-3} \sum_{m=l+4}^{j-1} U_{j;m(m+1)} U_{k;l(l+1)} A_\text{\tiny{LO}}^{I_1 \ldots I_{l}} A_\text{\tiny{LO}}^{I_{l+1} \ldots I_{m}} A_\text{\tiny{LO}}^{I_{m+1} \ldots j_l} A_\text{\tiny{LO}}^{I_{j+1} \ldots I_{k-1}} A_\text{\tiny{LO}}^{I_{k+1} \ldots I_n} \notag \\
& \phantom{\Biggl(} + \sum_{l=j}^{k-1} \sum_{m=0}^{j-1} U_{j;l(l+1)} U_{k;m(m+1)} A_\text{\tiny{LO}}^{I_1 \ldots I_{m}} A_\text{\tiny{LO}}^{I_{m+1} \ldots I_{j-1}} A_\text{\tiny{LO}}^{I_{j+1} \ldots I_l} A_\text{\tiny{LO}}^{I_{l+1} \ldots I_{k-1}} A_\text{\tiny{LO}}^{I_{k+1} \ldots I_n} \notag \\
& \phantom{\Biggl(} + \sum_{l=j}^{k-1} \sum_{m=k}^n U_{j;l(l+1)} U_{k;m(m+1)} A_\text{\tiny{LO}}^{I_1 \ldots I_{j-1}} A_\text{\tiny{LO}}^{I_{j+1} \ldots I_{l}} A_\text{\tiny{LO}}^{I_{l+1} \ldots k_l} A_\text{\tiny{LO}}^{I_{k+1} \ldots I_m} A_\text{\tiny{LO}}^{I_{m+1} \ldots I_n} \notag \\
& \phantom{\Biggl(}  + \sum_{l=j}^{k-3} \sum_{m=l+4}^{k-1} U_{j;l(l+1)} U_{k;m(m+1)} A_\text{\tiny{LO}}^{I_1 \ldots I_{j-1}} A_\text{\tiny{LO}}^{I_{j+1} \ldots I_{l}} A_\text{\tiny{LO}}^{I_{l+1} \ldots m} A_\text{\tiny{LO}}^{I_{m+1} \ldots I_{k-1}} A_\text{\tiny{LO}}^{I_{k+1} \ldots I_n} \notag \\
& \phantom{\Biggl(} + \sum_{l=0}^{j-1} \sum_{m=k}^n U_{j;l(l+1)} U_{k;m(m+1)} A_\text{\tiny{LO}}^{I_1 \ldots I_{l}} A_\text{\tiny{LO}}^{I_{l+1} \ldots I_{j-1}} A_\text{\tiny{LO}}^{I_{j+1} \ldots k_l} A_\text{\tiny{LO}}^{I_{k+1} \ldots I_m} A_\text{\tiny{LO}}^{I_{m+1} \ldots I_n} \notag \\
& \phantom{\Biggl(} + \sum_{l=0}^{j-1} \sum_{m=j}^{k-1} U_{j;l(l+1)} U_{k;m(m+1)} A_\text{\tiny{LO}}^{I_1 \ldots I_{l}} A_\text{\tiny{LO}}^{I_{l+1} \ldots I_{j-1}} A_\text{\tiny{LO}}^{I_{j+1} \ldots I_m} A_\text{\tiny{LO}}^{I_{m+1} \ldots I_{k-1}} A_\text{\tiny{LO}}^{I_{k+1} \ldots I_n}
\Biggr) \notag \\
& + \frac{\lambda^2}{4} \sum_{j=1}^{n-3} \sum_{k=j+2}^{n-1} \delta^{I_j I_k} I_{jk} \notag \\
& \times \Biggl( \sum_{l=k+1}^n \sum_{m=l}^n \delta^{I_l 6} U_{l;m(m+1)} A_\text{\tiny{LO}}^{I_1 \ldots I_{j-1}} A_\text{\tiny{LO}}^{I_{j+1} \ldots I_{k-1}} A_\text{\tiny{LO}}^{I_{k+1} \ldots I_{l-1}} A_\text{\tiny{LO}}^{I_{l+1} \ldots I_{m}} A_\text{\tiny{LO}}^{I_{m+1} \ldots I_n} \notag \\
& \phantom{\Biggl(} + \sum_{l=k+1}^n \sum_{m=k}^{l-1} \delta^{I_l 6} U_{l;m(m+1)} A_\text{\tiny{LO}}^{I_1 \ldots I_{j-1}} A_\text{\tiny{LO}}^{I_{j+1} \ldots I_{k-1}} A_\text{\tiny{LO}}^{I_{k+1} \ldots I_{m}} A_\text{\tiny{LO}}^{I_{m+1} \ldots I_{l-1}} A_\text{\tiny{LO}}^{I_{l+1} \ldots I_n} \Biggr) \notag \\
& + \frac{\lambda^2}{4} \sum_{j=2}^{n-2} \sum_{k=j+2}^n \delta^{I_j I_k} I_{jk} \notag \\
& \times \Biggl( \sum_{l=1}^{j-1} \sum_{m=l}^{j-1} \delta^{I_l 6} U_{l;m(m+1)} A_\text{\tiny{LO}}^{I_1 \ldots I_{l-1}} A_\text{\tiny{LO}}^{I_{l+1} \ldots I_{m}} A_\text{\tiny{LO}}^{I_{m+1} \ldots I_{j-1}} A_\text{\tiny{LO}}^{I_{j+1} \ldots I_{k-1}} A_\text{\tiny{LO}}^{I_{k+1} \ldots I_n} \notag \\
& \phantom{\Biggl(} + \sum_{l=1}^{j-1} \sum_{m=0}^{l-1} \delta^{I_l 6} U_{l;m(m+1)} A_\text{\tiny{LO}}^{I_1 \ldots I_{m}} A_\text{\tiny{LO}}^{I_{m+1} \ldots I_{l-1}} A_\text{\tiny{LO}}^{I_{l+1} \ldots I_{j-1}} A_\text{\tiny{LO}}^{I_{j+1} \ldots I_{k-1}} A_\text{\tiny{LO}}^{I_{k+1} \ldots I_n} \Biggr) \notag \\
& + \frac{\lambda^2}{4} \sum_{j=1}^{n-3} \sum_{k=j+2}^{n-1} \sum_{l=k+1}^n \sum_{m=0}^{j-1} \delta^{I_j I_k} \delta^{I_l 6} \notag \\
& \qquad\qquad \times I_{jk} U_{l;m(m+1)} A_\text{\tiny{LO}}^{I_1 \ldots I_{m}} A_\text{\tiny{LO}}^{I_{m+1} \ldots I_{j-1}} A_\text{\tiny{LO}}^{I_{j+1} \ldots I_{k-1}} A_\text{\tiny{LO}}^{I_{k+1} \ldots I_{l-1}} A_\text{\tiny{LO}}^{I_{l+1} \ldots I_{n}} \notag \\
& + \frac{\lambda^2}{4}  \sum_{j=3}^{n-2} \sum_{k=j+2}^{n} \sum_{l=1}^{j-1} \sum_{m=k}^{n} \delta^{I_j I_k} \delta^{I_l 6} \notag \\
& \qquad\qquad \times  I_{jk} U_{l;m(m+1)} A_\text{\tiny{LO}}^{I_1 \ldots I_{l-1}} A_\text{\tiny{LO}}^{I_{l+1} \ldots I_{j-1}} A_\text{\tiny{LO}}^{I_{j+1} \ldots I_{k-1}} A_\text{\tiny{LO}}^{I_{k+1} \ldots I_{m}} A_\text{\tiny{LO}}^{I_{m+1} \ldots I_{n}} \notag \\
& + \frac{\lambda^2}{4} \sum_{j=1}^{n-5} \sum_{k=j+2}^{n-3} \sum_{l=k+1}^{n-2} \sum_{m=l+2}^n \delta^{I_j I_k} \delta^{I_l I_m} \notag \\
& \qquad\qquad \times  I_{jk} I_{lm} A_\text{\tiny{LO}}^{I_1 \ldots I_{j-1}} A_\text{\tiny{LO}}^{I_{j+1} \ldots I_{k-1}} A_\text{\tiny{LO}}^{I_{k+1} \ldots I_{l-1}} A_\text{\tiny{LO}}^{I_{l+1} \ldots I_{m-1}} A_\text{\tiny{LO}}^{I_{m+1} \ldots I_{n}} \notag \\
& + \frac{\lambda^2}{4} \sum_{j=1}^{n-1} \sum_{k=j+1}^n \delta^{I_j 6} \delta^{I_k 6} \notag \\
& \qquad\qquad \times \Biggl( \sum_{l=k}^n U^{(2)}_{j;k;l(l+1)} A_\text{\tiny{LO}}^{I_1 \ldots I_{j-1}} A_\text{\tiny{LO}}^{I_{j+1} \ldots I_{k-1}} A_\text{\tiny{LO}}^{I_{k+1} \ldots I_{l}} A_\text{\tiny{LO}}^{I_{l+1} \ldots I_{n}} \notag \\
& \qquad\qquad \phantom{\Biggl(} + \sum_{l=j}^{k-1} U^{(2)}_{j;k;l(l+1)} A_\text{\tiny{LO}}^{I_1 \ldots I_{j-1}} A_\text{\tiny{LO}}^{I_{j+1} \ldots I_{l}} A_\text{\tiny{LO}}^{I_{l+1} \ldots I_{k-1}} A_\text{\tiny{LO}}^{I_{k+1} \ldots I_{n}} \notag \\
& \qquad\qquad \phantom{\Biggl(} + \sum_{l=0}^{j-1} U^{(2)}_{j;k;l(l+1)} A_\text{\tiny{LO}}^{I_1 \ldots I_{l}} A_\text{\tiny{LO}}^{I_{l+1} \ldots I_{j-1}} A_\text{\tiny{LO}}^{I_{j+1} \ldots I_{k-1}} A_\text{\tiny{LO}}^{I_{k+1} \ldots I_{n}} \Biggr) \,,
\label{eq:recursionevenNLOformal}
\end{align}
where the $U^{(2)}$-integrals are contained in the three last terms.

This formula is the one that was effectively implemented in the ancillary notebook and that has been used for producing the results of section \ref{sec:applications}.

\endgroup
\section{Integrals and regularization}
\label{sec:integrals}

In this appendix, we define and compute the \textit{bulk} and \textit{boundary} integrals that we encounter in this work.

\subsection{Bulk integrals}
\label{subsec:bulkintegrals}

In the computation of the Feynman diagrams at one loop, we encounter three-, four- and five-point massless Feynman integrals, which we define as follows:
\begin{subequations}
\begin{gather}
Y_{123} := \int d^4 x_4\ I_{14} I_{24} I_{34}\,, \label{subeq:Y123} \\
X_{1234} := \int d^4 x_5\ I_{15} I_{25} I_{35} I_{45}\,, \label{subeq:X1234} \\
H_{13,24} := \int d^4 x_{56}\ I_{15} I_{35} I_{26} I_{46} I_{56}\,, \label{subeq:H1324}
\intertext{with $I_{ij}$ the propagator function defined in (\ref{eq:I12}). In the last expression we have defined $d^4 x_{56} := d^4 x_5\ d^4 x_6$ for brevity. The letter assigned to each integral is evocative of the drawing of the propagators. We always encounter the $H$-integral in the following form:}
F_{13,24} := \frac{\left( \partial_1 - \partial_3 \right) \cdot \left(\partial_2 - \partial_4 \right)}{I_{13}I_{24}} H_{13,24}\,. \label{subeq:F1324}
\end{gather}
\end{subequations}
The notation presented above is standard and has already been used in e.g. \cite{Beisert:2002bb, Drukker:2008pi}. The three- and four-point massless integrals in Euclidean space are conformal and have been solved analytically (see e.g. \cite{tHooft:1978jhc, Usyukina:1992wz} and \cite{Drukker:2008pi,Kiryu:2018phb} for the modern notation). In $1d$ the $X$-integral is given by
\begin{equation}
\frac{X_{1234}}{I_{13} I_{24}} = - \frac{1}{8 \pi^2} \frac{\ell(\chi, 1)}{\chi(1-\chi)}\,,  \qquad\qquad \chi^2 := \frac{\tau_{12}^2 \tau_{34}^2}{\tau_{13}^2 \tau_{24}^2}\,,
\label{eq:X1234}
\end{equation}
with $\ell(\chi_1,\chi_2)$ defined in \eqref{eq:ell}.

The $Y$-integral can easily be obtained from this expression by taking the following limit:
\begin{align}
Y_{123} &= \lim_{x_4 \to \infty} (2\pi)^2 x_4^2\ X_{1234} \notag \\
&= \frac{I_{12}}{8\pi^2} \left( \frac{\tau_{12}}{\tau_{23}} \log |\tau_{13}| + \frac{\tau_{12}}{\tau_{31}} \log |\tau_{23}| + \frac{\tau_{12}^2}{\tau_{23}\tau_{31}} \log |\tau_{12}| \right)\,.
\label{eq:Y123}
\end{align}

The $H$-integral seems to have no known closed form so far, but (\ref{subeq:F1324}) can fortunately be reduced to a sum of $Y$- and $X$-integrals in the following way \cite{Beisert:2002bb}:
\begin{align}
F_{13,24} &= \frac{X_{1234}}{I_{12}I_{34}} - \frac{X_{1234}}{I_{14}I_{23}} + \left( \frac{1}{I_{14}} - \frac{1}{I_{12}} \right) Y_{124} + \left( \frac{1}{I_{23}} - \frac{1}{I_{34}} \right) Y_{234} \notag \\
& \qquad \qquad \qquad \qquad \qquad + \left( \frac{1}{I_{23}} - \frac{1}{I_{12}} \right) Y_{123} + \left( \frac{1}{I_{14}} - \frac{1}{I_{34}} \right) Y_{134}\,.
\label{eq:FXYidentity}
\end{align}

The integrals given above also appear in their respective pinching limits, i.e. when two external points are brought close to each other. The integrals simplify greatly in this limit, but they exhibit a logarithmic divergence which is tamed by using point-splitting regularization. For the $Y$-integral, we define
\begin{equation*}
Y_{122} := \lim_{x_3 \to x_2} Y_{123}\,, \qquad \qquad \qquad \lim_{x_3 \to x_2} I_{23} := \frac{1}{(2\pi)^2 \epsilon^2}\,.
\end{equation*}
%
Inserting this in (\ref{eq:Y123}) and expanding up to order $\mathcal{O} (\log \epsilon^2)$, we obtain
\begin{equation}
\diagYonetwotwo\ := Y_{112} = Y_{122} = - \frac{I_{12}}{16 \pi^2} \left( \log \frac{I_{12}}{I_{11}} - 2 \right)\,.
\label{eq:Y112}
\end{equation}
This result coincides with the expression given in e.g. \cite{Drukker:2008pi}.

For completion, we also give the pinching limit of the $X$- and $F$-integrals. The first one reads
\begin{equation}
\diagXoneonetwothree\ := X_{1123} = - \frac{I_{12} I_{13}}{16 \pi^2} \left( \log \frac{I_{12} I_{13}}{I_{11} I_{23}} - 2 \right)\,,
\label{eq:X1123}
\end{equation}
which is again the same as in \cite{Drukker:2008pi}.

Finally, the pinching limit $\tau_2 \to \tau_1$ of the $F$-integral gives
\begin{align}
F_{13,14} &= F_{14,13} = - F_{13,41} \notag \\
& = -\frac{X_{1134}}{I_{13} I_{14}} + \frac{Y_{113}}{I_{13}} + \frac{Y_{114}}{I_{14}} + \left(\frac{1}{I_{13}} + \frac{1}{I_{14}} - \frac{2}{I_{34}} \right) Y_{134}\,.
\label{eq:F1314}
\end{align}

\subsection{Boundary integrals}
\label{subsec:boundaryintegrals}

In the computations, we deal with two types of boundary integrals that we explain in detail below and that we name $T$- and $U$-integrals.

\subsubsection{$T$-integrals}
\label{subsubsec:Tintegrals}

In presence of the line, there is a new type of integral arising in addition to the bulk integrals of the previous appendix. We denote this integral by $T_{ij;kl}$\footnote{This class of integrals also appears in \cite{Kiryu:2018phb}, where they are defined slightly differently and labelled as $B_{ij;kl}$.}, and define it to be
\begin{equation}
T_{ij;kl} := \partial_{ij} \,\int_{\tau_k}^{\tau_l} d\tau_m\, \epsilon(ijm) \, Y_{ijm}\,,
\label{eq:Tdef}
\end{equation}
where $\epsilon(ijk)$ encodes the change of sign due the path ordering, formally defined as
\begin{equation}
\epsilon(ijk) := \text{sgn}\, \tau_{ij}\, \text{sgn}\, \tau_{ik}\, \text{sgn}\, \tau_{jk}\,.
\label{eq:epsdef}
\end{equation}

When the range of integration is the entire line, the integral is easy to perform and results in
\begin{equation}
T_{ij;(-\infty)(+\infty)} = - \frac{I_{ij}}{12}\,.
\label{eq:T1}
\end{equation}
In the case where $(i,j) = (k,l)$ it gives
\begin{equation}
T_{ij;ij} = \frac{I_{ij}}{12} \,.
\label{eq:T2}
\end{equation}

Let us now review some relations satisfied by the $T$-integrals. The following identity can be used in order to "swap" the limits of integration:
\begin{equation}
\left. T_{jk;il} \right|_{i<j<k<l} = -\frac{I_{jk}}{12} - T_{jk;li}\,,
\label{eq:T3}
\end{equation}
where the integration range $(li)$ on the right-hand side has to be understood as the union of segments $(l,+\infty) \cup (-\infty,i)$.

There also exist another relevant combination for the computations at one loop relating the $T$- and $Y$-integrals:
\begin{equation}
I_{ik} T_{jk;ki} + I_{jk} T_{ik;jk} = - \frac{I_{ik} I_{jk}}{12} +I_{ik} I_{jk}\left(\frac{1}{I_{ik}} + \frac{1}{I_{jk}}-\frac{2}{I_{ij}} \right)Y_{ijk}\,.
\label{eq:T5}
\end{equation}

In general the integrals can be performed explicitly for the different possible orderings of the $\tau$'s, and here we give the results assuming $\tau_1 < \tau_2 < \tau_3 < \tau_4$:
\begin{subequations}
\label{eq:Tint} %
\begin{gather}
T_{12;34} = \frac{1}{32 \pi^4 \tau_{12}^2} \left( 4 L_R \left( \frac{\tau_{12}}{\tau_{14}} \right) - 4 L_R \left( \frac{\tau_{12}}{\tau_{13}} \right) - C_{123} + C_{124} \right)\,, \\
T_{34;12} = \frac{1}{32 \pi^4 \tau_{34}^2} \left( 4 L_R \left( \frac{\tau_{34}}{\tau_{14}} \right) - 4 L_R \left( \frac{\tau_{34}}{\tau_{24}} \right) - C_{341} + C_{342} \right)\,, \\
T_{14;23} = \frac{1}{32 \pi^4 \tau_{14}^2} \left( 4 L_R \left( \frac{\tau_{24}}{\tau_{14}} \right) - 4 L_R \left( \frac{\tau_{34}}{\tau_{14}} \right) - C_{412} - C_{143} \right)\,, \\
T_{23;41} = \frac{1}{32 \pi^4 \tau_{23}^2} \left( -4 L_R \left( \frac{\tau_{23}}{\tau_{13}} \right) - 4 L_R \left( \frac{\tau_{23}}{\tau_{24}} \right) - C_{234} - C_{123} \right)\,,
\end{gather}
\end{subequations}
where we have defined the following help function:
\begin{equation}
C_{ijk} := - 32 \pi^4 \tau_{ij} (\tau_{ik} + \tau_{jk}) Y_{ijk}\,,
\end{equation}
and where the Rogers dilogarithm $L_R(x)$ is defined in \eqref{eq:Rogers}.

It is easy to take pinching limits of the integrals given above. For example, we can have
\begin{equation}
T_{12;23} = \frac{1}{32 \pi^4 \tau_{12}^2} \left( 4 L_R \left( \frac{\tau_{12}}{\tau_{13}} \right) - \frac{2 \pi^2}{3} + C_{123} \right) + Y_{112}\,,
\end{equation}
using the fact that $L_R(1) = \frac{\pi^2}{6}$. All the other pinching limits can be performed in the same way.

\subsubsection{$U$-integrals}
\label{subsubsec:Uintegrals}

Since the scalar $\phi^6$ couples directly to the Wilson line,  there is another class of integral that we have to consider. We denote this integral by $U_{a;ij}$ and it is defined as
\begin{equation}
U_{a;ij} := \int_{\tau_i}^{\tau_j} d\tau_n\,  I_{an}\,,
\label{eq:Uint}
\end{equation}
where $a$ is the insertion point of the scalar $\phi^6$ on the Wilson line and $ij$ indicate the range of integration. 
These integrals can be easily performed explicitly:
\begin{equation}
U_{a;ij}= \frac{1}{4\pi^2} \left( \frac{1}{\tau_i - \tau_a} +\frac{1}{\tau_a -\tau_j} \right)\,, 
\label{eq:defU}
\end{equation}
which is valid
both when $\tau_a < \tau_i <\tau_j$ and $\tau_i < \tau_j <\tau_a$. Though,  variations of this $U$-integral appear in the recursion, for example when $\tau_a = \tau_i$. In these cases, we just take the appropriate limit, regularizing the divergences in the integral with point-splitting regularization
as for the integrals of the previous sections. 
If $i,j=\pm\infty$, we again take the limit of the expression above. 

Given that we are interested in next-to-leading order computations,  we have to consider also integrals of (\ref{eq:Uint}), 
that arise when two scalars $\phi^6$ couple to the Wilson line.  We refer to these integrals as $U^{(2)}_{ab;ij}$ and it is defined as
\begin{equation}
U^{(2)}_{ab;ij}:=\int^{\tau_j}_{\tau_i}\, d\tau_n \,I_{an} U_{b;nj}= \int^{\tau_j}_{\tau_i}\, d\tau_n I_{an} \int^{\tau_j}_{\tau_n} d\tau_{m} I_{bm}  \,.
\label{eq:Utwo} 
\end{equation}
In the following, we perform the integrals of three different configurations, because all the other ones can be obtained from these by taking the appropriate limits and regularizing the integrals.  Assuming $\tau_1 < \tau_2 < \tau_3 < \tau_4$, we get:
\begin{subequations}
\begin{align}
 \Utwoint  =\frac{1}{16 \pi^4 } \Bigg( \frac{\tau_{ij}} {\tau_{ai}\tau_{aj}\tau_{jb}} +& \frac{1}{\tau_{ab}^2\tau_{ai}\tau_{aj}} \left( (\tau_a^2+ \tau_i \tau_j)  \log \frac{\tau_{ai}\tau_{bj}}{\tau_{bi}\tau_{aj}} \right.  \notag \\
&  \left. +\tau_a(\tau_i +\tau_j) \log \frac{\tau_{bi}\tau_{aj}}{\tau_{ai}\tau_{bj}} + \tau_{ba} \tau_{ij} \right) \Bigg)  \,,\\
\Utwointtwo= \frac{1}{16 \pi^4 } \Bigg( \frac{\tau_{ji}} {\tau_{ai}\tau_{ib}\tau_{aj}} +& \frac{1}{\tau_{ab}^2\tau_{ai}\tau_{aj}} \left( (\tau_a^2+ \tau_i \tau_j) \log \frac{\tau_{bi}\tau_{aj}}{\tau_{ai}\tau_{bj}}  \right.  \notag \\
& \left. +\tau_a(\tau_i +\tau_j) \log \frac{\tau_{ai}\tau_{bj}}{\tau_{bi}\tau_{aj}} + \tau_{ab} \tau_{ij} \right) \Bigg) \,,\\
\Utwointthree  = \frac{1}{16 \pi^4 } \Bigg( \frac{\tau_{ij}} {\tau_{bi}\tau_{bj}\tau_{ja}} +& \frac{1}{\tau_{ab}^2\tau_{bi}\tau_{bj}} \left((\tau_b^2+ \tau_i \tau_j)  \log \frac{\tau_{bi}\tau_{aj}}{\tau_{ai}\tau_{bj}} \right.  \notag \\
&  \left. +\tau_b(\tau_i +\tau_j)\log \frac{\tau_{ai}\tau_{bj}}{\tau_{bi}\tau_{aj}} + \tau_{ab} \tau_{ij} \right) \Bigg) \,, 
\end{align}
\end{subequations}
with $\tau_{ij}:= \tau_i -\tau_j$.
\section{Snowflake Casimir}
\label{app:casimir}

In this short appendix we explain how we calculated the snowflake-channel conformal blocks. Explicit expressions for these blocks have already appeared in the literature \cite{Fortin:2020zxw}, however here we use different cross-ratios that make the blocks symmetric in all its arguments.\footnote{We thank Lorenzo Quintavalle for sharing these formulae with us.}

Defining
\begin{equation} 
a_1 = \frac{1}{2}(\Delta_2 - \Delta_1)	\,, 
\quad a_2 = \frac{1}{2}(\Delta_4 - \Delta_3)\,,
\quad a_3 = \frac{1}{2}(\Delta_6 - \Delta_5),
\end{equation}
we can write the following Casimir operators:
\begin{align}
\begin{split}
\mathcal{C}_{2}^{(12)} & =
-(z_1-1)z_1^2 \partial_{z_1}^2 + (z_2-1) z_2 z_1^2 \partial_{z_1} \partial_{z_2} 
+ z_1^2 (-2 a_2 z_2 + 2 a_1 -1) \partial_{z_1}
\\
& -2a_1(z_2-1)z_2 z_1 \partial_{z_2} - 2 a_1 z_3 z_1 \partial_{z_3} + 4 a_1 a_2 z_2 z_1 +z_3 z_1^2 \partial_{z_1} \partial_{z_3} \,,
\end{split}
\\
\begin{split}
\mathcal{C}_{2}^{(34)} & =
-(z_2-1)z_2^2 \partial_{z_2}^2 + (z_3-1) z_3 z_2^2 \partial_{z_2} \partial_{z_3} 
+ z_2^2 (-2 a_3 z_3 + 2 a_2 -1) \partial_{z_2}
\\
& -2a_2(z_3-1)z_3 z_2 \partial_{z_3} - 2 a_2 z_1 z_2 \partial_{z_1} + 4 a_2 a_3 z_3 z_2 +z_1 z_2^2 \partial_{z_2} \partial_{z_1} \,,
\end{split}
\\
\begin{split}
\mathcal{C}_{2}^{(56)} & =
-(z_3-1)z_3^2 \partial_{z_3}^2 + (z_1-1) z_1 z_3^2 \partial_{z_3} \partial_{z_1} 
+ z_3^2 (-2 a_1 z_1 + 2 a_3 -1) \partial_{z_3}
\\
& -2a_3(z_1-1)z_1 z_3 \partial_{z_1} - 2 a_3 z_2 z_3 \partial_{z_2} + 4 a_3 a_1 z_1 z_3 +z_2 z_3^2 \partial_{z_3} \partial_{z_2}\,.
\end{split}
\end{align}
The conformal blocks are then eigenfunctions of the following Casimir equations:
\begin{align}
\mathcal{C}^{(12)}_2 g_{\Delta_a,\Delta_b,\Delta_c}(z_1,z_2,z_3) & = \Delta_a (\Delta_a-1) g_{\Delta_a,\Delta_b,\Delta_c}(z_1,z_2,z_3)\,,
\\
\mathcal{C}^{(34)}_2 g_{\Delta_a,\Delta_b,\Delta_c}(z_1,z_2,z_3) & = \Delta_b (\Delta_b-1) g_{\Delta_a,\Delta_b,\Delta_c}(z_1,z_2,z_3)\,,
\\
\mathcal{C}^{(56)}_2 g_{\Delta_a,\Delta_b,\Delta_c}(z_1,z_2,z_3) & = \Delta_c (\Delta_c-1) g_{\Delta_a,\Delta_b,\Delta_c}(z_1,z_2,z_3)\,.
\end{align}
In order to solve these equations, we give the ansatz
\begin{equation}
g_{\Delta_{a}, \Delta_b, \Delta_c}\left(z_1\,, z_2\,,z_3 \right) = z_1 ^{\Delta_a} z_2^{\Delta_b} z_3^{\Delta_c} \sum_{n_1, n_2, n_3} c_{n_1, n_2, n_3} z_1^{n_1} z_2^{n_2} z_3^{n_3} \,,
\end{equation}
and since we are only interested in extracting low-lying CFT data, we content ourselves with a handful of low-lying coefficients:
\begin{align} 
c_{0,0,0} & = 1\,,
\\
c_{1,0,0} & = \frac{(-2 a_1 + \Delta_a ) (\Delta_a + \Delta_b - \Delta_c)}{2 \Delta_a}\,,
\\
c_{0,1,0} & = \frac{(-2 a_2 + \Delta_b ) (\Delta_b + \Delta_c - \Delta_a)}{2 \Delta_b}\,,
\\
c_{0,0,1} & = \frac{(-2 a_3 + \Delta_c ) (\Delta_c + \Delta_a - \Delta_b)}{2 \Delta_c}\,,
\\
c_{1,1,0} & = -\frac{(-2 a_1 + \Delta_a)(-2 a_2 + \Delta_b)(1 + \Delta_a - \Delta_b - \Delta_c)(\Delta_a + \Delta_b - \Delta_c)}{4 \Delta_a \Delta_b}\,,
\\
c_{0,1,1} & = -\frac{(-2 a_2 + \Delta_b)(-2 a_3 + \Delta_c)(1 + \Delta_b - \Delta_c - \Delta_a)(\Delta_b + \Delta_c - \Delta_a)}{4 \Delta_b \Delta_c}\,,
\\
c_{1,0,1} & = -\frac{(-2 a_3 + \Delta_c)(-2 a_1 + \Delta_a)(1 + \Delta_c - \Delta_a - \Delta_b)(\Delta_c + \Delta_a - \Delta_b)}{4 \Delta_c \Delta_a}\,.
\end{align}
We refer the reader to \cite{Fortin:2020zxw} for a more detailed analysis of the snowflake channel, and for a closed-form expression for the $c_{n_1,n_2,n_3}$ coefficients (albeit in a different convention).

\bibliography{./auxi/Notes.bib}
\bibliographystyle{./auxi/JHEP}

\end{document}